\documentclass{aa} 
\usepackage{amsmath}
\usepackage{textcomp}
\usepackage{longtable}
\usepackage{pdfpages}
\usepackage{subfig}
\usepackage{graphicx}
\usepackage{lscape}
\usepackage{txfonts}
\usepackage{natbib}
\bibpunct{(}{)}{;}{a}{}{,}

\usepackage{color}

\begin{document}

   \title{Radii, masses, and ages of 18 bright stars using interferometry \\
and new estimations of exoplanetary parameters }

   \author{R.Ligi\inst{1}$^{,}$\thanks{Now at Aix Marseille Universit\'e, CNRS, LAM (Laboratoire d'Astrophysique de Marseille), UMR 7326, 13388, Marseille, France}, O. Creevey\inst{1,2}, D. Mourard\inst{1}, A. Crida\inst{1}, A.-M. Lagrange\inst{3}, N. Nardetto\inst{1}, K. Perraut\inst{3}, M. Schultheis\inst{1}, I. Tallon-Bosc\inst{4}, and T. ten Brummelaar\inst{5}.}

        \institute{Laboratoire Lagrange, Universit\'e C\^ote d'Azur, Observatoire de la C\^ote d'Azur, CNRS, Boulevard de l'Observatoire, CS 34229, 06304 Nice Cedex 4, France \\
    \email{roxanne.ligi@oca.eu} 
        \and  
        Institut d'Astrophysique Spatiale, CNRS, UMR 8617, Universit\'e Paris XI, B\^atiment 121, F-91405 Orsay Cedex
         \and
   Univ. Grenoble Alpes, IPAG, F-38000 Grenoble, France \\ CNRS, IPAG, F-38000 Grenoble, France
   	\and
   	 UCBL/CNRS CRAL, 9 avenue Charles Andr\'e, 69561 Saint-Genis-Laval Cedex, France
   	  \and
   Georgia State University, PO Box 3969, Atlanta GA 30302-3969, USA  
        }

   \date{Received 24 July 2015 / Accepted 16 October 2015}


  \abstract
   {Accurate stellar parameters are needed in numerous domains of astrophysics. The position of stars on the Hertzsprung-Russell diagram is an important indication of their structure and evolution, and it helps improve stellar models. Furthermore, the age and mass of stars hosting planets are required elements for studying exoplanetary systems. }
   {We aim at determining accurate parameters of a set of 18 bright exoplanet host and potential host stars from interferometric measurements, photometry, and stellar models.}
   {Using the VEGA/CHARA interferometer operating in the visible domain, we measured the angular diameters of 18 stars, ten of which host exoplanets. We combined them with their distances to estimate their radii. We used photometry to derive their bolometric flux and, then, their effective temperature and luminosity to place them on the H-R diagram. We then used the PARSEC models to derive their best fit ages and masses, with error bars derived from Monte Carlo calculations.}
   {Our interferometric measurements lead to an average of $1.9 \%$ uncertainty on angular diameters and $3\%$ on stellar radii. There is good agreement between measured and indirect estimations of angular diameters (either from SED fitting or from surface brightness relations) for main sequence (MS) stars, but not as good for more evolved stars. For each star, we provide a likelihood map in the mass-age plane; typically, two distinct sets of solutions appear (an \emph{\emph{old}} and a \emph{\emph{young}} age). The errors on the ages and masses that we provide account for the metallicity uncertainties, which are often neglected by other works. From measurements of its radius and density, we also provide the mass of 55 Cnc independently of models. From the stellar masses, we provide new estimates of semi-major axes and minimum masses of exoplanets with reliable uncertainties. We also derive the radius, density,  and mass of 55~Cnc~e, a super-Earth that transits its stellar host. Our exoplanetary parameters reflect the known population of exoplanets.}
        {This work illustrates how precise interferometric measurements of angular diameters and detailled modeling allow fundamental parameters of exoplanet host stars to be constrained at a level permiting analysis of the planet's parameters.}

               
\authorrunning{Ligi et al.}
   \maketitle
%

\section{Introduction}
\label{sec:Introduction}

Stellar parameters are essential for understanding stellar physics and evolution, and they are applied to many domains of astrophysics.
Several parameters are not directly measurable for single stars, such as the mass and age, and are thus derived from evolutionary models. These models are constrained by direct measurements that come from photometry, spectroscopy, or asteroseismology, for example. In particular, the radius is a fundamental parameter that takes part in the determination of many other parameters. 
\cite{Torres2010} highlight that the accuracy on stellar mass and radius should reach $1-3\%$ to constrain stellar models. Directly determining the radius together with seismic parameters brings strong constraints on ages \citep[$< 10\%$, see, e.g.,][]{Lebreton2014} and stellar masses \citep{Creevey2007}. It also removes the degeneracy brought by stellar models on derived parameters. For example, \cite{Guillot2011} calculated the parameters of the star CoRoT-2 without any direct measurement of the radius. This parameter thus becomes a model output, in the same way as for the mass and age.  

One of the most accurate ways to obtain stellar radii is to measure the angular diameter of stars (whose distance is known) using interferometry. Indeed, this method provides a direct measurement that reaches $1-2\%$ precision on the angular diameter. The radius is then determined in a way that is as independent of models as possible.
Knowing the distance and bolometric flux allows an inference of the stellar luminosity and effective temperature with little dependence on stellar atmospheres. These two quantities place stars on the Hertzsprung-Russell (H-R) diagram. Then, the generally accepted method used to determine the mass and age is to interpolate stellar evolutionary models or isochrones according to this location on the diagram. Also, stellar models and their improvement are based on the accurate location of stars on the H-R diagram.
Moreover, age and mass determination of stars for which the radius is known is more reliable than only using photometry. It constitutes an important resource for benchmark stars, which are also used, for example, to study age, rotation, atmospheres, and activity relations, and to determine stellar gravities. 

Fundamental parameters of stars are also essential for studying exoplanets. Indeed, the two main methods dedicated to exoplanet search are (i) the radial velocity (RV) method, which provides the mass function $(m_{\rm p} \sin(i))^3/(M_\star + m_{\rm p})^2$, and by supposing $m_p \ll M_\star$, it gives the ratio $m_{\rm p} \sin(i)/M_\star$, and (ii) the transit method, which provides the ratio of the planetary and stellar radii $R_{\rm p}/R_\star$ and the mean stellar density. Combining this density and RV data yields the surface gravity of the planet. Transit measurements and RV are known with very good accuracy, which is significantly better than the accuracy on stellar parameters. Thus, to better constrain exoplanetary parameters, better constraints on stellar parameters are needed. 

Unlike previous spatial missions, such as CoRoT and \textit{Kepler} which observed faint stars, the future will be marked by the discovery and characterization of exoplanets around bright stars, meaning those accessible with interferometry. For example, CHEOPS \citep{Broeg2013} will characterize exoplanets around already known stars brighter than $m_{\rm V}=13$, and TESS \citep{Ricker2014} will also discover transiting exoplanets around $4<m_{\rm V}<12$ stars. The PLATO 2.0 mission \citep{Rauer2012} will search for exoplanets around bright ($4<m_{\rm V}<11$) stars and characterize the host stars using asteroseismology. However, scientific exploitation of these missions will be greatly reinforced by independent measurements from the ground that allow the host star to be characterized. In particular, we will get a model-independent measure of the radius along with asteroseismic data, thus some parameters do not need to be fixed from stellar models, as for the mixing-length parameter that can then be calibrated. Furthermore, such measurements allow building very useful surface brightness (SB) relations.

When the number of angular diameters measured with interferometry increases, SB relations can be constrained with precisions on angular diameters of a few percent \citep[$1-2\%$, see e.g.][]{Kervella2004, Challouf2014,Boyajian2012a}. These relations still have to be improved to increase their precision (for example with homogeneous photometric measurements, in particular in the infrared domain) and accuracy (considering the stellar class for instance). Indeed, they are seldom calibrated for giants and supergiants or stars that do not have solar metallicity. However, more and more exoplanets are found around these stars. 

In this paper, we present an interferometric survey of 18 stars dedicated to the measurement of their angular diameters. Section~\ref{sec:Observations} describes the interferometric observations performed with VEGA/CHARA, and in Sect.~\ref{sec:StellarParameters}, we explain how we derived the stellar parameters (limb-darkened diameter, radius, bolometric flux, effective temperature, metallicity, and gravity) from direct measurements. In Sect.~\ref{sec:MassAge}, we explain how we estimate the mass and age of exoplanet host stars and discuss the results in Sect.~\ref{sec:Discussion}. Finally, using our estimations of stellar masses, we present new values of minimum masses of planets and also a new estimation of the parameters of the transiting super-Earth 55~Cnc~e in Sect.~\ref{sec:Exoplanets}.

\section{From visibilities to angular sizes}
\label{sec:Observations}

We performed interferometric measurements from 2010 to 2013 using the VEGA/CHARA instrument with three telescopes at medium spectral resolution and at observing wavelengths generally between 650 and 730~nm. We observed a spread of F, G, or K stars, including exoplanet hosts (see Table~\ref{tab:propertiesStars}). The selection of the host stars is described by \cite{Ligi2012SPIE}. In summary, we chose main sequence (MS), subgiant, and giant stars from the Extrasolar Planets Encyclopedia\footnote{\texttt{exoplanet.eu}} and sorted them according to their observability with VEGA (in term of coordinates and magnitude). We then selected stars with an expected accuracy on the angular diameter of less than $2\%$, that is with expected angular diameters between 0.3 and 3 milliarcseconds (mas). In total, ten exoplanet host stars were observed. The first four stars observed for this project (\object{14 And}, \object{$\upsilon$~And}, \object{42 Dra}, and \object{$\theta$~Cyg}) are presented by \citet[][hereafter Paper I]{Ligi2012}. Since the method used here to derive the stellar parameters differs from the one used in Paper I, we also include these stars in the present study. We also selected seven stars not known to host exoplanets and used the same criteria. The visibility curves are shown in Fig.~\ref{fig:VisHostStars} for exoplanet host stars (except for the first four stars, see Paper I) and in Fig.~\ref{fig:VisSingleStars} for the other stars.

\begin{table}
\caption{Properties of the stars of the studied sample. \hfill \ }
\centering
\begin{tabular}{l c c c c}
\hline
\hline
HD      &       Name    &       Sp. Class       &       $m_{\rm V}$     &       $m_{\rm K}$      \\
\hline
3651    &       \object{54 Psc}  &       K0V     &       5.88    &       4.00    \\
9826    &       \object{$\upsilon$ And}  &       F8V     &       4.09    &       2.86    \\
19994   &       \object{94 Cet}  &       F8V     &       5.07    &       3.75    \\
75732   &       \object{55 Cnc}  &       G8V     &       5.96    &       4.05    \\
167042$^\dagger$        &       -               &       K1IV    &       5.97    &       3.58    \\
170693  &       \object{42 Dra}  &       K1.5III &       4.82    &       2.09    \\
173416  &                       &       G8III   &       6.04    &       3.85    \\
185395  &       \object{$\theta$ Cyg}            &       F4V     &       4.49    &       3.54    \\
190360  &       -               &       G6IV    &       5.73    &       4.11    \\
217014  &       \object{51 Peg}  &       G2IV    &       5.45    &       3.91    \\
221345  &       \object{14 And}          &       K0III   &       5.23    &       2.33    \\
\hline
1367    &       \object{$\rho$ And}      &       K0II    &       6.18    &       4.23    \\
1671    &       -               &       F5III   &       5.15    &       4.07    \\
154633  &       -               &       G5V     &       6.11    &       3.93    \\
161178  &       -               &       G9III   &       5.87    &       3.66    \\
168151  &       -               &       K5V     &       4.99    &       3.94    \\
209369  &       \object{16 Cep}  &       F5V     &       5.03    &       3.96    \\
218560  &       -               &       K0III   &       6.21    &       3.70    \\
\hline
\end{tabular}
\tablefoot{The eleven first ones (above the line) are known exoplanet hosts \citep[except $\theta$~Cyg, see][]{Ligi2012}. Photometry is given in the Johnson system (see Sect.~\ref{sec:Observations}). $^\dagger$HD167042 was classified as a K1III giant by Hipparcos, but \cite{Sato2008} classify it as a less evolved subgiant because of its position on the H-R digram.}
\label{tab:propertiesStars}
\end{table}

VEGA \citep{Mourard2009, Ligi2013} is a spectro-interferometer working at visible wavelengths ([450-850] nm) at medium (6000) or high (30000) spectral resolution. It takes advantage of the CHARA \citep{tenBrummelaar2005} baselines, which range from 34 to 331 m, to reach a maximum spatial resolution of $\sim 0.3$ mas. VEGA is able to recombine the light coming from two to four of the six one-meter telescopes hosted by the CHARA array. The telescopes are arranged in a Y shape, which allows a wide range of orientations and thus a rich ($u, v$) coverage.

Interferometry is a high angular resolution technique that measures the Fourier transform of the brightness distribution of a source, called the complex visibility. Its argument, the visibility, depends on the source size. The simplest representation of a star is the uniform disk (UD), where its intensity is considered uniform over the stellar disk. In this case, the corresponding squared visibility ($V^2$) can be written as\,

\begin{equation}
V^2(x) = \left| \frac{2J_{1}(x)}{x}\right| ^2 \ ,
\end{equation}
where $J_{1}(x)$ represents the first-order Bessel function and 
\begin{equation}
x = \frac{\pi B \theta_{UD}}{ \lambda} \ ,
\end{equation}
where $B$ is the projected baseline length, $\lambda$ the observing wavelength, and $\theta_{\rm UD}$ the UD diameter (in rad) \citep[see, for instance,][]{Ligi2015}.

Each measurement was calibrated by stars found in the \texttt{SearchCal} tool\footnote{Available at \texttt{http://www.jmmc.fr/searchcal}} \citep{Bonneau2006} and followed the sequence $C-T-C$ ($C$ referring to the calibrator and $T$ to the science target), several times in a row when possible. Calibrators were chosen by excluding variable stars and stars in multiple systems, and we used their estimated spectrophotometric UD diameter in the R band (Johnson-Cousin system) to calibrate the raw squared visibility of the targets. We chose stars ideally unresolved by VEGA, i.e. whose $V^2$ is close to unity, which allows an accurate measurement of the transfer function of the instrument. The properties of the calibrators are given in Table~\ref{tab:calibrators}, and we label them by a number to clarify the observing sequences given in the journal of observations (available online). For example, the first observing sequence of HD3651 used Stars $\sharp$1 and $\sharp$2 as calibrators. A summary of the observations is given in Table~\ref{tab:resumeObsLog}. For the observing journal of $\upsilon$~And, 42~Dra, 14~And, and $\theta$~Cyg, please refer to Paper I.

\begin{table}
\caption{Properties of the calibrators used for VEGA observations, from the \texttt{SearchCal} catalog.\hfill \ }
\centering
\begin{tabular}{l l c c c}
\hline \hline
$\sharp$ & Star         &       $m_{\rm V}$     &       $m_{\rm K}$     &       $\theta_{\rm UD,R}$  \\
\hline
1 & HD560               &       5.535   &       5.700   &       0.185   $\pm$   0.013   \\
2 & HD14191             &       5.571   &       5.390   &       0.251   $\pm$   0.018   \\
3 & HD10982             &       5.852   &       5.891   &       0.180   $\pm$   0.013   \\
4 & HD7804              &       5.137   &       4.921   &       0.322   $\pm$   0.023   \\
5 & HD12573             &       5.423   &       4.995   &       0.339   $\pm$   0.024   \\
6 & HD21790             &       4.727   &       4.886   &       0.270   $\pm$   0.019   \\
7 & HD88960             &       5.485   &       5.387   &       0.246   $\pm$   0.018   \\
8 & HD54801             &       5.747   &       5.366   &       0.268   $\pm$   0.019   \\
9 & HD177003    &       5.376   &       5.895   &       0.130   $\pm$   0.009   \\
10 & HD196740   &       5.051   &       5.401   &       0.176   $\pm$   0.012   \\
11 & HD190993   &       5.068   &       5.566   &       0.158   $\pm$   0.011   \\
12 & HD204414   &       5.382   &       5.290   &       0.250   $\pm$   0.018   \\
13 & HD13869    &       5.245   &       5.228   &       0.247   $\pm$   0.018   \\
14 & HD10390    &       5.629   &       5.777   &       0.175   $\pm$   0.012   \\
15 & HD214680   &       4.882   &       5.498   &       0.150   $\pm$   0.011   \\
16 & HD149212   &       4.959   &       4.961   &       0.278   $\pm$   0.020   \\
17 & HD145454   &       5.436   &       5.431   &       0.227   $\pm$   0.016   \\
18 & HD149681   &       5.552   &       4.989   &       0.353   $\pm$   0.025   \\
19 & HD204770   &       5.408   &       5.602   &       0.168   $\pm$   0.012   \\
20 & HD219485   &       5.882   &       5.872   &       0.175   $\pm$   0.012   \\
\hline
\end{tabular}
\tablefoot{ Angular diameters are expressed in mas, and the error is given for the limb-darkened diameter. Magnitudes correspond to Johnson filters (see Sect.~\ref{sec:Observations}).}
\label{tab:calibrators}
\end{table}

From the measured visibilities, we derived $\theta_{\rm UD}$ for each star using the \texttt{LITpro} software\footnote{Available at \texttt{http://www.jmmc.fr/litpro$\_$page.htm}} \citep{Tallon2008}. These determinations are based on a large number of data points for most of the stars. Most importantly, they correspond to low $V^2$ and are thus very constraining for angular diameter adjustments. For the other stars, data of bad quality or difficulties getting optimal observing strategies prevented us from getting more data. The errors on angular diameters were also calculated by \texttt{LITpro}. To correctly account for instrumental noises in addition to the photon noise and to prevent bias in the calibration process, we decided to fix the lower limit on the uncertainty on the raw squared visibility measurement to 0.05. This method is conservative as shown by \cite{Mourard2012,Mourard2015} but provides reliable uncertainties on angular diameters.
We obtained an average accuracy of $1.9\%$ on angular diameters, which is in good agreement with the requirements made when choosing the targets (see all the results in Table~\ref{tab:diametres}).

\begin{table}
\centering
\caption{Summary of the observing journal (see Sect.~\ref{sec:Observations}).\hfill \ }
\label{tab:resumeObsLog}
\begin{tabular}{l  c c}
\hline \hline
Star            & Nbr. of       & Cal. \\
                        & data pts. &  \\
\hline
\object{HD3651}          & 26    & 1, 2, 3 \\
\object{HD19994}         & 27    & 4, 5, 6 \\
\object{HD75732}         & 20    & 7, 8 \\
\object{HD167014}        & 15    &        9 \\
\object{HD173416}        & 10    &        10 \\
\object{HD190360}        & 19    & 10, 11, 12 \\
\object{HD217014}        & 13    & 12, 1 \\
\object{HD1367}          & 10    & 4, 5 \\
\object{HD1671}          &  24   &       13 ,14, 15 \\
\object{HD154633}        & 11    & 16, 17 \\
\object{HD161178}        & 7             & 16, 18 \\
\object{HD168151}        & 12    & 16, 17 \\
\object{HD209369}        & 11    & 19, 20 \\
\object{HD218560}        & 11    & 19 \\
\hline
\end{tabular}
\end{table}

\section{Stellar parameters from direct measurements}
\label{sec:StellarParameters}

\subsection{Bolometric flux}
\label{sec:Fbol}

\begin{figure*}
\centering
\includegraphics[scale=0.7]{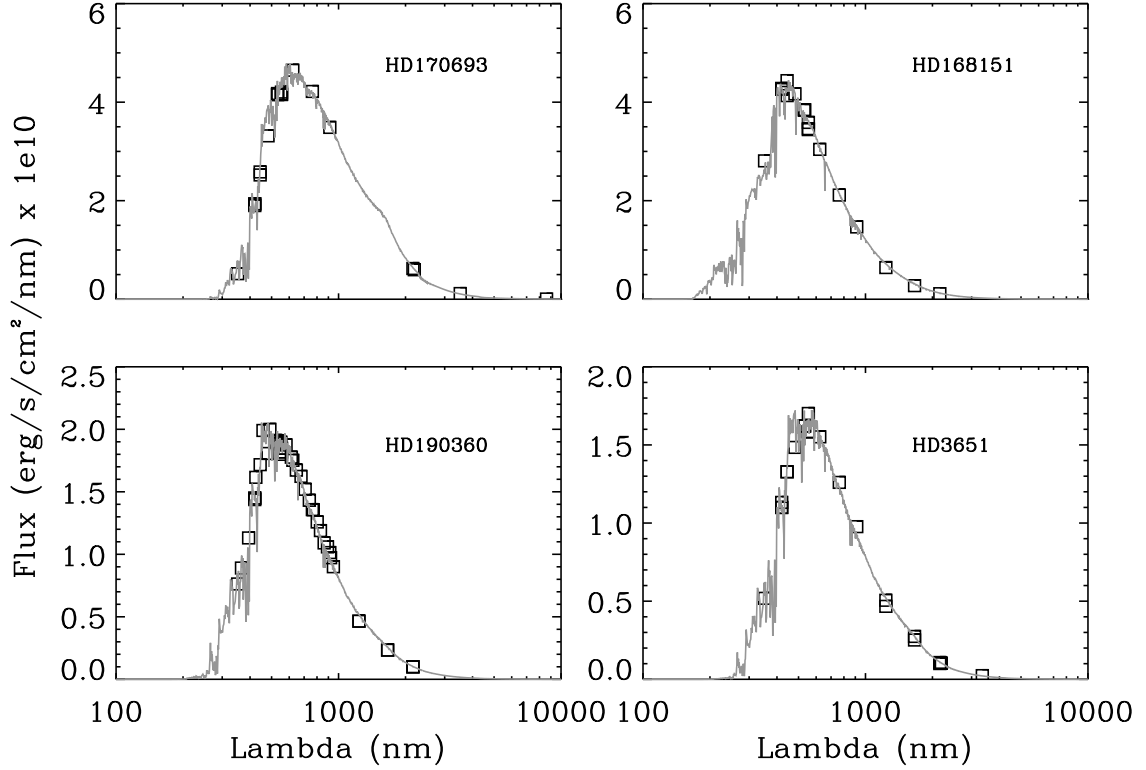}
\caption{Photometric energy distribution of four stars of the sample. Squares represent the photometric points and the gray curve represents the fitted spectrum (see Sect.~\ref{sec:Fbol}).}
\label{fig:flux}
\end{figure*}

We determined the bolometric flux $F_{\rm bol}$ by fitting libraries of model spectra to the literature photometry-converted-to-flux measurements. Figure~\ref{fig:flux} gives an example of the photometric energy distribution for four stars from our sample with their fits. The data were obtained using the VizieR Photometry viewer\footnote{\texttt{http://vizier.u-strasbg.fr/vizier/sed/}}  (data available on request).
We used the BASEL empirical library of spectra \citep{lej97}, which covers a 
wavelength range between 9.1 and 160~000 nm.
The data-fitting method incorporates a nonlinear least-squares 
minimization algorithm (Levenberg-Marquardt) 
to find the optimal scaled interpolated spectrum that fits a set of  
observed flux points. The method is described in detail in \cite{Creevey2015}
and validated by comparing it with a second approach, originally described in \citet{vanBelle2008}, and using different stellar libraries.
The fitting method requires on input a set of 
parameters --- $T_{\rm eff}$, [Fe/H], $\log(g)$,
A$_V$, and a scaling factor --- 
which define the characteristics of 
the spectra, the extinction to apply, and ratio of the stellar radius to 
the star's distance, which we denote as $\theta_{\rm SED}$ to imply it comes from the models.  

\begin{table}[h]
\caption{Angular diameters of our targets (in mas). \hfill \ }
\centering
\begin{tabular}{l c c c c }
\hline \hline
HD              & $\theta_{\rm UD} \pm \sigma \theta_{\rm UD}$  &       $\mu_\lambda$   &       $\theta_{\rm LD} \pm \sigma \theta_{\rm LD} (\%) $           &       $\chi^2_{\rm red}$ \\
\hline
3651    & 0.687 $\pm$   0.007   &       0.537   &       0.722   $\pm$   0.007   (0.97)  &       0.97    \\
9826    & 1.119 $\pm$   0.026   &       0.425   &       1.161   $\pm$   0.027   (2.34)  &       6.95    \\
19994   & 0.731 $\pm$   0.010   &       0.448   &       0.761   $\pm$   0.011   (1.41)  &       0.67    \\
75732   & 0.687 $\pm$   0.011   &       0.561   &       0.724   $\pm$   0.012   (1.64)  &       0.36    \\
167042  & 0.998 $\pm$   0.013   &       0.616   &       1.056   $\pm$   0.014   (1.28)  &       0.30    \\
170693  & 1.965 $\pm$   0.009   &       0.634   &       2.097   $\pm$   0.009   (0.41)  &       0.20    \\
173416  & 0.937 $\pm$   0.033   &       0.608   &       0.995   $\pm$   0.034   (3.45)  &       0.59    \\
185395  & 0.726 $\pm$   0.007   &       0.355   &       0.749   $\pm$   0.008   (1.01)  &       8.47    \\
190360  & 0.596 $\pm$   0.006   &       0.480   &       0.622   $\pm$   0.007   (1.08)  &       1.00    \\
217014  & 0.624 $\pm$   0.013   &       0.458   &       0.650   $\pm$   0.014   (2.14)  &       2.27    \\
221345  & 1.404 $\pm$   0.029   &       0.614   &       1.489   $\pm$   0.032   (2.16)  &       2.73    \\
1367    & 0.719 $\pm$   0.013   &       0.505   &       0.754   $\pm$   0.014   (1.84)  &       0.44    \\
1671    & 0.582 $\pm$   0.006   &       0.359   &       0.600   $\pm$   0.006   (0.92)  &       0.42         \\
154633  & 0.763 $\pm$   0.011   &       0.569   &       0.804   $\pm$   0.012   (1.44)  &       0.33    \\
161178  & 0.897 $\pm$   0.040   &       0.545   &       0.944   $\pm$   0.043   (4.50)  &       1.89    \\
168151  & 0.642 $\pm$   0.014   &       0.386   &       0.664   $\pm$   0.015   (2.20)  &       0.61    \\
209369  & 0.601 $\pm$   0.017   &       0.380   &       0.621   $\pm$   0.018   (2.85)  &       1.72    \\
218560  & 0.875 $\pm$   0.020   &       0.600   &       0.927   $\pm$   0.022   (2.38)  &       0.64    \\
\hline
\end{tabular}
\tablefoot{Errors in $\%$ are given in parenthesis (see Sect.~\ref{sec:LDDandTeff}).}
\label{tab:diametres}
\end{table}

\begin{figure}
\includegraphics[scale=0.5]{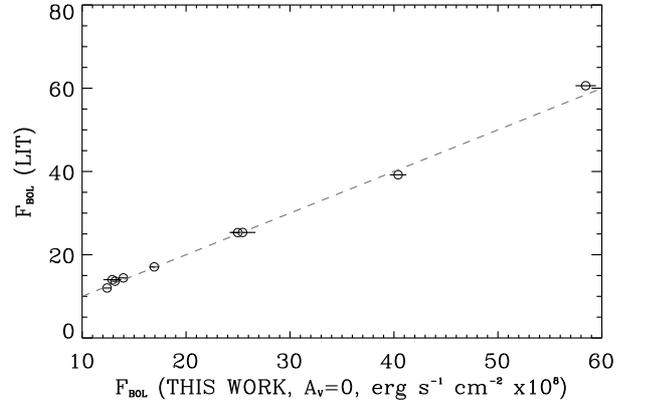}
\caption{Comparison between $F_{\rm bol}$ found in the literature and $F_{\rm bol}$ calculated in this work, for A$_V$ = 0 (see Sect.~\ref{sec:Fbol}).}
\label{fig:compfbol}
\end{figure}

We compiled the $\log(g)$ and [Fe/H] from the literature and used the mean of these values as fixed input parameters to the fitting method. The errors in $\log(g)$ are taken as the standard deviations of the values found in the literature or a minimum of 0.1 dex.
Because the stars are relatively close, the A$_V$ were all set to zero in a first round of analysis. However, for our subsequent analysis we adopted a value of A$_V$ determined using the method described in \cite{Schultheis2014}.  
The bolometric flux is then calculated by integrating the optimal fitted spectrum. 
Table~\ref{tab:InputParam} gives the fixed input $\log(g)$, [Fe/H], and A$_V$, along with the fitted
$T_{\rm eff}$ and $\theta_{\rm SED}$ (scaled to mas). The bolometric
flux used (with A$_V \ne$ 0) is given in the second-to-last column, and the bolometric flux with zero extinction is given in the last column for comparison.    
The errors on the fitted parameters are obtained by adding the formal error from the fit in quadrature, along with the difference in this parameter when changing the fixed parameters by $1\sigma$, and for [Fe/H] we used an error of 0.1 dex.

In Fig.~\ref{fig:compfbol}, we compare the bolometric flux determined using our method assuming A$_V$=0 with that found in the literature, also assuming A$_V$ = 0. The error bars are 2$\sigma$ error bars from our method (for visual purpose), showing good agreement with previous determinations. The literature values were found in \cite{vanBelle2009} and \cite{Boyajian2013b}.

\begin{table*}
\caption{Fixed input parameters to determine the bolometric flux.\hfill \ }
\centering
\begin{tabular}{l|c  c c c c c c}
\hline
\hline
                & \multicolumn{3}{c}{Fixed parameters} & \multicolumn{2}{c}{Fitted parameters} & \multicolumn{2}{c}{Calculated parameters} \\
HD              &  A$_V$                & [Fe/H] & $\log(g)$                    & $T_{\rm eff}$ &         $\theta_{\rm SED}$ & $F_{\rm bol}$              &         $F_{\rm bol}$           \\
                &                       &                &  [cm  s$^2$]  &         [K]                     &               [mas]                   &                                         &               (A$_V$ = 0)             \\      
        \hline
3651    &       0.060   & 0.1   &       4.4 $\pm$ 0.17  &       5297 $\pm$      27      & 0.715 $\pm$ 0.014       &       13.409 $\pm$ 0.236      &       13.163 $\pm$ 0.169           \\
9826    &       0.185   & 0.1   &       4.2     $\pm$ 0.14 &    6494 $\pm$      39      & 1.073 $\pm$ 0.016       &       68.200 $\pm$ 2.310      &       58.448 $\pm$ 0.493           \\
19994   &       0.090   & 0.2   &       4.2     $\pm$ 0.14 &    6039 $\pm$      26      & 0.767 $\pm$ 0.011       &       25.798 $\pm$ 0.654      &       24.980 $\pm$ 0.291           \\
75732   &       0.0075  & 0.3   &       4.4     $\pm$ 0.12 &    5219 $\pm$      26      & 0.709 $\pm$ 0.012       &       12.435 $\pm$ 0.168      &       12.399 $\pm$ 0.168           \\
167042  &       0.103   & -0.1  &       3.2     $\pm$ 0.10 &    4774 $\pm$      33      & 0.958 $\pm$ 0.028       &       15.886 $\pm$ 0.551      &       12.927 $\pm$ 0.429           \\
170693  &       0.052   & -0.5  &       2.1     $\pm$ 0.54 &    4460 $\pm$      24      & 1.933 $\pm$ 0.023       &       49.180 $\pm$ 0.600      &       49.723 $\pm$ 0.102           \\
173416  &       0.047   & -0.2  &       2.5     $\pm$ 0.10 &    4735 $\pm$      23      & 0.917 $\pm$ 0.013       &       13.179 $\pm$ 0.265      &       13.733 $\pm$ 0.148           \\
185395  &       0.328   & 0.0   &       4.3     $\pm$ 0.15 &    7181 $\pm$      28      & 0.775 $\pm$ 0.010       &       49.400 $\pm$ 0.460      &       40.372 $\pm$ 0.403           \\
190360  &       0.044   & 0.2   &       4.3     $\pm$ 0.09 &    5577 $\pm$      26      & 0.669 $\pm$ 0.011       &       14.405 $\pm$ 0.195      &       13.987 $\pm$ 0.213           \\
217014  &       0.078   & 0.2   &       4.3     $\pm$ 0.11 &    5804 $\pm$      27      & 0.689 $\pm$ 0.011       &       17.965 $\pm$ 0.238      &       16.939 $\pm$ 0.241           \\
221345  &       0.046   & -0.3  &       2.4     $\pm$ 0.29 &    4692 $\pm$      25      & 1.359 $\pm$ 0.023       &       27.983 $\pm$ 0.447      &       27.055 $\pm$ 0.418           \\
1367    &       0.588   & 0.0   &       3.0     $\pm$ 0.10 &    5488 $\pm$      23      & 0.725 $\pm$ 0.009       &       15.959 $\pm$ 0.432      &       9.750  $\pm$ 0.060           \\
1671    &       0.473   & -0.1  &       3.7     $\pm$ 0.10 &    7047 $\pm$      27  & 0.619 $\pm$ 0.007    &       31.473 $\pm$ 0.259      &       21.401 $\pm$ 0.185           \\
154633  &       0.046   & -0.1  &       3.0     $\pm$ 0.10 &    4934 $\pm$      24      & 0.788 $\pm$ 0.010       &       12.243 $\pm$ 0.211      &       11.937 $\pm$ 0.087           \\
161178  &       0.408   & -0.2  &       2.4     $\pm$ 0.25 &    5158 $\pm$      26      & 0.885 $\pm$ 0.018       &       19.799 $\pm$ 0.343      &       15.748 $\pm$ 0.078           \\
168151  &       0.129   & -0.3  &       4.1     $\pm$ 0.50 &    6563 $\pm$      38      & 0.679 $\pm$ 0.016       &       28.519 $\pm$ 0.674      &       25.442 $\pm$ 0.625           \\      
209369  &       0.116   & -0.2  &       3.8     $\pm$ 0.10 &    6447 $\pm$      41      & 0.682 $\pm$ 0.017       &       26.737 $\pm$ 0.686      &       24.166 $\pm$ 0.560           \\
218560  &       0.059   & 0             &       1.5     $\pm$ 0.10 &    4631 $\pm$   24      & 0.929 $\pm$ 0.014     &       13.375 $\pm$ 0.138      &       12.800 $\pm$ 0.134             \\
\hline
\end{tabular}
\tablefoot{ $F_{\rm bol}$ is expressed in $10^8$~erg  s$^{-1}$ cm$^{-2}$, and the error adopted in the rest of the study on [Fe/H] is 0.1 dex. We adopt a minimum of 0.1 dex for the error in $\log(g)$ (see Sect.~\ref{sec:Fbol}).}
\label{tab:InputParam}
\end{table*}

\subsection{Limb-darkened diameters, effective temperatures, and radii}
\label{sec:LDDandTeff}

Instead of UD, a more realistic representation of stars is to consider them as limb-darkened (LD) disks, which means that the intensity at the center of the stellar disk is higher than at the limb. The visibility of such objects is expressed by
\begin{equation}
\begin{aligned}
V^2(x) &=  \left(  \frac{1-\mu_\lambda}{2} +  \frac{\mu_\lambda}{3} \right)^{-2}        \\
& \times \left[ \left( 1-\mu_\lambda \right) \frac{J_1(x)}{x} + \mu_\lambda \left( \frac{\pi}{2} \right)^{1/2}  \frac{J_{3/2}(x)}{x^{3/2}} \right]^2 \ ,
\end{aligned}
\end{equation}
where $J_i$ represents the Bessel function at $i$th order, and $\mu_\lambda$ the limb-darkening (LD) coefficient that defines the LD diameter $\theta_{\rm LD}$. 
We used the LD coefficients $\mu_\lambda$ from \cite{Claret2011}, which depend on the metallicity [Fe/H], gravity $\log(g)$, effective temperature $T_{\rm eff,\star}$ of the star, and the considered wavelength. Since we measure very accurate uniform angular diameters, the estimation of the limb-darkening has to be very precise as well, which requires a detailed estimation of the LD coefficients. The coefficients are given by steps of 250~K on $T_{\rm eff}$, 0.5 on $\log(g)$ (where g is in cm/s$^2$), and non-uniform steps in [Fe/H]. Since we observed around 720~nm, we had to consider both R and I filters (in the Johnson-Cousin system).

We first computed linear interpolations over the coefficients corresponding to [Fe/H] and $\log(g)$ surrounding the stellar parameters for each filter R and I and each temperature surrounding the initial photometric temperature (determined from $F_{\rm bol}$) by $\pm 250$~K. (We took the closest
values to our stars available on the tables.) Then, we averaged the resulting LD coefficients on the filters to have one coefficient per temperature. Finally, we computed linear interpolations until the derived $\theta_{\rm LD}$ calculated with the LD coefficient converge with the values of $T_{\rm eff,\star}$ and $F_{\rm bol}$. The final interferometric parameters are given in Table~\ref{tab:diametres}. We used the final LD coefficient to estimate the final $\theta_{\rm LD}$ using the \texttt{LITpro} software. Then, the final $T_{\rm eff,\star}$ is directly derived from the LD diameter and $F_{\rm bol}$\,: 
\begin{equation}
T_{\rm eff,\star} = \left( \frac{4 \times F_{\rm bol}} {\sigma_{\rm SB} \theta_{\rm LD}^2} \right)^{0.25} \ ,
\label{eq:Teff}
\end{equation}
where $\sigma_{\rm SB}$ is the Stefan-Boltzmann constant.

The stellar radius is obtained by combining the LD diameter and the distance $d$ \citep[from \textit{Hipparcos} parallaxes,][]{vanLeeuwen2007}\,:
\begin{equation}
R_{\star_{[R_\odot]}} = \frac{\theta_{\rm LD_{[\rm mas]}} \times d_{[\rm pc]}}{9.305} \ . 
\label{eq:Rayon}
\end{equation}
To determine the errors on $T_{\rm eff,\star}$ and $R_\star$, we consider that the parameters on the righthand side of each equation are independent random variables with Gaussian probability density functions. For any quantity $X$, the uncertainty on its estimate is noted $\sigma_X$, and the relative uncertainty $\sigma_X/X$ is noted $\tilde\sigma_X$. Then, the standard deviation of each parameter that we want to estimate is given analytically to first order by a classical propagation of errors, following the formula\,:
\begin{equation}
\begin{aligned}
\tilde\sigma_{T_{\rm eff,\star}}  &= \sqrt{  \left( (1/2) \times \tilde\sigma_{\theta_{\rm LD}} \right)^2 +  \left( (1/4) \times \tilde\sigma_{F_{\rm bol}} \right)^2  } \\
\tilde\sigma_{R_\star} &= \sqrt{ \tilde\sigma_{\theta_{\rm LD}}^2 +  \tilde\sigma_d ^2 } \ ,
\end{aligned}
\end{equation} 
where $\sigma_{\theta_{\rm LD}}$, $\sigma_{F_{\rm bol}}$, and $\sigma_d$ are the errors on the LD diameter, bolometrix flux, and distance, respectively. Then, we calculate the stellar luminosity $L_\star$ by combining the bolometric flux and the distance\,:
\begin{equation}
L_\star = 4\pi d^2 F_{\rm bol} \ ,
\label{eq:Lum}
\end{equation}
and its error
\begin{equation}
\tilde\sigma_{L_\star} =  \sqrt{  \left( 2 \times \tilde\sigma_d \right)^2 +   \tilde\sigma_{F_{\rm bol}}  ^2  } \ .
\end{equation}
Finally, we calculate the gravitational mass $M_{\rm grav,\star}$ using $\log(g)$ and $R_\star$
\begin{equation}
M_{\rm grav,\star} =  \frac{R_{\star}^2 \times 10^{\log(g)}}{G}
\end{equation}
and its error
\begin{equation}
\tilde \sigma_{M_{\rm grav,\star}} = \sqrt{ \left( 2 \times  \tilde\sigma_{R_\star} \right) ^2   + \left( \sigma_{\log(g)} \times \ln(10) \right)^2 }
.\end{equation}
The parameters and their errors are shown in Table~\ref{tab:finalParameters}.

\subsection{General results}
\label{sec:Results}

\begin{table*}
\caption{Stellar parameters calculated from interferometric measurements and photometry. \hfill \ }
\centering
\begin{tabular}{l c c c c c}
\hline \hline
HD              & $T_{\rm eff,\star} \pm \sigma T_{\rm eff,\star}$      &       $d      \pm     \sigma d       (\%)$   &       $R_\star        \pm     \sigma R_\star  (\%)$   &       $L_\star        \pm     \sigma L_\star$        &       $M_{\rm grav,\star} \pm \sigma M_{\rm grav,\star}$ \\
                &[K]                                                                    & [pc]                                            &               [$R_{\odot}$]                   & [$L_{\odot}$]           &[$M_{\odot}$] \\
\hline
3651    &       5270    $\pm$   34      & 11.06 $\pm$   0.04    (0.35)  &       0.8584  $\pm$   0.0088  (1.03)  &       0.5103  $\pm$   0.0097  &       0.68 $\pm$ 0.26              \\
9826    &       6243    $\pm$   74      & 13.49 $\pm$   0.03    (0.26)  &       1.684   $\pm$   0.040   (2.36)  &       3.866   $\pm$   0.030   &       1.64 $\pm$ 0.53              \\
19994   &       6063    $\pm$   45      & 22.58 $\pm$   0.14    (0.63)  &       1.847   $\pm$   0.028   (1.54)  &       4.138   $\pm$   0.066   &       1.97 $\pm$ 0.64              \\
75732   &       5165    $\pm$   46      & 12.34 $\pm$   0.11    (0.93)  &       0.960   $\pm$   0.018   (1.89)  &       0.589   $\pm$   0.014   &       0.85 $\pm$ 0.24      \\
167042  &       4547    $\pm$   49      & 50.23 $\pm$   0.66    (1.31)  &       5.70    $\pm$   0.10    (1.83)  &       12.47   $\pm$   0.54    &       1.88 $\pm$ 0.07              \\
170693  &       4280    $\pm$   16      & 96.53 $\pm$   1.86    (1.93)  &       21.76   $\pm$   0.43    (1.97)  &       142.55  $\pm$   5.77    &       2.17 $\pm$ 2.70      \\
173416  &       4544    $\pm$   79      & 139.5 $\pm$   5.5             (3.91)  &       14.92   $\pm$   0.78    (5.21)  &       85.17   $\pm$   6.72    &       2.57 $\pm$ 0.27              \\
185395  &       7305    $\pm$   42      & 18.34 $\pm$   0.05    (0.28)  &       1.475   $\pm$   0.015   (1.05)  &       5.564   $\pm$   0.065   &       1.58 $\pm$ 0.55              \\
190360  &       5781    $\pm$   37      & 15.86 $\pm$   0.09    (0.54)  &       1.061   $\pm$   0.013   (1.21)  &       1.127   $\pm$   0.019   &       0.82 $\pm$ 0.17              \\
217014  &       5978    $\pm$   67      & 15.61 $\pm$   0.09    (0.59)  &       1.090   $\pm$   0.024   (2.22)  &       1.362   $\pm$   0.024   &       0.86 $\pm$ 0.22              \\
221345  &       4483    $\pm$   50      & 79.18 $\pm$   1.69    (2.14)  &       12.67   $\pm$   0.39    (3.04)  &       58.18   $\pm$   2.55    &       1.47 $\pm$ 0.99              \\
1367    &       5382    $\pm$   51      & 117.4 $\pm$   5.8             (4.93)  &       9.51    $\pm$   0.50    (5.26)  &       68.14   $\pm$   6.74    &       3.30 $\pm$ 0.35              \\
1671    &       7133    $\pm$   36      & 48.54 $\pm$   0.49    (1.02)  &       3.151   $\pm$   0.043   (1.37)  &       23.07   $\pm$   0.51    &       1.82 $\pm$ 0.05              \\
154633  &       4883    $\pm$   36      & 115.7 $\pm$   3.5             (3.01)  &       10.00   $\pm$   0.33    (3.34)  &       51.02   $\pm$   3.09    &       3.65 $\pm$ 0.24              \\
161178  &       4992    $\pm$   115     & 111.2 $\pm$   3.1             (2.78)  &       11.29   $\pm$   0.60    (5.29)  &       71.03   $\pm$   4.16    &       1.17 $\pm$ 0.68              \\
168151  &       6638    $\pm$   83      & 22.92 $\pm$   0.09    (0.39)  &       1.635   $\pm$   0.037   (2.23)  &       4.66    $\pm$   0.12    &       1.23 $\pm$ 0.15              \\
209369  &       6755    $\pm$   106     & 36.72 $\pm$   0.23    (0.62)  &       2.45    $\pm$   0.072   (2.92)  &       11.22   $\pm$   0.32    &       1.38 $\pm$ 0.08              \\
218560  &       4637    $\pm$   56      & 324.7 $\pm$   33.7    (10.39) &       32.39   $\pm$   3.45    (10.66) &       434             $\pm$   90              &       1.21 $\pm$ 0.26              \\
\hline
\end{tabular}
\tablefoot{Errors in $\%$ are given in parenthesis (see Sect.~\ref{sec:LDDandTeff}).}
\label{tab:finalParameters}
\end{table*}

\begin{figure}
\centering
\includegraphics[scale=0.5]{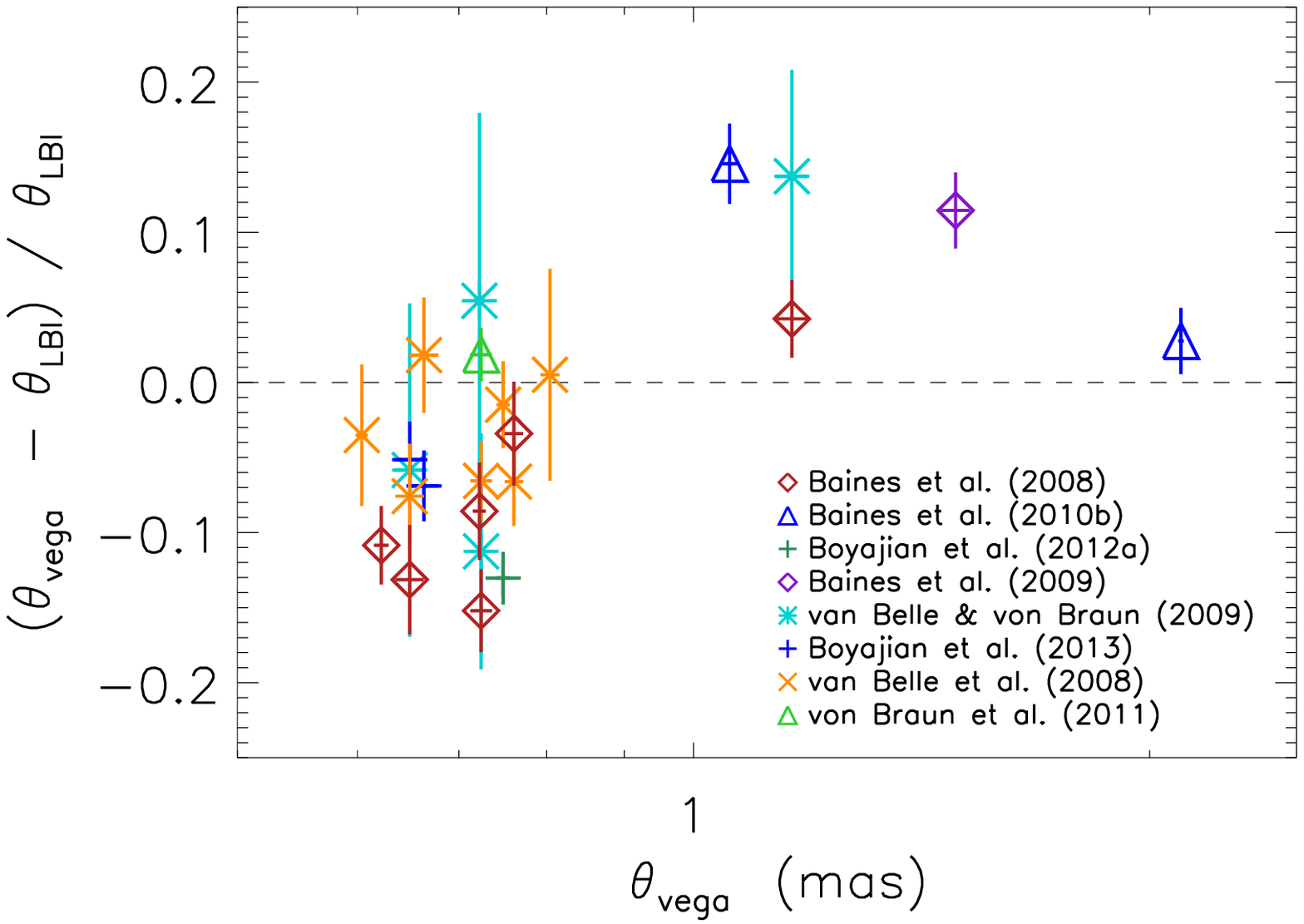} \\
\includegraphics[scale=0.5]{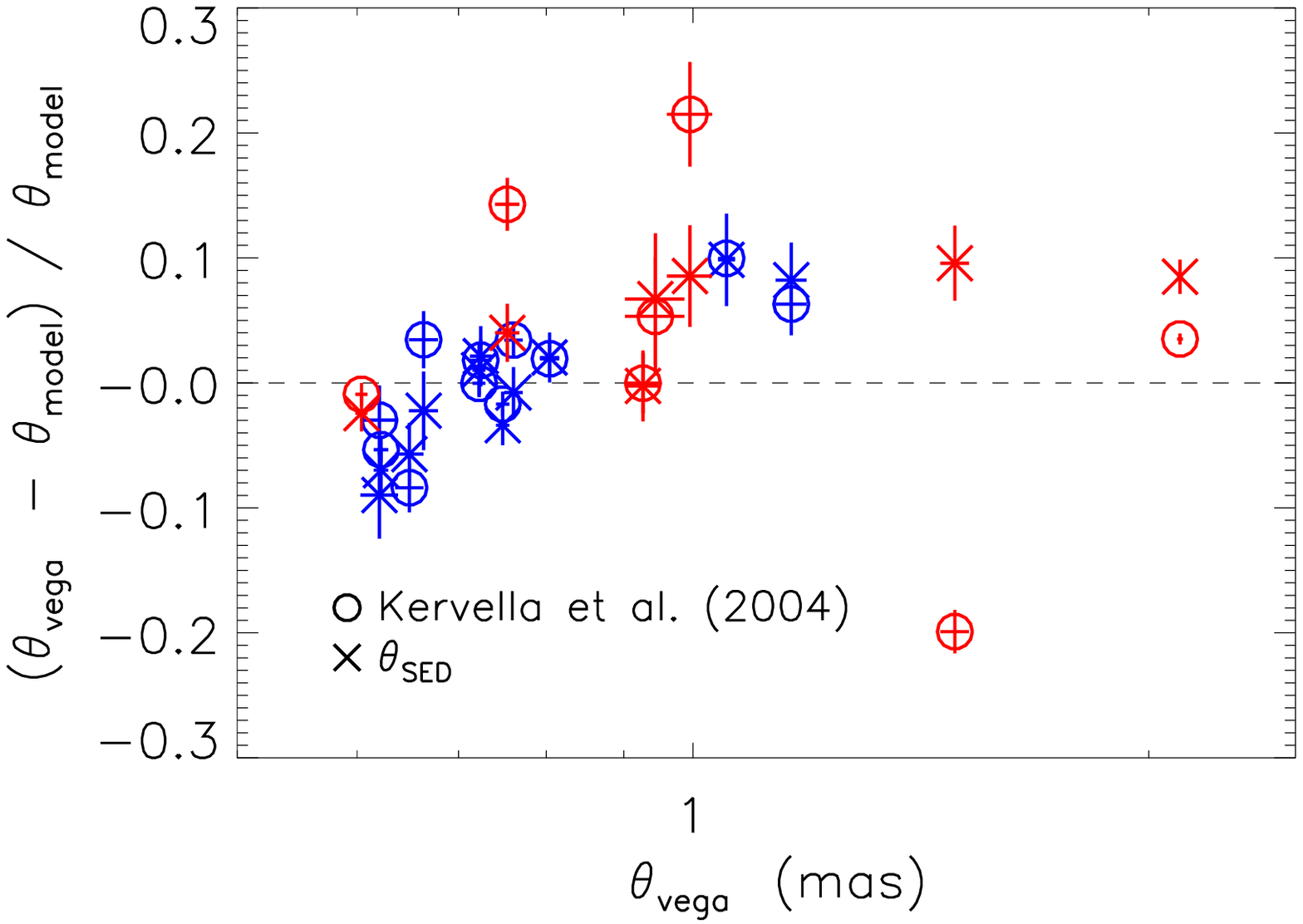}
\caption{\textit{Upper panel}\,: Comparison between angular diameters measured with VEGA and with other instruments. \textit{Bottom panel}\,: Estimation of empirically determined angular diameters versus angular diameters measured with VEGA. Dwarfs and subgiants stars are plotted in blue, and giants and bright giants in red (see Sect.~\ref{sec:Results}).}
\label{fig:diamLD}
\end{figure}

\begin{figure}
\centering
\includegraphics[scale=0.5]{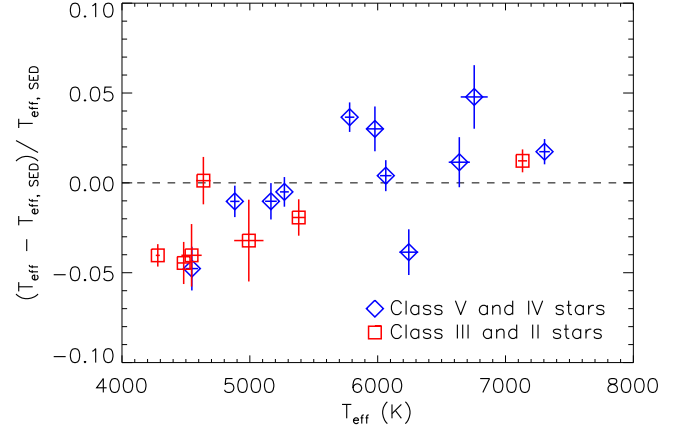}
\caption{Comparison between interferometric temperatures and temperatures derived from SED. Dwarfs and subgiants stars are plotted with blue diamonds, and giants and bright giants with red squares (see Sect.~\ref{sec:Results}).}
\label{fig:teff}
\end{figure}

Our interferometric measurements led to precise $\theta_{\rm LD}$, $T_{\rm eff,\star}$, and radii ($1.9\%$, $57$~K, and $3\%$ average accuracies with medians of $1.8 \%$, $1.1 \%$, and $2.2 \%$, respectively). 
Some of the exoplanet host stars presented here, along with a few non-host stars, have already been observed with other long baseline interferometers (LBI), including the CLASSIC/CHARA \citep{Sturmann2003} and the Palomar Testbed Interferometer \cite[PTI,][]{Colavita1999}. In Fig.~\ref{fig:diamLD} (upper panel), we compare our estimation of $\theta_{\rm LD}$ with previous interferometric measurements. The LD diameter is achromatic (unlike the UD diameter) and should ideally be identical when estimated from different measurements and instruments. However, in practice, the estimations are not always similar, so many measurements are needed to find consistent values. Here, our values of $\theta_{\rm LD}$ are in very good agreement with previous estimations: the average relative difference $(\theta_{\rm LD}-\theta_{\rm LBI})/\theta_{\rm LBI}$ is -0.03 $\pm$ 0.08 ($\theta_{\rm LBI}$ being the interferometric angular diameter measured with another instrument than VEGA), and the median is -0.04. These differences might be explained by the calibrators used for the observations or the method used to determine the LD coefficients. However, using the visible domain allows the stars to be better resolved compared to the infrared one, thus the angular diameters are better constrained. The values measured here are also in good agreement with the ones given in Paper I. The angular diameters are slightly smaller in the present analysis, but they fit within the error bars, which are also a bit larger here. The differences come from the method used to derive the LD coefficients since we kept the same data, and these coefficients are important when we reach high precisions.

We also compared our measurements with other indirect estimations of angular diameters (Fig.~\ref{fig:diamLD}, bottom panel), including angular diameters directly determined from $F_{\rm bol}$ ($\theta_{\rm SED}$) and from an empirical law determined by \cite{Kervella2004}. These laws are based on photometry and calibrated by interferometry, and they allow predictions of angular diameters of dwarfs and subgiant stars. We chose to use the V and K bands and the associated relation $\log(\theta_{\rm LD}) = 0.0755(m_{\rm V}-m_{\rm K}) + 0.5170 - 0.2m_{\rm K}$. We consider an error of $2\%$ in the final $\theta_{\rm LD}$, while the best estimation is $1\%$ in their paper. The apparent magnitudes are expressed in the Johnson system, thus we had to keep the same system to calculate the angular diameters. We found the photometry of our stars in the JSDC catalog \citep{Bonneau2006}. While all the V magnitudes have already been expressed in this system, for some stars, the K band comes from the 2MASS reference. Thus, we converted the 2MASS magnitudes $m_{\rm Ks}$ into the Johnson system magnitudes $m_{\rm K}$ using $m_{\rm K} = m_{\rm Ks} + 0.032$\footnote{\texttt{http://www.pas.rochester.edu/$\sim$emamajek/memo$\_$BB88$\_$ \\
Johnson2MASS$\_$VK$\_$K.txt}}. They are given in Table~\ref{tab:propertiesStars}.
Again, we found good agreement between our measurements and the empirically determined angular diameters. This agreement is slightly better between angular diameters of dwarfs and subgiants determined with the \cite{Kervella2004} relation and interferometric measurements, which is expected since this law is calibrated specifically for Class V and IV stars. In both comparisons, small angular diameters measured with VEGA are generally smaller, and large angular diameters ($\theta_{\rm LD} >$ 0.9 mas) are a bit larger than previous values.

Concerning the effective temperatures, the average difference between $T_{\rm eff}$ (directly determined from SED fitting) and $T_{\rm eff,\star}$ (from interferometric measurements) $(T_{\rm eff,\star} - T_{\rm eff}) / T_{\rm eff}$ is $-0.007 \pm 0.030$. This is shown in Fig.~(\ref{fig:teff}). It is not surprising to find such differences between the temperatures since they are correlated with the angular diameters and have been obtained with the same $F_{\rm bol}$. As for angular diameters, the largest discrepancies are found for stars not on the MS: the relative difference is on average $-0.023 \pm 0.022$ for giants and bright giants, whereas it is of $0.003 \pm 0.030$ for dwarfs and subgiants. Our estimations of temperatures are slightly lower than the estimations from SED for low temperatures and slightly higher for high temperatures.

\section{Ages and masses of stars}
\label{sec:MassAge}

Once a star is placed on the H-R diagram, we determine its mass and
age using stellar evolution models.  This is crucial to providing
benchmark stars to stellar physicists and also to better
understanding planetary systems. In this section, we perform this
for the 18 stars of our sample with careful attention given to
determining the errors.

We used the recently published PARSEC stellar models \citep{Bressan2012} to determine the masses and ages of the 18 stars. The details of these models are well documented in \citet{Bressan2012}, but here we give a brief summary.  
Models are initiated on the pre-main sequence phase and evolve beyond the horizontal branch, which is sufficient for our purposes. High temperature opacity tables \citep[OPAL,][]{Iglesias1996} are used in conjunction with those calculated from their own code \citep[Aesopus,][]{Marigo2009} for lower temperatures. The models make use of the FREEEOS code\footnote{\texttt{http://freeeos.sourceforge.net/}} to calculate the equation of state, and the nuclear reaction rates comprise the p-p, Ne-Na, and Mg-Al chains, the CNO cycle, and some alpha-capture reactions.

Energy transport in the convective regions is described by the mixing-length theory of \cite{BohmVitense1958}, and the mixing-length parameter found for the Sun is 1.74. Convective overshoot from the convective core and below the convective envelope is a variable parameter that depends on stellar mass and chemical composition. Microscopic diffusion is included following the implementation of \citet{Salasnich1999}.
The reference distribution of heavy elements is given by \citet{Grevesse1998} except for some species where the \citet{Caffau2011} ones are used, and this gives a present solar metallicity of Z$_\odot$ = 0.01524 and Z$_\odot$/X$_\odot$ = 0.0207.  A chemical enrichment law is derived from the solar value using the primordial helium abundance (0.2485), and this is given as Y = 0.2485 + 1.78 Z.  The approximation [M/H] = log (Z/Z$_\odot$) is used to determine the metallicity.

The isochrones span $\log(age)$
from $6.6$ to $10.13$ in steps of $0.01$ and [M/H] from 0.5 to -0.8
dex in steps of $\sim$~0.015. We assume that [M/H]=[Fe/H] because no
additional information is available to differentiate them. 

For this data to be appropriate, the points on one single isochrone
should not be separated on the H-R diagram by a large distance
compared to $\sigma_{L_\star}$ and $\sigma_{T_{\rm eff,\star}}$. As this
is generally not the case, we performed spline interpolations of each
isochrone to produce a refined table for each star around $L_\star$
and $T_\star$, except for HD1367 and HD218560 due to their complex
position on the H-R diagram. For these two stars, we did not build any interpolation, which gives more consistent results.

\subsection{Best fit (least squares)}

To find the mass and age of a star, we perform a least squares
algorithm, looking for the parameter combination in our table that
minimizes the quantity:
\begin{equation}
\chi^2 = \frac{(L-L_\star)^2}{{\sigma_{L_\star}}^2} 
+ \frac{(T_{\rm eff}-T_{\rm eff,\star})^2}{{\sigma_{T_{\rm eff,\star}}}^2} + \frac{([M/H]-[M/H]_\star)}{{\sigma_{[M/H]_\star}}^2}\ .
\label{eq:Chi2}
\end{equation}

Although not intrinsically degenerate (because the number of
constraints equals that of parameters to be determined given a fixed set of parameters), this problem
does not have a unique solution, especially in some parts of the H-R
diagram, where the isochrones cross, so that a given luminosity and
temperature may correspond to two stars of different ages and
masses. Typically, there is a young ($<400$ Myrs) and an
old ($>400$ Myrs) solution. This is described particularly well
by \citet[][Fig.2]{Bonfanti2015}, who also show that two solutions are
possible when also using the PARSEC tables, one in the Gyr range
and the other in the Myr range. They show that without knowledge of the stellar mass, it is not possible to establish the
evolutionary stage of the star.  Additional stellar properties may
allow one to rule out one of the two solutions (e.g. chromospheric
activity, Lithium abundance, gyrochronology, or independent measure of the stellar mass, see discussion about HD75732 below), but we choose to report
both here for completeness, in contrast to many authors who keep
only one solution without justification\footnote{That some of
  our stars have planets is not a suggestion to constrain their age
  either. These stars no longer have protoplanetary disks
  \citep[inside which the planet forms, e.g.,][]{Armitage2013}, so
  they are older than $\sim 5$~Myrs \citep[e.g.,][]{Mamajek2009}, but
  nothing can exclude their being as young as 10~Myrs.}.

The two solutions are given in Table~\ref{tab:massesandages},
  with the uncertainties computed in next section. The value of
$\chi^2$ is provided, and generally $\ll 1$, showing very good
agreement. If $\chi^2$ is large, it means that no model corresponds to
this luminosity, temperature, and metallicity within the acceptable
uncertainty. We discuss these particular cases in
Sect.~\ref{sec:Discussion}. The isochrones corresponding to the
  old and young ages are shown in Fig.~\ref{fig:isochrones} for each
  star with $L_\star$, $T_{\rm eff,\star}$ and their error bars as a
  red cross.

\subsection{Likelihood maps and calculation of uncertainties}

The likelihood function $\mathcal{L}$ gives the probability of
  getting the observed data for a given set of stellar
  parameters. Generally, it is easy to express it as a function of the
  observables, but it is tricky to express it as a function of the physical
  parameters one wants to determine. Here, we want the probability density function of the stellar mass
  and age, which is an expression of $\mathcal{L}$ as a function of
  $M_\star$ and age\footnote{More precisely, the probability distribution function (PDF) of the mass and
    age is proportionnal to $\mathcal{L}\times f_0$, where $f_0$ is
    the prior. Assuming a constant star formation rate in the history
    of the Galaxy, $f_0$ should be independent of the age\,; although
    the IMF is not flat, our uncertainties on the masses are small
    enough that a flat prior in $M_\star$ is acceptable. In the end,
    $f_0$ can be taken uniform, and the PDF is simply proportional to
    $\mathcal{L}(M_\star,age)$.}, not of $L_\star$, $T_{\rm eff,\star}$, and [M/H]. For more
  details on the likelihood and a Bayesian approach, the reader is
  referred to \citet{Pont-Eyer-2004} and
  \citet{Jorgensen-Lindegren-2005}.

To produce a simplified map of $\mathcal{L}$ in the ($M_\star,
  {\rm age}$) plane, we pick random points in the H-R diagram such that each of the three terms on the righthand side of Eq.~\ref{eq:Chi2} is less than 1, 2, and 3. And we find the corresponding mass and age by least
  squares, as described above, and plot them in the ($M_\star,
  {\rm age}$) plane in red, yellow, or blue. The resulting maps are
  displayed in Fig.~\ref{fig:patates}. These crude contour plots of the
  likelihood show two important
  features: (i) the mass and age are far from being independent, and
  they show a clear negative correlation in the old solution, as expected; (ii) for a
  majority of stars in our sample, the likelihood has two distinct
  peaks, and the marginal PDF of the age is far from Gaussian. These two
  peaks correspond to the \emph{\emph{young}} and \emph{\emph{old}} solutions
  discussed previously, which are the local modes of $\mathcal{L}$ in
  these two regions.  

Maps of the joint PDF of the mass and age are the most
  information-rich way of presenting the results. However, they are
  not very convenient to use. Below, we therefore estimate the standard
  deviation in age and mass for each of the two solutions. We do not
  have a precise and detailed map of the PDF, but a Monte-Carlo (MC) method
  is appropriate, as shown by \citet{Jorgensen-Lindegren-2005}.  

Simply drawing independent Gaussian sets of $L_\star$ and $T_{\rm
  eff,\star}$ would erase the natural correlation between $L_\star$
and $T_{\rm eff,\star}$, which occurs because they are both
increasing functions of $F_{\rm bol}$. (The covariance between
$L_\star$ and $T_{\rm eff,\star}$ is proportional to the variance of
$F_{\rm bol}$, see Appendix A.) To have a more realistic and
smaller cloud of points, we proceed as follows:
\begin{enumerate}
\item draw 1500 quadruplets \{$F_{\rm bol}$, $d$, $\theta$, $[M/H]$\}, as
  the occurrences of four independent Gaussian random variables of
  appropriate mean and standard deviation;
\item combine them into 1500 triplets:\\ \{$L_\star=4\pi\ F_{\rm
  bol}\ d^2$\ , $T_{\rm eff,\star}=\left(4\,F_{\rm bol}\,/\,\sigma_{\rm SB}\theta^2\right)^{1/4}$\ , $[M/H]$\};
\item apply the least squares procedure described above to find the corresponding 1500 \{$M_\star$, age\} pairs;
\item compute the mean and standard deviation of the masses and ages
  found and their correlation.
\end{enumerate}
The clouds of points are shown in Fig.~\ref{fig:isochrones}. In the top lefthand panel (HD3651), the cloud of points is particularly elongated. It shows that this procedure captures the expected correlations because it starts from the most basic parameters.
We stress that in the literature, error bars are usually poorly described and derived, even though they are of outmost importance for astrophysical interpretation.

As clearly explained in \citet{Pont-Eyer-2004}, the estimator of the ages and masses provided by the MC approach may be biased.
Thus, we do not report the average ages and masses obtained  in Table~\ref{tab:massesandages}, but we have checked that they are very close to the best fit value in most of the cases (within a quarter of the standard deviation).

\subsection{Discussion}
\label{sec:Discussion}

\begin{table*}
\caption{Masses and ages derived from PARSEC isochrones (see Sect.~\ref{sec:MassAge}).\hfill \ }
\centering
\label{tab:massesandages}
\begin{tabular}{l | c c c c | c c c c}
\hline \hline
Star            &                                       \multicolumn{4}{|c}{Old solution}                                       & \multicolumn{4}{|c}{Young solution}                \\

                        &       $M_\star\pm \sigma (\%)$        &       Age     $\pm \sigma (\%)$    & $\log(g)$     &       $\chi^2$ & $M_\star\pm \sigma (\%)$     &         Age     $\pm \sigma (\%)$       & $\log(g)$     &       $\chi^2$ \\
                        &       [$M_{\odot}$]                           &       [Gyrs]                                  &       [cm s$^2$]                &                        &               [$M_{\odot}$]                           &       [Myrs]          &       [cm s$^2$]                &                       \\
                        \hline
HD3651          &       0.848$\pm$ 0.040 (4.66)         &       10.00$\pm$      3.52 (35)    &       4.6     &       0.0009   &      0.889$\pm$ 0.024 (2.67)         &       38.90$\pm$ 3.686 (9)       &       3.2             &       0.027 \\
HD9826          &       1.312$\pm$      0.075 (5.72)    &       3.02$\pm$       0.92 (30)    &       4.2     &       0.0008   &      1.359$\pm$ 0.027 (1.99)         &       14.79$\pm$ 1.446 (10)      &       1.3                     &       0.020 \\        
HD19994         &       1.317$\pm$ 0.079 (5.96)         &       3.80$\pm$  0.84 (22)      &       4.8     &       0.0064   &      1.459$\pm$ 0.023 (1.58)          &       12.30$\pm$ 1.326 (11)   &       4.3                     &       0.0095  \\
HD75732         &       0.874$\pm$      0.013 (1.44)    &       13.19$\pm$      1.18  (9)    &       4.3     &       0.919    &      0.968$\pm$ 0.018 (1.83)         &       30.90$\pm$ 3.028 (10)      &       4.3                     &       0.0124   \\
HD167042        &       1.646$\pm$      0.311 (18.88)   &       1.82$\pm$       5.23 (287)   &       4.3     &       7.766    &              -                                               &       -                                               &       -                       &       -                \\
HD170693        &       0.879$\pm$ 0.050 (7.86)         &       13.19$\pm$  1.92 (15)      &       4.7     &       2.309    &              -                                               &       -                                               &       -                       &       -                \\
HD173416        &       1.336$\pm$ 0.258 (19.3)         &       3.47$\pm$  3.18 (92)      &       3.2     &       0.0003   &              -                                               &       -                                               &       -                       &       -                \\
HD185395        &       1.49$\pm$ 0.060 (4.01)          &       0.50$\pm$  0.40 (79)      &       4.0     &       0.0002   &      1.519$\pm$      0.037 (2.42)  &       18.62$\pm$      4.436 (24)      &       4.4                     &       0.013    \\
HD190360        &       1.04$\pm$ 0.059 (5.71)          &       4.79$\pm$  2.51 (52)      &       4.6     &       0.0011   &      1.073$\pm$  0.045 (4.22)  &       29.51$\pm$ 2.11 (7)             &       4.2                     &       0.049    \\
HD217014        &       1.12$\pm$ 0.063 (5.66)          &       1.996$\pm$  2.24 (112)     &       1.5     & $\sim$0.0      &      1.138$\pm$      0.073 (6.39)  &       27.54$\pm$      0.177 (6)       &       3.0                     &       0.019    \\
HD221345        &       0.898$\pm$ 0.069 (7.63)         &       13.19$\pm$  2.10 (16)      &       4.3     &       0.1569   &      -                                                       &       -                                               &       -                       &       -        \\
HD1367          &       2.560$\pm$      0.089 (3.46)    &       0.575$\pm$      0.05 (8)     &       2.9     &       0.254    &      -                                                       &       -                                               &       -                       &       -        \\
HD1671          &       1.875$\pm$      0.075 (4.00)    &       1.203$\pm$ 0.14 (11)       &       0.3     &       0.1834   &      2.074$\pm$      0.037 (1.78)  &       4.47$\pm$ 0.341 (8)             &       0.4                     &       0.0091 \\
HD154633        &       1.962$\pm$ 0.207 (10.5)         &       1.175$\pm$ 0.52 (45)       &       3.1     &       0.0051   &      -                                                       &       -                                               &       -                       &       -        \\
HD161178        &       1.936$\pm$      0.212 (11.0)    &       1.446$\pm$ 0.49 (34)       &       0.7     &       0.0052   &      -                                                       &       -                                               &       -                       &       -        \\
HD168151        &       1.189$\pm$      0.068 (5.68)    &       3.468$\pm$ 0.76 (22)       &       3.8     &  $\sim$0.0 &  1.294$\pm$      0.025 (1.95)    &       12.30$\pm$      1.082 (9)   & 2.4                     &  0.0126 \\
HD209369        &       1.531$\pm$      0.061 (3.99)    &       1.95$\pm$ 0.24 (12)               &       3.5     &       0.0067   &      1.666$\pm$ 0.033 (1.96)            &       6.918$\pm$      0.708 (10)      &       4.3                     &       0.0002    \\
HD218560        &       3.638$\pm$ 0.441 (12.13)                &       0.263$\pm$      0.12 (44)    &       2.0     &       0.0225   &              -                                               &       -                                               &       -                       &       -                \\
\hline
\end{tabular}
\end{table*}

The errors on masses are generally $\sim 7.6\%$ for the \emph{\emph{old}} solution, but the errors on the ages are much larger. The \emph{\emph{young}} solutions give errors of $\sim 10\%$ because there are generally fewer possibilities at $1\sigma$ for younger ages.

Concerning MS stars, we always find two solutions thanks to their position on the H-R diagram (except for \object{HD154633}), unlike for giant stars. The best fit results give $\chi^2 \ll 1$, the ones corresponding to \emph{\emph{old}} solutions being smaller than the ones for \emph{\emph{young}} solutions. However, these quantities remain very low, which means that without additional constrains, it is difficult to choose one solution over the other, as explained in the previous section. 
Only for two stars (\object{HD170693} and \object{HD167042}) are the results not consistent with the models ($\chi^2 \gg 1$). One possibility for explaining inconsistency with the models is that the mixing-length parameter in the models should be adjusted, \citep[see, e.g.,][]{Creevey2012a,Bonaca2012}. One would require a downward revision of this parameter to make the models fit. Including an extra free parameter gives another degree of freedom, which clearly enlarges the uncertainties. However, with only a few constraints, one is required to use stellar models with a fixed mixing-length parameter.

The masses derived in this paper are in good agreement with the ones derived for the four stars in Paper I, except for \object{HD221345}. The mass estimated here is more than twice lower than the first estimation, but agrees  more with other values in the literature (see below). This is directly due to the $\log(g)$ used in Paper I, which is close to the one used in the present study but is not sufficient to describe a star at an evolved stage. Besides, the gravitational masses (Table~\ref{tab:finalParameters}) are not in good agreement with the masses derived from models because of the $\log(g)$ used to calculate them. Indeed, the $\log(g)$ derived from the best fit method is generally different from the one shown in Table~\ref{tab:InputParam}, and this is particularly true for giants.
In what follows, we discuss the results for each star and compare them with other previous results.

\paragraph{\object{HD167042}} - 
When comparing angular diameters, one of the most important discrepancies is found for HD167042. \cite{Baines2010a} find $\theta_{\rm Baines} = 0.92 \pm 0.02$~mas, which gives  $\theta_{\rm LD}/\theta_{\rm Baines} = 1.145$. This directly affects the determination of other parameters such as the luminosity and temperature (there is a difference of $4.8\%$ between $T_{\rm eff}$ and $T_{\rm eff,\star}$) that determine the position of the star in the H-R diagram. However, our angular diameter measurement is close to the estimation by \cite{Kervella2004}. 
Concerning the age, no solution better than $3 \sigma$ is found, which translates into $\chi ^2 = 7.8$, thus we cannot conclude anything about the true age of the star. 

\paragraph{\object{HD75732}/\object{55 Cnc}} - 
The star HD75732 is a MS star but shows a $\chi^2$ slightly higher than the other stars of the same spectral type. We found an angular diameter that does not agree with the \cite{Baines2008} measurement, which is much larger (0.854 mas) than the one we found. However, our angular diameter of 55~Cnc is similar at $1.02\%$ to the one derived by \citet[][$\theta = 0.711 \pm 0.004$ mas]{vonBraun2011}, who did an extensive study of the system of 55~Cnc. Thus, we compare our results with those found by \cite{vonBraun2011}. When setting the values of $\theta$, $F_{\rm bol}$, $T_{\rm eff}$, [M/H] \citep[from][]{Valenti2005}, and $L_\star$ given in their paper as input parameters in our code, we find a mass of 0.87~$M_\odot$ and an age of 13.2 Gyrs, in agreement with our results. For the \emph{\emph{young}} solution, we obtain 0.96~$M_\odot$ and 33.8~Myrs, which is slightly older than our solution.
The smaller errors can be explained by the lower error on the metallicity (0.04 dex), as shown in Sect.~\ref{sec:Metallicity}. These results are very consistent with the masses and ages we determined from the parameters derived in the present paper. However, we find an older star than that with the Yonsei-Yale model used by \citet[][10.2 $\pm$ 2.5~Gyrs]{vonBraun2011}, and we find two solutions. It is not clear though whether the Yonsei-Yale method provides only one or two solutions.

\cite{Maxted2015} give a mean stellar density of 55 Cnc directly from the analysis of the transit of \object{55~Cnc~e} and Kepler's law: $\rho_\star = 1.084^{+0.040}_{-0.036} \rho_\odot$. Using this value and the radius which we measured ($R_{\star} = 0.96 \pm 0.02 R_\odot$), we obtain a model-independent stellar mass of 0.960 $\pm$ 0.067~$M_\odot$ (taking a mean error of 0.038~$\rho_{\odot}$ on the stellar density). This mass is very consistent with the one derived from isochrones and that corresponds to the young solution ($0.968\pm0.018 M_\odot$). Thus, this direct estimate of the mass favors the \emph{\emph{young}} age (31 Myrs) derived from isochrones. And since this mass comes from a direct estimation, we use it to derive planetary parameters in the rest of the paper.

\paragraph{\object{HD3651}/\object{54 Psc}} - 
The PARSEC models reproduce this star well. The radius we measured is slightly larger than the one measured by \citet[][0.81 $R_\odot$]{vanBelle2009} but in very good agreement with the \cite{Baines2008} value (0.85 $R_\odot$). 
From chromospheric activity, \cite{Wright2004} derived an age of 5.9 Gyrs, which is much younger than our solution, but this would tempt us to adopt our old solution rather than the young one. In contrast, \cite{Boyajian2013b} find an older age (14.9 Gyrs) using Yonsei-Yale isochrones, but they do not take the error on the metallicity into account, and they specify that their method can lead to unrealistic age determinations, especially for the lowest luminosity stars. 

\paragraph{\object{HD9826}/\object{$\upsilon$ And}} - 
This star has been studied extensively because it hosts at least four exoplanets \citep{Curiel2011}. Its age was determined several times, as were its mass, radius, and temperature; they are fully summarized in \cite{McArthur2010}. Our estimations of these parameters are in good agreement with the previous determinations. For example, the age and mass fall well within the previous estimations that are set between 2.3 and 5 Gyrs for the age and 1.24 and 1.31 $M_\odot$ for the mass, and these were determined using different methods.

\paragraph{\object{HD19994}/\object{94 Cet}} - 
Many age determinations have been performed concerning this star. \cite{Wright2004} found an age of 3.55 Gyrs, in good agreement with our estimation. Using the \cite{Baines2008} radius, \cite{Boyajian2013b} found an older star of 4.8 Gyrs along with a mass of 1.275 $M_\odot$, also in good agreement. In the catalog of the Geneva Copenhagen Survey \citep[CGS;][]{Holmberg2009}, the age of 3 Gyrs is given for HD19994. \cite{Saffe2005} used the chromospheric activity to determine the age of HD19994 and found a much larger estimation (13.01 Gyrs), which is also older than the other estimations made by them using isochrones (4.7 Gyrs) and lithium abundances (1.4 Gyrs). The latter would favor our older solution.

\paragraph{\object{HD170693}/\object{42 Dra}} - 
We only find an \emph{\emph{old}} solution for HD170693 with a high $\chi^2$. \cite{Bonfanti2015} used the same isochrones as we did (PARSEC tables) to characterize this star and found an age and mass of 9 Gyrs and 1.0$\pm$0.1 $M_\odot$. However, the input parameters were not the same in their study, and we stress here that, unlike them, we bring a direct determination of the radius that is a free parameter for them and is equal to 20.9$\pm$0.6 $R_\odot$. 
Our error on $T_{\rm eff,\star}$ is very small, and it is most likely this parameter that dominates our solution.
A better precision on the parallax would allow verification of the very small error on the temperature.

\paragraph{\object{HD173416}} - 
There are not many studies concerning this star. \cite{Bonfanti2015} find that it is younger than our estimation (1.5$\pm$0.6 Gyrs), and this is to our knowledge the only determination of the age before ours. They found a mass of 1.8 $M_\odot$, which is closer to \citet[][2.0$\pm$0.3 $M_\odot$]{Liu2009}  and \citet[][2.37 $M_\odot$]{Liu2010} estimations using Yonse-Yale isochrones. None of them used a direct angular diameter measurement as input in their model.

\paragraph{\object{HD185395}/\object{$\theta$ Cyg}} - 
This star has long interested scientists for the unusual radial velocity variations it presents and make it a complex system not fully understood yet, suspecting several planets around the star \cite[see][and references therein for additional information]{Ligi2012}. \cite{Guzik2011} discuss the solar-like oscillations it shows and the probability of having $\gamma$-Dor pulsations by considering two stellar masses, 1.38 and 1.29 $M_\odot$ with different metallicities. They state that for solar metallicity (as we consider in our study), $\gamma$ Dor g-mode pulsators, expected masses are higher, on the order of 1.6 $M_\odot$, which is a value close to the estimation of the mass we found. 

\paragraph{\object{HD190360}} - 
A wide range of ages has been found for HD190360; 11.3 Gyrs \citep{Boyajian2013b} using the \cite{Baines2008} radius\,; 6.7 Gyrs along with a mass of 0.96 $M_\odot$ \citep{Naef2003}\,; \cite{Ibukiyama2002} gave 12.11 Gyrs; \cite{Valenti2005} gave 7.2 Gyrs; and \cite{Bensby2014} estimated 4.9 Gyrs with an upper limit of 9.4 Gyrs and a lower limit of 2.8 Gyrs, along with a mass of $0.99^{+0.05}_{-0.06}$ $M_\odot$, among others. This is a good example that shows how difficult the age and mass determination is and how it depends on input parameters and model. Our mass estimation is in very good agreement with the \cite{Fuhrmann1998b} (1.04 $M_\odot$) and \cite{Bensby2014} ($0.99^{0.05}_{-0.06} M_\odot$) estimations.

\paragraph{\object{HD217014}/\object{51 Peg}} - 
This star is known as the first solar-like star around which an exoplanet has been found \citep{Mayor1995}. We find masses consistent with a solar type star, but younger. Our estimation of the age is closer than that of \citet[3.3$\pm$1.2 Gyrs,][]{Bonfanti2015}, in particular concerning the mass (1.1$\pm$0.02 $M_\odot$). The reanalysis of the GCS data in \cite{Casagrande2011} uses two sets of models, and they find a median age and mass of 5.33 Gyrs and 1.06 $M_\odot$, and 7.4 Gyrs and 1.02 $M_\odot$ with PADOVA and BASTI, respectively.

\paragraph{\object{HD221345}/\object{14 And}} - 
The only previous age and mass determinations we found for this star are those from \citet[][3.20 $\pm$ 2.10 Gyrs and 1.40 $\pm$ 0.20 $M_\odot$]{Bonfanti2015} and \citet[][4.5 $\pm$ 1.9 Gyrs and 1.1 $\pm$ 0.2 $M_\odot$]{Baines2009}, and this last mass is in good agreement with our estimation. As shown previously, the estimation of these parameters is difficult, and one can find as many values as there are estimations. \cite{Baines2009} measured a smaller angular diameter than we did, which translates into a higher luminosity and might explain the difference. However, there is a strong discrepancy between empirically determined angular diameters and our measurement: we found $1.49\pm 0.03$ mas but $\theta_{\rm SED}$ = 1.359 $\pm 0.023 $ mas and  $\theta_{\rm Kervella}$  = 1.859 mas. This might be explained by the fact that 14 And is a giant, so the \cite{Kervella2004} relation is not appropriate for this star. Also, the infrared photometry is not homogeneous with the visible part. There also are discrepancies between $T_{\rm eff}$ and $T_{\rm eff,\star}$ ($4.5\%$ difference). \\

The comparisons for the non-host stars are more difficult since for most of them, we bring here the first estimation of their mass and age. \cite{Casagrande2011} provide an estimation of the age and mass of \object{HD1671}, \object{HD168151}, and \object{HD209369}. 
For \object{HD1671}, we find a slightly younger star than \cite{Casagrande2011}, but the masses are consistent, particularly when comparing with those obtained with the PADOVA code (1.82 $M_\odot$ for both median and most probable masses). 
Concerning \object{HD168151}, our age estimation is between the \citet[][5 Gyrs]{Boyajian2013b} and \citet[][$\sim$~2.5 Gyrs]{Casagrande2011} results, and our mass estimation is a bit lower. Finally, \cite{Casagrande2011} give very similar results to ours for \object{HD209369}.

\subsection{On the role of metallicity}
\label{sec:Metallicity}

Taking the uncertainty on the
metallicity into account significantly increases the range of the
distributions of the masses and ages, \emph{id est} the final
error bars. To quantify the error bugdet due to the metallicity, we take the case of HD3651
as an example, for which we have reasonable errors and low $\chi^2$. When setting the error on the metallicity to $\sigma$([M/H])=0.001 dex (instead of 0.1 dex, see Sect.~\ref{sec:Fbol}), we get errors of 22$\%$ and 2.1$\%$ on the age and mass, respectively, for the \emph{\emph{old}} solution. Thus, the error on the metallicity contributes to one third of the total error of the age and to half of the error on the mass. It is even more significant for the young solution, where the errors reduce to 3$\%$ and 0.43$\%$ for the age and mass.
If we only consider the uncertainty on the metallicity (reducing the errors on $F_{\rm bol}$, $\theta$, and $d$ by a factor $10^{-5}$), we get very similar errors on the age and mass than the ones shown in Table~\ref{tab:massesandages}: the errors are of 31$\%$ and 4.37$\%$ on the age and mass for the old solution, and 9$\%$ and 2.5$\%$ for the young solution. This emphasizes the significant contribution of the error on the metallicity.
Standard deviations assuming a fixed metallicity are therefore underestimated and should be considered as a lower limit.

\section{Exoplanetary parameters}
\label{sec:Exoplanets}

\subsection{Planetary masses, semi-major axes, and habitable zone}
\label{sec:PlanetMasses}

\begin{table*}
\caption{Orbital parameters of planets (see Sect.~\ref{sec:PlanetMasses}). \hfill \ }
\label{tab:OrbitalParameters}
\centering
\begin{tabular}{l l c c c c l}
\hline \hline
Star    &       Planet          &               $\omega$         &      P                                        &       K                                &      e                                               &       Ref. \\
                &                               &               [deg]           &       [days]                          &               [m$\cdot$ s$^{-1}$]       &                                                       & \\
\hline
HD3651  &       b               &       233.3 $\pm$     7.4      &      62.21   $\pm$ 0.02   &        16.0  $\pm$ 1.2  &      0.62 $\pm$ 0.05                 &        \cite{Butler2006} \\
HD9826  &       b               &       324.9 $\pm$ 3.8  &      4.617   $\pm$ 2.3e-5 &  70.51 $\pm$ 0.45 & 0.0215     $\pm$ 7.e-4     &       \cite{Curiel2011} \\
        &       c       &       258.8 $\pm$     0.43 &  1276.46 $\pm$ 0.57          &      68.14 $\pm$     0.45 &  0.2987  $\pm$ 0.0072    &        \\
        &       d                       &       241.7 $\pm$     1.6      &       241.26  $\pm$ 0.06   &  56.26 $\pm$ 0.52 &      0.2596  $\pm$ 0.0079    &        \\
        &       e                       &       367.3 $\pm$     2.3      &       3848.86 $\pm$ 0.74   &  11.54 $\pm$ 0.31 &      0.00536 $\pm$ 0.00044         &        \\
HD19994 &       b           &   41.0  $\pm$     8.0  &  535.7   $\pm$ 3.1    &    36.2  $\pm$     1.9  &  0.300 $\pm$ 0.040               &        \cite{Butler2006}, \cite{Mayor2004} \\ 
HD75732.        &       b               &       110   $\pm$     54   &  14.65   $\pm$ 0.0001 &  71.11 $\pm$   0.24 &  0.004 $\pm$ 0.003               &       \cite{Endl2012}\\
        &       c                &      356   $\pm$     22   &  44.38   $\pm$ 0.007  &        10.12 $\pm$ 0.23 &      0.07  $\pm$ 0.02                &         \\
        &       d                 &     254   $\pm$     32   &  4909    $\pm$ 30     &        45.2  $\pm$     0.4  &  0.02  $\pm$ 0.008               &         \\
        &       e                 &     90    $\pm$     0        &      0.736546   $\pm$ 3.e-6  &        6.30  $\pm$     0.21 &  0.        $\pm$ 0.                      &        \\
        &       f                 &     139   $\pm$ 8    &      261.2   $\pm$ 0.4    &        6.2   $\pm$     0.3  &  0.32  $\pm$ 0.05                &        \\
HD167042        &       b                       &       82        $\pm$ 52          &      417.6   $\pm$ 4.5        &      33.3  $\pm$     1.6      &       0.101 $\pm$0.066                &       \cite{Sato2008}\\
HD170693        &       b                       &       218.7 $\pm$ 10.6 &       479.1   $\pm$ 6.2        &      110.5 $\pm$     7.       &      0.38  $\pm$  0.06            &       \cite{Dollinger2009} \\ 
HD173416        &       b                       &       254  $\pm$ 11    &       323.6   $\pm$ 2.2        & 51.8   $\pm$ 2.0      &      0.21  $\pm$0.04                 &       \cite{Liu2009} \\ 
HD190360        &       b                       &       12.4  $\pm$ 9.3  &       2891    $\pm$ 85.        &      23.5  $\pm$     0.5  &  0.36  $\pm$     0.03                 & \cite{Vogt2005}\\
        &       c     & 153.7 $\pm$ 32   &      17.10   $\pm$ 0.015  &  4.6   $\pm$ 1.1      &      0.01  $\pm$ 0.1                 &        \\ 
HD217014        &       b                  &    58    $\pm$ 0    &      4.23    $\pm$ 3.6e-5 &       55.94 $\pm$ 0.69 &      0.013 $\pm$ 0.012               &        \cite{Butler2006}  \\ 
HD221345        &       b               &       0         $\pm$ 0   &   185.84  $\pm$ 0.23   & 100.0 $\pm$     1.3  &  0         $\pm$ 0                       &        \cite{Sato2008} \\
\hline
\end{tabular}
\end{table*}

\begin{table*}
\caption{Semi-major axes and minimum masses of exoplanets, and habitable zones of host stars derived from orbital parameters (found in the literature) and stellar parameters estimated in this work from isochrones (see Sect.~\ref{sec:PlanetMasses}). }
\centering
\begin{tabular}{l l | c c | c c | c}
\hline \hline
                &               &       \multicolumn{2}{| c |}{Old solution}                                     & \multicolumn{2}{c|}{Young solution}   &                               \\
Star    &Planet & $a$   & $m_{\rm p}\sin(i)$ &  $a$  & $m_{\rm p}\sin(i)$ &               HZ      \\      
                &               & [au]                  &   [$M_{\rm Jup}$]             & [au]                    &   [$M_{\rm Jup}$]                     & [AU] \\
\hline

HD3651          &       b       & 0.2908 $\pm$ 0.0045                   &       0.220 $\pm$ 0.021     &       0.2955 $\pm$ 0.0026     &       0.227 $\pm$ 0.021       &       0.62 -       1.23                            \\      
        
HD9826          &       b       &       0.0594 $\pm$ 0.0116             &       0.692 $\pm$ 0.027     &       0.06010 $\pm$ 0.00040   &       0.708 $\pm$ 0.010       &       1.56 - 3.14                          \\      
                &       c$^\dagger$     &       2.521 $\pm$ 0.049               &       4.16 $\pm$ 0.16              &       2.551 $\pm$ 0.017       &       4.257 $\pm$ 0.064           &                                       \\      
                &       d       &       0.830 $\pm$ 0.016       &       1.994 $\pm$ 0.079             &       0.8401 $\pm$ 0.0056     &       2.041 $\pm$ 0.033           &                                       \\      
                &       e       &       5.26 $\pm$ 0.10 &       1.066 $\pm$ 0.050           &       5.324 $\pm$ 0.035       &       1.091 $\pm$ 0.033               &                                       \\      
                
HD19994         &       b       &       1.415 $\pm$ 0.029               &       1.658 $\pm$ 0.111     &       1.4639 $\pm$ 0.0096     &       1.775 $\pm$ 0.098               &       1.64 - 3.29                                  \\

HD167042&       b       &       1.281 $\pm$ 0.089       &       1.67 $\pm$ 0.24            &       -               &                       -                       &       3.24 - 6.42                          \\      

HD170693        &       b       &       1.148 $\pm$ 0.031       &       3.61 $\pm $0.31              &               -       &                       -                       &       11.15 -       22.14                           \\      

HD173416        &       b       &       1.016 $\pm$ 0.066               &       2.08 $\pm$ 0.28              &       -               &                       -                       &       8.46 - 16.79                         \\      

HD190360        &       b       &       4.02 $\pm$ 0.11         &       1.573 $\pm$ 0.073             &       4.067 $\pm$ 0.098       &       1.61 $\pm$ 0.06            &       0.88 - 1.76                             \\      
                &       c       &       0.1314 $\pm$ 0.0025     &       18.97$\pm$4.58$^*$      &       0.1331 $\pm$ 0.0019    &       19.42$\pm$4.69$^*$ &    \\      
                                
HD217014        &       b       &       0.0532 $\pm$ 0.0010     &       0.480 $\pm$ 0.019     &       0.0534 $\pm$ 0.0011     &       0.485 $\pm$ 0.021       &       0.95 - 1.90                          \\      

HD221345        &       b       &       0.615 $\pm$ 0.015       &       2.61 $\pm$ 0.14              &       -               &                       -                       &               7.02 - 13.9                  \\      
\hline
\end{tabular}
\tablefoot{We took $M_{\rm Jup}=1.8986 \cdot 10^{27}$~kg. $^*$Expressed in Earth mass, with $M_\oplus = 5.9736 \cdot 10^{24}$~kg. $^\dagger$Lies in the HZ.}
\label{tab:PlanetaryMasses}
\end{table*}

\begin{figure}[ht]
\includegraphics[scale=0.5]{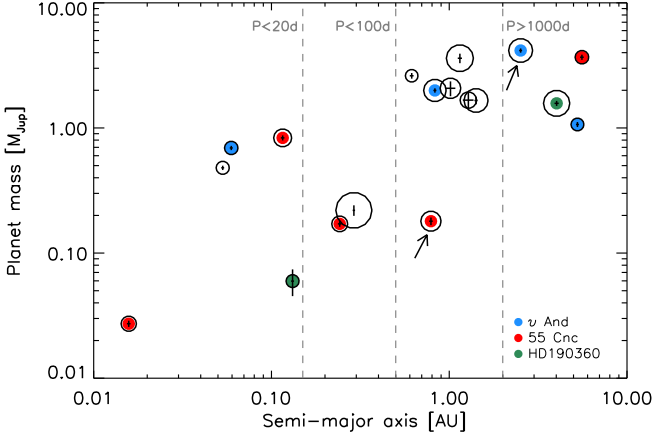}
\caption{Minimum masses of exoplanets versus their distance to the star from the \emph{\emph{old}} solution for the stellar mass (except for 55~Cnc, whose planetary minimum masses come from a direct estimate of the mass). Symbol sizes are an increasing function of the eccentricity of planets. The arrows indicate the planets in the HZ, and multiple planet systems are shown in color (see Sect.~\ref{sec:PlanetMasses}). }
\label{fig:ExoplanetParam}
\end{figure}

Radial velocity measurements constitute one of the two most prolific methods used to discover exoplanets. It gives the minimum mass of the exoplanet $m_{\rm p}\sin(i)$\,:
\begin{equation}
m_{\rm p}\sin(i) = \frac{ M_\star^{2/3} P^{1/3} K (1-e^2)^{1/2}}{(2 \pi G)^{1/3}} \ ,
\end{equation}
where $m_{\rm p}$ and M$_\star$ are the planetary and stellar masses, P and K are the period and the semi-amplitude of the radial velocity signal, $e$ is the eccentricity of the planet, and G is the gravitational constant.
Thus, to determine the minimum mass of the exoplanet itself, one has to know the stellar mass. In the previous section, we give the stellar masses of ten exoplanet host stars (see Table~\ref{tab:massesandages}), which yields the semi-major axes $a$ and the masses of their exoplanets using the observables $P$ and $K$ given in Table~\ref{tab:OrbitalParameters}. For half of the stellar sample, there are two solutions concerning the age and mass (an old one and a young one), thus we give the corresponding semi-major axes and planetary masses for each solution. Our errors on $a$ account for the uncertainty in the stellar mass (which is not always the case in the literature) derived from the MC method. The planetary parameters are given in Table~\ref{tab:PlanetaryMasses}. The old and young sets of planetary parameters are generally very close to each other, sometimes almost identical, because the young and old stellar masses are not dramatically different. Thus, Fig.~\ref{fig:ExoplanetParam} only shows the solutions derived from the old solution for the stellar masses. However, a planet of a given mass has a different structure after a few dozen Myrs or a few Gyrs of evolution, so the fact that a young solution exists matters. 
The system of 55~Cnc does not appear in Table~\ref{tab:PlanetaryMasses} since it has a direct determination of the mass that does not correspond to either a young or an old solution. The parameters of this system are thus given in Table~\ref{tab:55CncSystem}.

\begin{table}
\caption{Semi-major axes and minimum masses of exoplanets of the 55~Cnc system derived from orbital parameters (found in the literature, see Table~\ref{tab:OrbitalParameters}) and a direct estimation of the stellar mass (see Sect.~\ref{sec:Discussion}).}
\centering
\begin{tabular}{l | c c c}
\hline \hline
Planet                  & $a$   & $m_{\rm p}\sin(i)$    \\      
                                & [au]                  &   [$M_{\rm Jup}$]              \\
                        \hline
        b                       & 0.1156 $\pm$ 0.0027           & 0.833  $\pm$ 0.039                                                             \\      
        c                       & 0.2420 $\pm$ 0.0056           &       0.1711 $\pm$ 0.0089                                                    \\
        d                       & 5.58 $\pm$ 0.13               &       3.68 $\pm$ 0.17                                                      \\
        e                       & 0.01575 $\pm$ 0.00037         &       8.66 $\pm$ 0.50$^*$                  \\                                      
        f$^\dagger$     &  0.789$\pm$ 0.018             & 0.180 $\pm$ 0.012                                                         \\
\hline
\end{tabular}
\tablefoot{$^*$Expressed in Earth mass, with $M_\oplus = 5.9736 \cdot 10^{24}$~kg. $^\dagger$Lies in the HZ, located between 0.67 and 1.33 au.}
\label{tab:55CncSystem}
\end{table}

The habitable zone (HZ) is defined as a range of distances where liquid water can be found on an exoplanet. We used the method described by \cite{Jones2006} to calculate it. We first calculate the critical flux at the inner boundary
\begin{equation}
S_{b,i}(T_{\rm eff,\star}) = (4.190 \times 10^{-8} T_{\rm eff,\star}^2) - (2.139 \times 10^{-4} T_{\rm eff,\star}) + 1.296
\end{equation}
and at the outer boundary
\begin{equation}
S_{b,o}(T_{\rm eff,\star}) = (6.190 \times 10^{-9} T_{\rm eff,\star}^2) - (3.319 \times 10^{-5} T_{\rm eff,\star}) + 0.2341 \ ,
\end{equation}
where $S_{b}(T_{\rm eff,\star})$ is given in units of the solar constant and $T_{\rm eff,\star}$ in K. We can then calculate the inner and outer distances of the HZ in au:
\begin{equation}
\begin{aligned}
r_i &= \left[  \frac{L_\star}{S_{b,i}(T_{\rm eff,\star})} \right] \\
r_o &= \left[  \frac{L_\star}{S_{b,o}(T_{\rm eff,\star})} \right] \ ,
\end{aligned}
\end{equation}
where $L_\star$ is the luminosity of the star in $L_{\odot}$ from Table~\ref{tab:finalParameters}. The resulting values are given in Table~\ref{tab:PlanetaryMasses} for each star. \cite{Jones2006} specify that this method is based on a simplified model that neglects enhanced cloud formation and the formation of CO$_2$ clouds, which results in a conservative HZ. Thus, the HZ could in reality be wider. 
As expected, the values of HZ found by \cite{Jones2006} are close to our estimations when the stellar parameter estimations are in good agreement. This is the case for HD9826, HD217014, and HD19994. For HD75732, HD3651, and HD190360, we found HZ to be closer to their star than what is given by \cite{Jones2006}. It is the same for the planetary masses, which depend on the stellar masses and thus explain differences between different estimations. As noted in Sect.~\ref{sec:Discussion}, for example, our estimation of the mass of HD221345 is lower than what is estimated in Paper I. This directly translates into a lower minimum mass for HD221345 b.

According to our values, only HD9826~c and HD75732~f lie in their HZ. They are large exoplanets (of the Jupiter type), thus life as we know it could hardly been found on them. However, their moons could be terrestrial bodies with water on their surface and possibly an atmosphere, if these planets have a system similar to those of the solar system giant planets (think of Titan and Europa).

In Fig.~\ref{fig:ExoplanetParam}, we see that small exoplanets lie closer to their stellar host than large planets. This is of course due to an instrumental bias, but our sample is quite representative of the population of known exoplanets.

\subsection{The case of 55 Cnc e}
\label{sec:55Cnc}

The system of 55~Cnc holds a transiting super-Earth, 55~Cnc~e, which was independently discovered by \cite{Winn2011} and \cite{Demory2011}. The transit method provides the ratio of the planetary to the stellar radius and the density of the star. Thus, to correctly determine the planetary radius $R_{\rm p}$, one has to know the stellar radius. This method also provides the inclination of the system. If RV measurements are also performed, which is the case for the system of 55~Cnc, the true planetary mass $M_{\rm p}$ can then be derived, contrary to the minimum mass that is currently found. Then, the density $\rho_{\rm p}$ of the planet can be derived. Von Braun et al. (2011) give a complete review of this system using at first interferometric measurements to determine 55 Cnc's radius. Here, we consider our interferometric measurement for the radius and our direct determination of the mass to derive 55~Cnc~e's radius, mass, and density.

\begin{table}[t]
\caption{Parameters of 55~Cnc~e derived from this work using the transit values given by \cite{Dragomir2014} (see Sect.~\ref{sec:55Cnc}).}
\label{tab:55Cnc}
\centering
\begin{tabular}{c c}
\hline \hline
\multicolumn{2}{c}{55 Cnc e} \\
\hline
 R$_{\rm p}$ [$R_\oplus$]       & 2.031$^{+0.091}_{-0.088}$ \\
M$_{\rm p}$     [$M_\oplus$]    & 8.631 $\pm$ 0.495 \\
$\rho_{\rm p}$ [g.cm$^{-3}$]    & 5.680$^{+0.709}_{-0.749}$ \\
\hline
\end{tabular}
\end{table}
        
The results are given in Table~\ref{tab:55Cnc}. We calculated them using the transit parameters given by \cite{Dragomir2014}. For the planetary mass, we do not consider the error on the inclination $i$ since it is negligible (it implies a variation on the order of $1 \text{\textperthousand}$ on the error on the mass).
Since the stellar radius and density are known, we can express the planetary density as\,
\begin{equation}
\rho_p = \frac{3^{1/3}}{2 \pi^{2/3} G^{1/3}} \rho_\star^{2/3}\,R_\star^{-1}\,TD^{-3/2}\,P^{1/3}\,K\,(1-e^2)^{1/2}  \ ,
\end{equation}
where TD refers to the transit depth caused by the planet. This expression of $\rho_{\rm p}$ is independent of $M_\star$ and directly linked to measured quantities. It therefore allows for a precise estimate of the planetary density with small uncertainties from a standard propagation of errors.
The mass we find (8.631 $\pm$ 0.495 $M_\oplus$) places 55~Cnc~e just below the no-iron line in Fig.7 of \cite{Demory2011} and between the $50 \%$ water and the Earth-like lines of Fig.3 in \cite{Winn2011}. Our results are also in good agreement with the radius and density given by \cite{Dragomir2014} and \cite{Winn2011}, but are more accurate thanks to an accurate and direct determination of the stellar radius and density, since the error bar on $\rho_p$ is dominated by the error on TD. We thus confirm that 55~Cnc~e can be classified as a super-Earth or a mini-Neptune.

These results illustrate that the knowledge of exoplanet characteristics pass through the knowledge of stellar parameters. Their accuracy are decisive in detecting exoplanets potentially hosting life.

\section{Conclusion}
\label{sec:Conclusion}

We performed interferometric measurements with the VEGA/CHARA instrument in visible wavelength to measure the angular diameter of 18 stars. Our measurements are very constraining for adjustments as we reach low $V^2$, and we got many data points. We thus reached an average of $1.9\%$ accuracy on angular diameters. These angular diameters are generally consistent with previous interferometric measurements or with the estimations using the \cite{Kervella2004} empirical law. However, a bigger discrepancy is found toward giant stars and stars with angular diameters larger than 1 mas. Using photometry, we derived the luminosity and effective temperatures with an average precision of $57$~K, which allowed us to place the stars in the H-R diagram. Then, we used PARSEC models to derive stellar masses and ages. To do so, we used a best fit approach and a MC approach to estimate the error bars, which give consistent results except for the few stars whose $L_{\star}$ and $T_{\rm eff, \star}$ appear to not be consistent with the models. For those stars, a better estimation of the metallicity or a different mixing-length parameter should lead to a better match with the models. 

We showed that for the same luminosity and temperature, several solutions can be found, especially for MS stars. Each time, an old solution and a young solution gave ages in the Myrs and the Gyrs range, respectively, in agreement with \cite{Bonfanti2015}. However, the masses are generally similar for the two solutions. In any case, finding the age of a star is not an easy exercise because this parameter depends on the model, and the problem is degenerate. We also showed that the error on the metallicity is very decisive and that not taking an error on the metallicity into account yields a dramatic underestimation of the errors on the age and mass. 

Using orbital parameters given by radial velocity measurements and the newly determined stellar masses, we characterized the ten exoplanetary systems included in our sample; that is, we computed the minimum masses and semi-major axes of exoplanets with reliable error bars. These results reproduce the population of known exoplanets well. We also derived the habitable zone of these systems and verified that two planets lie in their HZ, although they are not Earth-like. We looked in more detail at the super-Earth 55~Cnc~e because it transits its star. We therefore gave a new estimation of its radius, mass, and density, which we find to be more precise than previous measurements.

The progression from interferometric observations of host stars to planetary parameters is complex. It should be realized with care, paying attention to the propagation of errors at each step. This is the price to pay to get accurate values with realistic uncertainties, which benefit from the direct measurement of the radius of the stars.

\begin{acknowledgements}

We acknowledge the anonymous referee, whose constructive and relevant remarks led to a significant improvement of our paper.
R.L. acknowledges the support from the Observatoire de la C\^ote d'Azur after her Ph.D. We acknowledge all the CHARA team and the VEGA team for the interferometric observations.
This research made use of The Extrasolar Planets Encyclopaedia at exoplanet.eu and of the VizieR catalog access tool, CDS, Strasbourg, France.

\end{acknowledgements}

\bibliographystyle{aa}
\bibliography{These}

\newpage

\begin{appendix}

\section{On the distributions of the random variables}
\label{sec:AppendixA}


\subsection{Expected values of $L_\star$ and $T_{\rm eff}$}

Let $F_{\rm bol}$ be the random variable ``bolometric flux of the star'',
$\theta$ the random variable ``angular diameter of the star'' and $d$
the random variable ``distance to the star''. We assume that they are
all independent and Gaussian with respective means and standard
deviations $\bar{F_{\rm bol}}$, $\bar{\theta}$, $\bar{d}$, and $\sigma_{F_{\rm bol}}$,
$\sigma_\theta$, $\sigma_d$.

Let $L_\star=4\pi\,F_{\rm bol}\,d^2$ and $T_{\rm eff}=\beta\,F_{\rm bol}^{1/4}\,\theta^{-1/2}$ be two
other random variables (with $\beta=(4/\sigma_{SB})^{1/4}$).  We note
that $d^2$, hence $L_\star$, do not have a Gaussian distribution. To
compute the expected values of $L_\star$ and $T_{\rm eff}$, let us
note $F_{\rm bol}=\bar{F_{\rm bol}}(1+\tilde{F_{\rm bol}})$, $d=\bar{d}(1+\tilde{d})$, and
$\theta=\bar{\theta}(1+\tilde{\theta})$ so that $\tilde{F_{\rm bol}}$,
$\tilde{d}$ and $\tilde{\theta}$ are three independent Gaussian
variables of mean $0$ and respective standard deviations
$\sigma_{\tilde{F_{\rm bol}}}=\sigma_{F_{\rm bol}}/\bar{F_{\rm bol}}$,
 $\sigma_{\tilde{d}}=\sigma_d/\bar{d}$ and
$\sigma_{\tilde{\theta}}=\sigma_\theta/\bar{\theta}$, assumed to be small:
\begin{eqnarray}
E(L_\star)\!\! & = & E(4\pi\,F_{\rm bol}\,d^2) = E\left(4\pi\,\bar{F_{\rm bol}}\,\bar{d}^2\,(1+\tilde{F_{\rm bol}})(1+\tilde{d})^2\right) \nonumber\\
 & = & 4\pi\,\bar{F_{\rm bol}}\,\bar{d}^2\ E\left(1+\tilde{F_{\rm bol}}+2\tilde{d}+2\tilde{F_{\rm bol}}\tilde{d}+\tilde{d}^2+\tilde{F_{\rm bol}}\tilde{d}^2\right) \nonumber\\
 & = & 4\pi\,\bar{F_{\rm bol}}\,\bar{d}^2\ \big(1 + \underbrace{E(\tilde{F_{\rm bol}})}_{0} + 2\underbrace{E(\tilde{d})}_{0}+2\underbrace{E(\tilde{F_{\rm bol}})}_{0}\underbrace{E(\tilde{d})}_{0} \nonumber\\
 & & \ \hspace{2cm}+E(\tilde{d}^2)+\underbrace{E(\tilde{F_{\rm bol}})}_{0}E(\tilde{d}^2)\big) \nonumber\\
E(L_\star)\!\! & = & 4\pi\,\bar{F_{\rm bol}}\,\bar{d}^2\times\left(1+{\sigma_{\tilde{d}}}^2\right)
,\end{eqnarray}
where we used the linearity of the expected value and the fact that if
$X$ and $Y$ are independent random variables, $E(XY)=E(X)E(Y)$.

One can show similarly that, to second order in $\sigma_{\tilde{F_{\rm bol}}}$ and $\sigma_{\tilde{\theta}}$, 
\begin{equation}
E(T_{\rm eff}) = \beta\,\bar{F_{\rm bol}}^{1/4}\,\bar\theta^{-1/2}\times\left(1+\frac{12(\sigma_{\tilde{\theta}})^2-3(\sigma_{\tilde{F_{\rm bol}}})^2}{32}\right)
.\end{equation}

\subsection{Covariance and correlation of $L_\star$ and $T_{\rm eff}$}

Now that we have $E(L_\star)$ and $E(T_{\rm eff})$, we can compute their covariance\,:
$Cov(L_\star,T_{\rm eff}) = E\left([L_\star-E(L_\star)][T_{\rm eff}-E(T_{\rm eff})]\right)$. Using similar arguments
to those above and expanding to second order again, one finds
\begin{equation}
Cov(L_\star,T_{\rm eff}) = \frac{1}{4}\bar{L_\star}\bar{T_{\rm eff}}\,{\sigma_{\tilde{F_{\rm bol}}}}^2\ ,\\ \\ 
\end{equation}
noting
$\bar{L_\star}=4\pi\bar{F_{\rm bol}}\bar{d}^2 < E(L_\star)$ and
$\bar{T_{\rm eff}}=\beta\,\bar{F_{\rm bol}}^{1/4}\,\bar\theta^{-1/2}\neq
E(T_{\rm eff})$\,.

From this, we see that if $\sigma_{F_{\rm bol}}=0$, i.e. in the ideal case where
$F_{\rm bol}$ were known exactly without uncertainty, then $L_\star$ and
$T_{\rm eff}$ are not correlated\footnote{which does not mean that they are
  independent\,!}. In the $L_\star-T_{\rm eff}$ plane, the cloud of points drawn by our
Monte-Carlo algorithm will be fitted by ellipses with vertical and
horizontal axes.  In contrast
$\sigma_{L_\star}=\bar{L_\star}\,\sigma_{\tilde{F_{\rm bol}}}$ and
$\sigma_{T_{\rm eff}}=\frac14\bar{T_{\rm eff}}\sigma_{\tilde{F_{\rm bol}}}$ in the case where
$\sigma_d=\sigma_\theta=0$, so that in the end the
correlation between $L_\star$ and $T_{\rm eff}$ is
$Corr(L_\star,T_{\rm eff})=Cov(L_\star,T_{\rm eff})/\sigma_{L_\star}\sigma_{T_{\rm eff}}=1$\,; $L_\star$ and $T_{\rm eff}$ are perfectly
correlated and in the $L_\star-T_{\rm eff}$ plane, the cloud of points will
look like a thin diagonal line
this time. From these two extreme cases, one can
have a feeling of how important the correlation is by comparing
$\sigma_{\tilde{F_{\rm bol}}}$ with $\sigma_{\tilde{d}}$ and
$\sigma_{\tilde{\theta}}$\,. In general, they are of the same order of
magnitude, showing the importance of taking this correlation into
account by using our algorithm instead of drawing $L_\star$ and $T_{\rm eff}$
independently.



\onecolumn

\centering
\section{Additional table and figures}
\begin{longtable}{lllllll}

\caption{\label{tab:ObsJournal} Journal of observations}\\
\hline\hline

\textbf{HD} & Seq. & B & $V^2$ & $eV^2$ & $\lambda$ & $\Delta\lambda$ \\
MJD                     &               &       (m) &           &               &               (nm)    &       (nm) \\
\hline
\endfirsthead
\caption{continued.}\\
\hline\hline 

\textbf{HD} & Seq. & B & $V^2$ & $eV^2$ & $\lambda$ & $\Delta\lambda$ \\
MJD                     &               &       (m) &           &               &               (nm)    &       (nm) \\

\hline
\endhead
\hline
\endfoot

          \textbf{HD3651} &             &                       &                       &                       &               &               \\      
                        56163.5 &C1-T-C2&       155.64  &       0.252   &       0.050   &       707     &       15      \\      
                        56163.5 &                                       &       155.64  &       0.279   &       0.050   &       735     &       15      \\      
                        56164.5 &C1-T-C1&       147.23  &       0.210   &       0.031   &       718     &       15      \\      
                        56164.5 &                                       &       210.43  &       0.008   &       0.050   &       718     &       15      \\      
                        56495.5 &T-C3   &       140.08  &       0.273   &       0.035   &       700     &       20      \\      
                        56495.5 &                                       &       200.88  &       0.032   &       0.050   &       700     &       20      \\      
                        56495.5 &                               &       140.08  &       0.347   &       0.067   &       710     &       20      \\      
                        56495.5 &C3-T   &       149.52  &       0.280   &       0.018   &       700     &       20      \\      
                        56495.5 &                                       &       149.52  &       0.221   &       0.033   &       710     &       20      \\      
                        56495.5 &                                       &       149.52  &       0.273   &       0.031   &       730     &       20      \\      
                        56495.5 & T-C3& 151.92  &       0.219   &       0.017   &       700     &       20      \\      
                        56495.5 &                                       &       216.60  &       -0.028  &       0.045   &       700     &       20      \\      
                        56495.5 &                                       &       151.92  &       0.232   &       0.027   &       710     &       20      \\      
                        56495.5 &                                       &       151.92  &       0.296   &       0.038   &       730     &       20      \\      
                        56495.5 &                                       &       216.60  &       -0.010  &       0.050   &       730     &       20      \\      
                        56502.5 &C2-T-C1&       33.18   &       0.893   &       0.187   &       700     &       20      \\      
                        56502.5 &                                       &       170.03  &       0.190   &       0.040   &       700     &       20      \\      
                        56502.5 &                                       &       202.70  &       -0.010  &       0.065   &       700     &       20      \\      
                        56502.5 &                                       &       170.03  &       0.168   &       0.038   &       710     &       20      \\      
                        56502.5 &                                       &       202.70  &       0.014   &       0.072   &       710     &       20      \\      
                        56502.5 &C1-T-C3        &       33.22   &       0.944   &       0.143   &       710     &       20      \\      
                        56502.5 &                                       &       171.41  &       0.084   &       0.083   &       710     &       20      \\      
                        56502.5 &C3-T-C1        &       33.33   &       0.932   &       0.195   &       700     &       20      \\      
                        56502.5 &                                       &       172.99  &       0.119   &       0.027   &       700     &       20      \\      
                        56502.5 &                                       &       172.99  &       0.226   &       0.037   &       710     &       20      \\      
                        56502.5 &                                       &       205.75  &       -0.001  &       0.061   &       710     &       20      \\      
                                        &                                       &                       &                       &                       &               &               \\      
                \textbf{HD19994}&                                       &                       &                       &                       &               &               \\      
                        56197.5 &C4-T   &       28.42&  0.940   &       0.129   &       710     &       20      \\      
                        56197.5 &                                       &       150.12  &       0.090         &       0.114   &       710     &       20      \\      
                        56197.5 &                                       &       177.90  &       0.044         &       0.050   &       710     &       20      \\      
                        56197.5 &                                       &       28.42   &       0.890         &       0.143   &       730     &       20      \\      
                        56227.5 & C5-T- C6&     153.58  &       0.158   &       0.050   &       705     &       15      \\      
                        56227.5 &                                       &       219.09  &       0.034         &       0.050   &       705     &       15      \\      
                        56227.5 &                                       &       153.58  &       0.155         &       0.050   &       720     &       15      \\      
                        56227.5 &                                       &       219.09  &       -0.013  &       0.050   &       720     &       15      \\      
                        56227.5 &                                       &       153.58  &       0.201         &       0.050   &       735     &       15      \\      
                        56229.5 & C4-T-C6  &    29.66   &       0.852   &       0.050   &       643     &       15      \\      
                        56229.5 &                                       &       29.66   &       0.913         &       0.050   &       658     &       15      \\      
                        56229.5 &                                       &       29.66   &       0.934         &       0.050   &       672     &       13      \\      
                        56529.5 & C6-T-C6 &     152.81  &       0.214   &       0.054   &       730     &       20      \\      
                        56529.5 &                                       &       215.65  &       0.013         &       0.075   &       730     &       20      \\      
                        56534.5 &C6-T   &       28.10   &       1.000   &       0.195   &       710     &       20      \\      
                        56534.5 & C6-T-C6 &     146.53  &       0.297   &       0.050   &       710     &       20      \\      
                        56534.5 &                                       &       151.20  &       0.197         &       0.050   &       710     &       20      \\      
                        56534.5 &                                       &       179.10  &       0.069         &       0.067   &       710     &       20      \\      
                        56533.5 &                                       &       142.17  &       0.304         &       0.050   &       730     &       25      \\      
                        56533.5 &                                       &       169.32  &       0.054         &       0.053   &       730     &       25      \\      
                        56534.5 &                                       &       146.53  &       0.242         &       0.050   &       730     &       25      \\      
                        56534.5 &                                       &       151.20  &       0.246         &       0.050   &       730     &       25      \\      
                        56534.5 &                                       &       179.10  &       0.100         &       0.050   &       730     &       25      \\      
                        56533.5 &                                       &       27.82   &       0.896         &       0.079   &       748     &       25      \\      
                        56533.5 &                                       &       169.32  &       0.123         &       0.053   &       748     &       25      \\      
                        56534.5 &                                       &       151.20  &       0.326         &       0.050   &       748     &       25      \\      
                        56534.5 &                                       &       179.10  &       0.083         &       0.050   &       748     &       25      \\      
                                        &                                       &                       &                       &                       &               &               \\      
        \textbf{HD75732}        &                                       &                       &                       &                       &               &               \\      
                        56034.5 & C7-T-C7&      56.96   &       0.843   &       0.050   &       708     &       15      \\      
                        56034.5 &                                       &       276.00  &       0.021         &       0.056   &       708     &       15      \\      
                        56034.5 &                                       &       276.05  &       -0.029  &      0.050   &       720     &       16      \\      
                        56034.5 &                                       &       56.96   &       0.857         &       0.050   &       733     &       15      \\      
                        56034.5 &                               &       276.00  &       -0.063  &      0.093   &       733     &       15      \\      
                        56033.5 & C7-T-C7&      56.33   &       0.853   &       0.050   &       720     &       16      \\      
                        56033.5 &                                       &       56.33   &       0.858         &       0.050   &       733     &       15      \\      
                        56033.5 &                                       &       244.57  &       0.044         &       0.052   &       733     &       15      \\      
                        55887.5 & C8-T-C8&      174.23  &       0.151   &       0.050   &       708     &       15      \\      
                        55887.5 &                                       &       174.70  &       0.204         &       0.069   &       708     &       15      \\      
                        55888.5 & C8-T& 107.87  &       0.525   &       0.095   &       708     &       15      \\      
                        55887.5 & C8-T-C8&      174.23  &       0.158   &       0.050   &       720     &       16      \\      
                        55887.5 &                                       &       174.70  &       0.115         &       0.061   &       720     &       16      \\      
                        55888.5 & C8-T& 107.87  &       0.459   &       0.050   &       720     &       16      \\      
                        55888.5 &                                       &       177.20  &       0.131         &       0.086   &       720     &       16      \\      
                        55887.5 & C8-T-C8&      99.01   &       0.595   &       0.078   &       733     &       15      \\      
                        55887.5 &                                       &       174.69  &       0.079         &       0.073   &       733     &       15      \\      
                        55887.5 &                                       &       238.14  &       0.050         &       0.077   &       733     &       15      \\      
                        55888.5 & C8-T& 107.89  &       0.584   &       0.097   &       733     &       15      \\      
                        55888.5 &                                       &       177.22  &       0.219         &       0.086   &       733     &       15      \\      
                                        &                                       &                       &                       &                       &               &               \\      
        \textbf{HD167042}       &                                       &                       &                       &                       &               &               \\      
                56190.5 &       C9-T                    &       64.77   &   0.579   &  0.050    & 710     &       20    \\         
                56191.5 &       C9-T-C9                 &       65.48   &   0.639   &  0.050    & 710     &       20    \\          
                56190.5 &       C9-T                    &       64.78   &   0.420   &  0.033    & 539.5 & 15    \\          
                56190.5 &       C9-T                    &       151.34  &   0.081   &  0.125    & 539.5 & 15    \\          
                56190.5 &       C9-T                    &       64.77   &   0.607   &  0.050    & 730     &       20    \\          
                56190.5 &       C9-T                    &       215.89  &   -0.057  &  0.115    & 730     &       20    \\          
                56191.5 &       C9-T-C9                 &       65.48   &   0.617   &  0.050    & 730   & 20    \\         
                56191.5 &       C9-T-C9                 &       219.74  &   -0.064  &  0.138    & 730   & 20    \\         
                56191.5 &       C9-T-C9                 &       65.48   &   0.405   &  0.151    & 552.5 & 15    \\        
                56190.5 &       C9-T                    &       64.78   &   0.410   &  0.050    & 539.5 & 15    \\          
                56190.5 &       C9-T                    &       151.34  &   0.086   &  0.093    & 539.5 &         15    \\          
                56190.5 &       C9-T                    &       64.77   &   0.639   &  0.050    & 747.5 &         25    \\         
                56190.5 &       C9-T                    &       215.89  &   0.097   &  0.135    & 747.5 &         25    \\          
                56501.5 &       C9-T-C9                 &       64.81   &   0.590   &  0.050    & 710   &         20    \\          
                56501.5 &       C9-T-C9                 &       64.81   &   0.553   &  0.083    & 727.5 & 25    \\          
                                        &                                       &                       &                       &                       &               &               \\      
        \textbf{HD173416} &                                     &                       &                       &                       &               &               \\      
                        56533.5 & C10-T-C10             &       65.58   &       0.714         &       0.072   &       710     &       20      \\      
                        56533.5 &                                       &       155.80  &       0.073         &       0.070   &       710     &       20      \\      
                        56533.5 &                                       &       221.07  &       0.019         &       0.082   &       710     &       20      \\      
                        56533.5 & C10-T-C10                     & 64.17 &       0.586         &       0.050   &       710     &       20      \\      
                        56533.5 &                                       &       152.36  &       0.020         &       0.083   &       710     &       20      \\      
                        56533.5 &C10-T-C10                      & 65.58 &       0.596         &       0.082   &       730     &       20      \\      
                        56533.5 &                                       &       155.80  &       0.086         &       0.050   &       730     &       20      \\      
                        56533.5 &                                       &       221.07  &       -0.044  &      0.050   &       730     &       20      \\      
                        56533.5 & C10-T-C10     & 64.17 &       0.698   &       0.089   &       730     &       20      \\      
                        56533.5 &                                       &       152.36  &       0.046         &       0.050   &       730     &       20      \\      
                                        &                                       &                       &                       &                       &               &               \\      
        \textbf{HD190360} &                                     &                       &                       &                       &               &               \\      
                        56496.3 &        C10-T-C10              &       175.50  &       0.277   &       0.034   &       700     &       20      \\      
                        56496.3 &                                       &       33.97   &       0.917   &       0.040   &       710     &       20      \\      
                        56496.3 &                                       &       175.50  &       0.241   &       0.050   &       710     &       20      \\      
                        56496.3 &                                       &       175.50  &       0.291   &       0.028   &       727.5   &       25      \\      
                        56496.3 &       C10-T-C10               &       33.98   &       0.879   &       0.062   &       700     &       20      \\      
                        56496.3 &                                       &       176.00  &       0.172   &       0.052   &       700     &       20      \\      
                        56496.3 &                                       &       33.98   &       0.927   &       0.074   &       710     &       20      \\      
                        56501.4 &       C11-T-C12               &       176.13  &       0.214   &       0.023   &       700     &       20      \\      
                        56501.4 &                                       &       243.67  &       -0.040  &       0.105   &       700     &       20      \\      
                        56501.4 &                                       &       176.13  &       0.226   &       0.016   &       710     &       20      \\      
                        56501.4 &                                       &       176.13  &       0.163   &       0.039   &       730     &       20      \\      
                        56501.4 &       C12-T-C12               &       174.19  &       0.262   &       0.020   &       700     &       20      \\      
                        56501.4 &                                       &       174.11  &       0.197   &       0.058   &       710     &       20      \\      
                        56501.4 &                                       &       174.11  &       0.281   &       0.026   &       730     &       20      \\      
                        56123.3 &       C11-T-C11               &       218.43  &       0.124   &       0.060   &       730     &       20      \\      
                        56124.3 &       C11-T                   &       221.80  &       0.050   &       0.070   &       730     &       20      \\      
                        56130.4 &        C11-T                  &       152.66  &       0.410   &       0.098   &       707.5   &       15      \\      
                        56130.4 &                                       &       152.73  &       0.416   &       0.110   &       730     &       20      \\      
                        56130.4 &                                       &       248.07  &       -0.001  &       0.094   &       730     &       20      \\      
                                          &                                     &                       &                       &                       &               &               \\      
        \textbf{HD217014}       &                                       &                       &                       &                       &               &               \\      
                        56164.5 & C12-T-C1&     64.55   &       0.870   &       0.035   &       708     &       15      \\      
                        56164.5 &                                       &       64.55   &       0.842         &       0.027   &       720     &       20      \\      
                        56164.5 &                                       &       64.55   &       0.886         &       0.024   &       733     &       15      \\      
                        56164.5 &                                       &       150.70  &       0.270         &       0.025   &       720     &       20      \\      
                        56164.5 &                                       &       150.70  &       0.300         &       0.047   &       733     &       15      \\      
                        56164.5 &                                       &       215.02  &       0.009         &       0.037   &       720     &       20      \\      
                        56164.5 &                                       &       215.02  &       0.049         &       0.064   &       733     &       15      \\      
                        56495.5 & C12-T-C12&    168.42  &       0.189   &       0.039   &       700     &       20      \\      
                        56495.5 &                                       &       168.42  &       0.289         &       0.022   &       710     &       20      \\      
                        56495.5 &                                       &       168.42  &       0.236         &       0.050   &       730     &       20      \\      
                        56495.5 & C12-T-C12&    168.07  &       0.257   &       0.036   &       700     &       20      \\      
                        56495.5 &                                       &       168.07  &       0.276         &       0.058   &       710     &       20      \\      
                        56495.5 &                                       &       168.07  &       0.281         &       0.046   &       730     &       20      \\      
                                        &                                       &                       &                       &                       &               &               \\      
        \textbf{HD1367}         &                                       &                       &                       &                       &               &               \\      
                        56196.5 &C4-T-C4        &       63.25   &       0.740         &       0.050   &       727     &       30      \\      
                        56196.5 &                                       &       153.10  &       0.190         &       0.050   &       727     &       30      \\      
                        56226.5 & C4-T-C5       &       65.10   &       0.721         &       0.051   &       705     &       15      \\      
                        56226.5 &                                       &       155.92  &       0.228         &       0.050   &       705     &       15      \\      
                        56226.5 &                                       &       220.87  &       0.000         &       0.073   &       705     &       15      \\      
                        56227.5 & C4-T-C5       &       154.51  &       0.204         &       0.050   &       705     &       15      \\      
                        56227.5 &                                       &       154.51  &       0.209         &       0.050   &       720     &       15      \\      
                        56227.5 &                                       &       218.42  &       0.051         &       0.050   &       720     &       15      \\      
                        56227.5 &                                       &       154.51  &       0.228         &       0.050   &       735     &       15      \\      
                        56227.5 &                                       &       218.42  &       0.000         &       0.050   &       735     &       15      \\      
                                        &                                       &                       &                       &                       &               &               \\      
        \textbf{HD1671} &                                       &                       &                       &                       &               &               \\      
                        56198.5 &T-C13 &        210.97  &       0.111   &       0.050   &       710     &       20      \\      
                        56198.5 &                                       &       271.69  &       0.007         &       0.050   &       730     &       20      \\      
                        56198.5 & C13-T-C13     &       210.97  &       0.167         &       0.050   &       710     &       20      \\      
                        56198.5 &                                       &       276.04  &       -0.015  &      0.050   &       710     &       20      \\      
                        56198.5 &                                       &       210.97  &       0.120         &       0.050   &       730     &       20      \\      
                        56198.5 &                                       &       276.04  &       0.044         &       0.050   &       730     &       20      \\      
                        56528.5 & C13-T-C14     &       150.46  &       0.426         &       0.050   &       710     &       20      \\      
                        56528.5 &                                       &       213.65  &       0.093         &       0.050   &       710     &       20      \\      
                        56528.5 &                                       &       150.46  &       0.436         &       0.050   &       730     &       20      \\      
                        56528.5 &                                       &       63.53   &       0.873         &       0.050   &       748     &       25      \\      
                        56529.5 & C14-T-C14     &       177.42  &       0.259         &       0.050   &       710     &       20      \\      
                        56529.5 &                                       &       210.90  &       0.131         &       0.050   &       710     &       20      \\      
                        56529.5 &                                       &       177.42  &       0.351         &       0.050   &       730     &       20      \\      
                        56533.5 &                                       &       210.95  &       0.091         &       0.050   &       710     &       20      \\      
                        56533.5 & C15-T-C15     &       138.25  &       0.466         &       0.050   &       710     &       20      \\      
                        56533.5 &                                       &       197.37  &       0.120         &       0.050   &       710     &       20      \\      
                        56533.5 &                                       &       138.25  &       0.488         &       0.050   &       730     &       20      \\      
                        56533.5 &                                       &       197.37  &       0.191         &       0.050   &       730     &       20      \\      
                        56533.5 &                                       &       59.44   &       0.888         &       0.050   &       748     &       25      \\      
                        56533.5 &                                       &       138.25  &       0.451         &       0.050   &       748     &       25      \\      
                        56533.5 & C13-T-C13     &       177.44  &       0.273         &       0.050   &       730     &       20      \\      
                        56533.5 &                                       &       210.95  &       0.097         &       0.050   &       730     &       20      \\      
                        56533.5 &                                       &       177.44  &       0.290         &       0.050   &       748     &       25      \\      
                        56533.5 &                                       &       210.95  &       0.103         &       0.050   &       748     &       25      \\      
                                        &                                       &                       &                       &                       &               &               \\      
        \textbf{HD154633}       &                                       &                       &                       &                       &               &               \\      
                        55722.5 & C16-T-C16     &       202.47  &       0.035         &       0.050   &       715     &       20      \\      
                        55722.5 &                                       &       202.47  &       0.079         &       0.050   &       735     &       20      \\      
                        55721.5 & C16-C17-T-C16 &       105.13  &       0.425         &       0.066   &       711     &       18      \\      
                        55721.5 &                                       &       140.77  &       0.220         &       0.050   &       711     &       18      \\      
                        55721.5 &                                       &       233.81  &       0.056         &       0.052   &       711     &       18      \\      
                        55721.5 & C16-C17-T-C16 &       105.13  &       0.437         &       0.050   &       735     &       20      \\      
                        55721.5 &                                       &       140.77  &       0.243         &       0.050   &       735     &       20      \\      
                        55721.5 & C16--T-C16&   107.74  &       0.458   &       0.050   &       711     &       18      \\      
                        55721.5 &                                       &       247.24  &       -0.003         &       0.050   &       711     &       18      \\      
                        55721.5 & C16--T-C16&   107.74                  &       0.466         &       0.050   &       735     &       20      \\      
                        55721.5 &                                       &       151.27  &       0.199         &       0.050   &       735     &       20      \\      
                                        &                                       &                       &                       &               &               &               \\      
        \textbf{HD161178}       &                                       &                       &                       &                       &               &               \\      
                        56165.5 &C16-T-C16&     65.26   &       0.582   &       0.050   &       710     &       20      \\      
                        56165.5 &                                       &       155.92  &       0.010         &       0.050   &       710     &       20      \\      
                        56165.5 &                                       &       220.96  &       -0.032         &       0.050   &       710     &       20      \\      
                        56165.5 &C16-T-C16&     65.26   &       0.600   &       0.050   &       730     &       20      \\      
                        56168.5 &C16-T-C18&     101.66  &       0.391   &       0.050   &       710     &       20      \\      
                        56168.5 &                                       &       101.66  &       0.474         &       0.050   &       730     &       20      \\      
                        56168.5 &                                       &       155.97  &       0.042         &       0.050   &       730     &       20      \\      
                                        &                                       &                       &                       &                       &               &               \\      
        \textbf{HD168151} &                                     &                       &                       &                       &               &               \\      
                        55722.5 &C16-T-C16&     105.05  &       0.672   &       0.080   &       735     &       20      \\      
                        55722.5 &                                       &       199.04  &       0.141         &       0.050   &       735     &       20      \\      
                        55722.5 &                                       &       288.56  &       -0.043         &       0.063   &       735     &       20      \\      
                        55722.5 & C16-T-C16     &       107.72  &       0.522         &       0.067   &       735     &       20      \\      
                        55722.5 &                                       &       214.20  &       0.110         &       0.050   &       735     &       20      \\      
                        55722.5 &                                       &       306.93  &       -0.054         &       0.065   &       735     &       20      \\      
                        55721.5 &C16-T-C17-C16  &       104.07  &       0.529         &       0.050   &       735     &       20      \\      
                        55721.5 &                                       &       137.01  &       0.410         &       0.055   &       735     &       20      \\      
                        55721.5 &                                       &       228.95  &       0.048         &       0.050   &       735     &       20      \\      
                        55721.5 &C16-T-C17-C16& 107.36  &       0.579   &       0.051   &       735     &       20      \\      
                        55721.5 &                                       &       149.40  &       0.351         &       0.057   &       735     &       20      \\      
                        55721.5 &                                       &       244.86  &       0.070         &       0.050   &       735     &       20      \\      
                                        &                                       &                       &                       &                       &               &               \\      
        \textbf{HD209369} &                                     &                       &                       &                       &               &               \\      
                        55767.4 &T-C19& 106.06  &       0.592   &       0.05    &       710     &       16      \\      
                        55767.4 &                               &       144.00  &       0.272   &       0.05    &       710     &       16      \\      
                        55774.3 &C19-T-C19      &       107.47  &       0.597   &       0.05    &       645.5   &       15      \\      
                        55774.3 &                               &       144.82  &       0.416   &       0.05    &       645.5   &       15      \\      
                        55802.5 &C19-T  &       103.21  &       0.732   &       0.05    &       737.5   &       15      \\      
                        55802.5 &                               &       155.75  &       0.371   &       0.05    &       737.5   &       15      \\      
                        55802.5 & C20-T-C19     &       102.08  &       0.660   &       0.05    &       737.5   &       15      \\      
                        55802.5 &                               &       155.77  &       0.341   &       0.05    &       737.5   &       15      \\      
                        55803.4 &C19-T-C20      &       104.83  &       0.620   &       0.05    &       737.5   &       15      \\      
                        55803.4 &                               &       155.41  &       0.293   &       0.05    &       737.5   &       15      \\      
                        55803.4 &                               &       251.17  &       0.079   &       0.05    &       737.5   &       15      \\      
                                                &                               &                       &                       &                       &                       &                 \\      %
        \textbf{HD218560}       &                                       &                       &                       &                       &               &               \\      
                        55768.5 &C19-T-C19      &       64.18   &       0.704         &       0.060   &       710     &       16      \\      
                        55768.5 &                                       &       153.18  &       0.076         &       0.050   &       710     &       16      \\      
                        55768.5 &                                       &       217.05  &       -0.004         &       0.050   &       710     &       16      \\      
                        55768.5 &C19-T& 65.48   &       0.768   &       0.054   &       710     &       16      \\      
                        55768.5 &                                       &       155.86  &       0.109         &       0.050   &       710     &       16      \\      
                        55802.5 &C19-T-C19      &       107.79  &       0.283         &       0.050   &       710     &       16      \\      
                        55802.5 &                                       &       151.98  &       0.082         &       0.050   &       710     &       16      \\      
                        55802.5 &                                       &       247.99  &       0.049         &       0.060   &       710     &       16      \\      
                        55802.5 &C19-T-C19      &       107.79  &       0.274         &       0.050   &       710     &       16      \\      
                        55802.5 &                                       &       151.98  &       0.076         &       0.050   &       710     &       16      \\      
                        55802.5 &                                       &       247.99  &       0.046         &       0.057   &       710     &       16      \\      
\hline

\end{longtable}
\twocolumn

\begin{figure*}
\begin{center}$
\begin{array}{cc}
\vspace*{-1.cm}
\hspace*{1cm}
\includegraphics[scale=0.5]{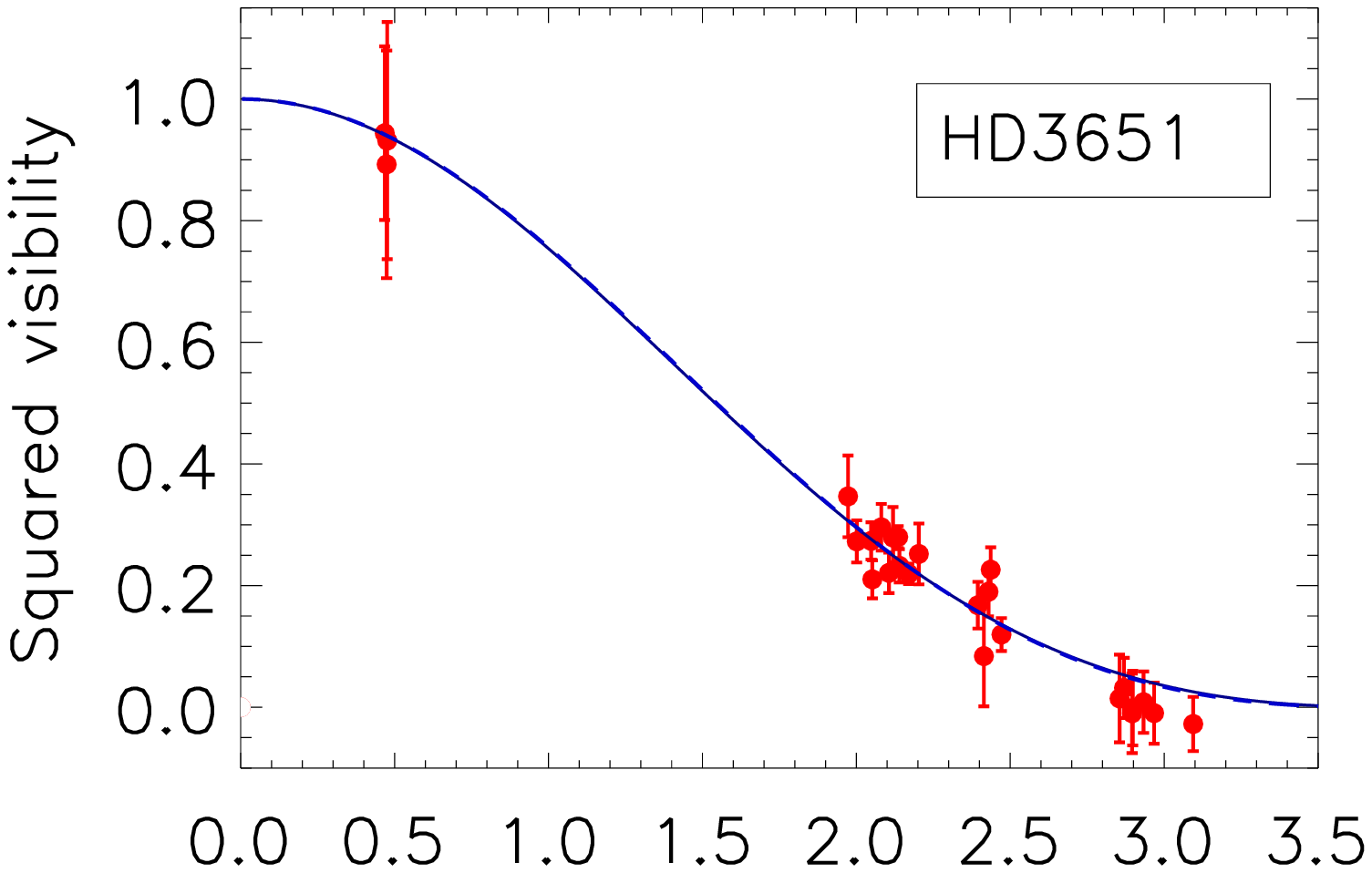} &
\hspace*{-2cm}
\includegraphics[scale=0.5]{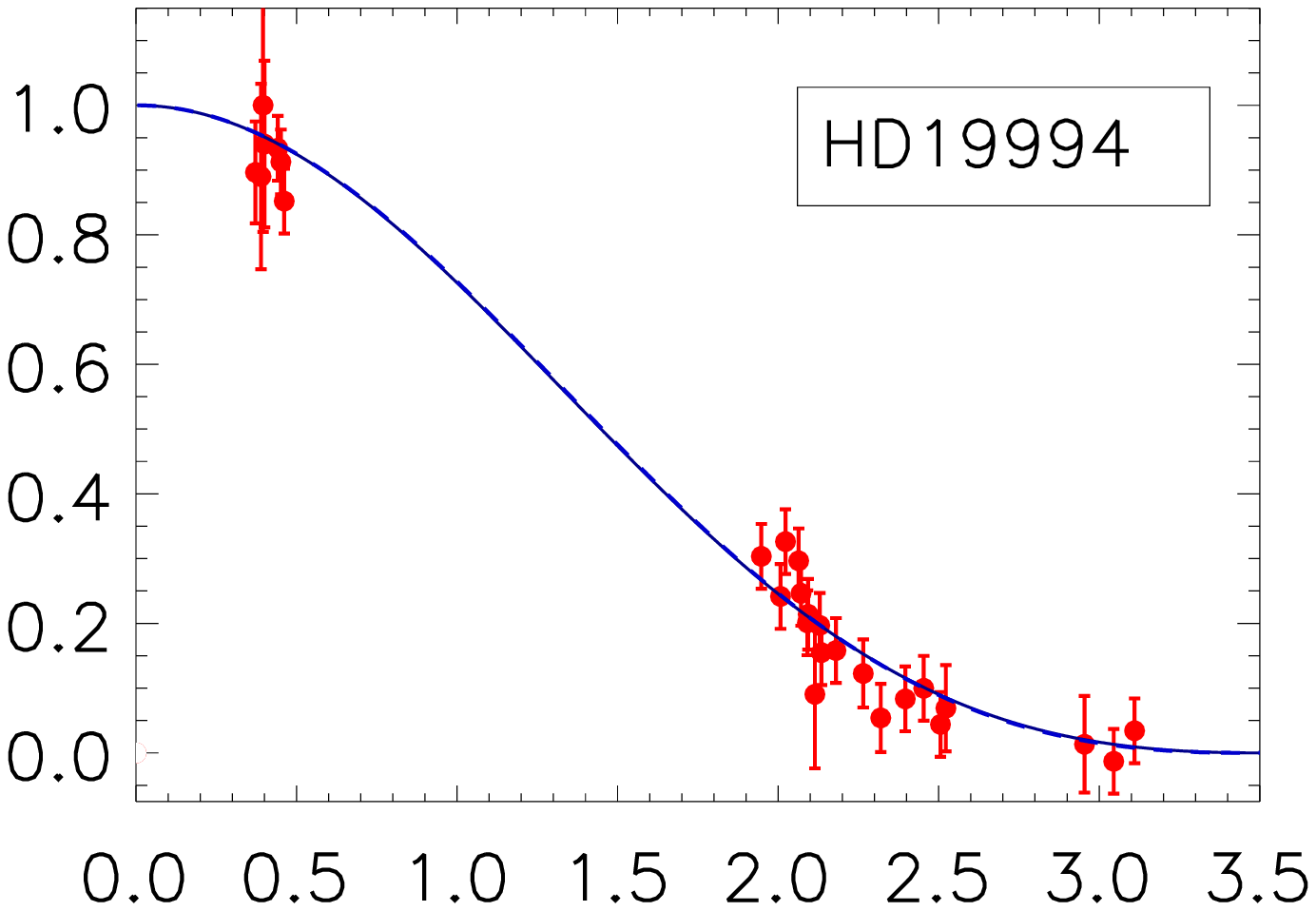} \\
\vspace*{-1.4cm}
\hspace*{1cm}
\includegraphics[scale=0.5]{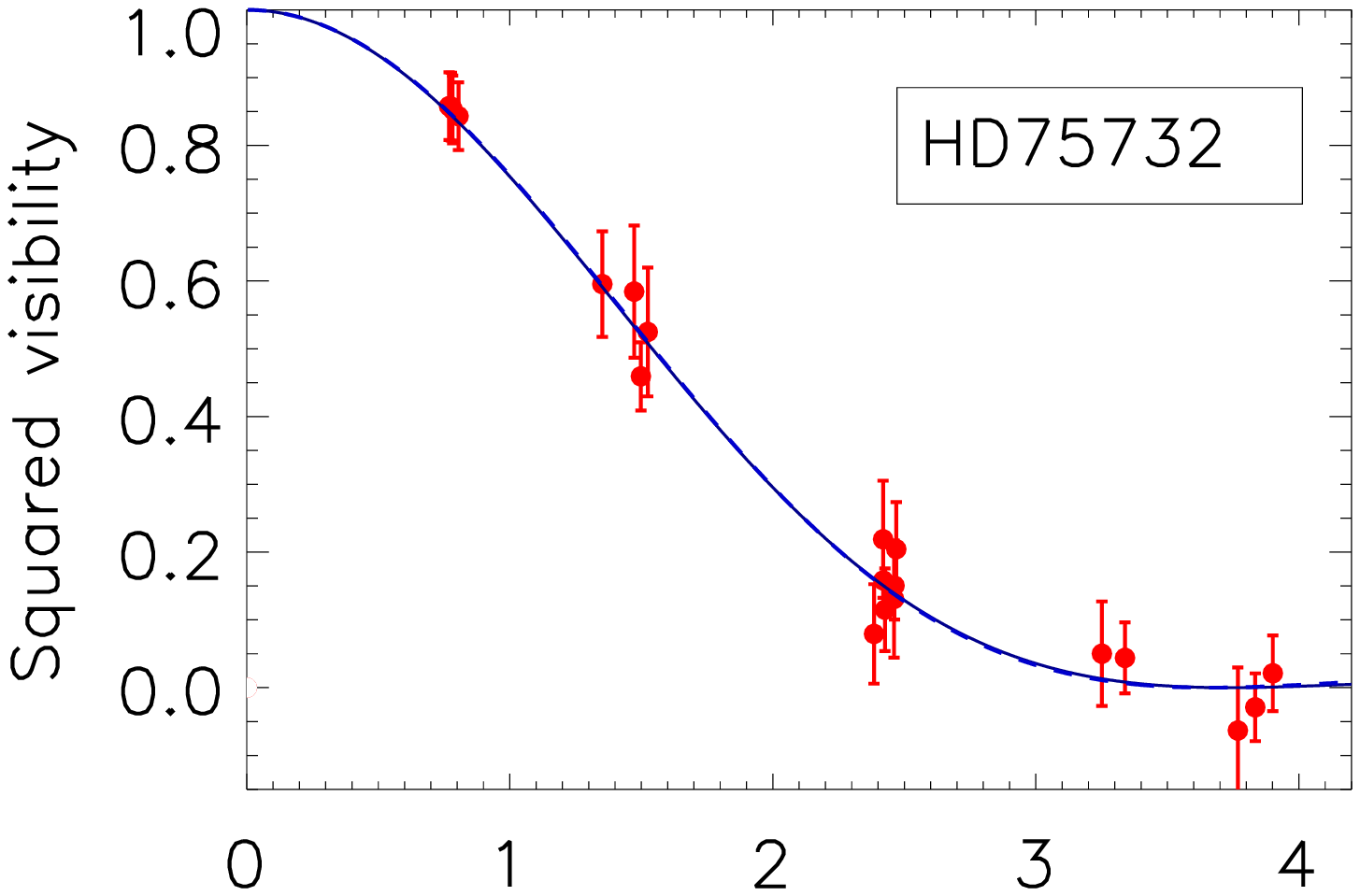} &
\hspace*{-2cm}
\includegraphics[scale=0.5]{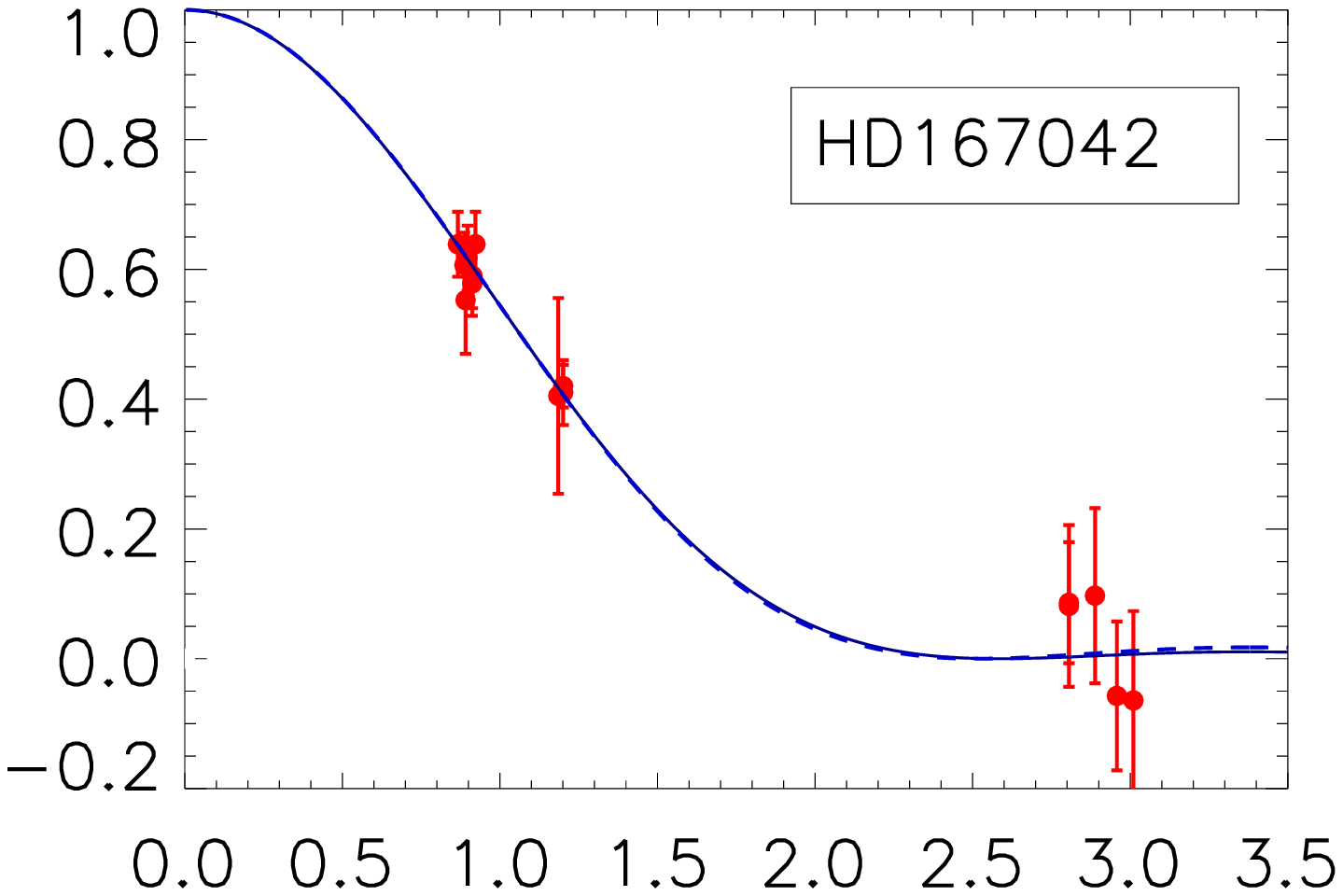}\\
\hspace*{1cm}
\vspace*{-1.4cm}
\includegraphics[scale=0.5]{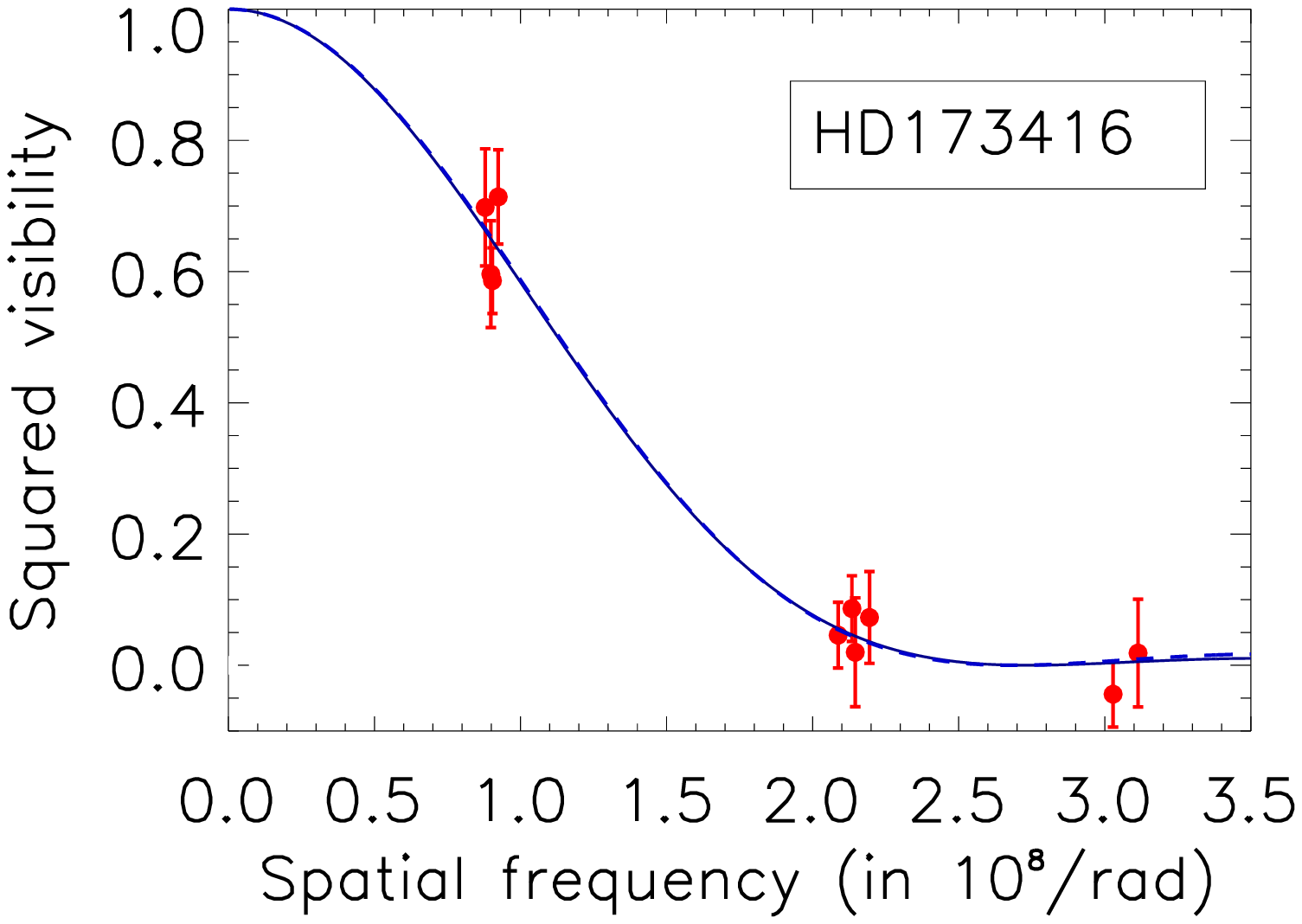}&
\hspace*{-2cm}
\includegraphics[scale=0.5]{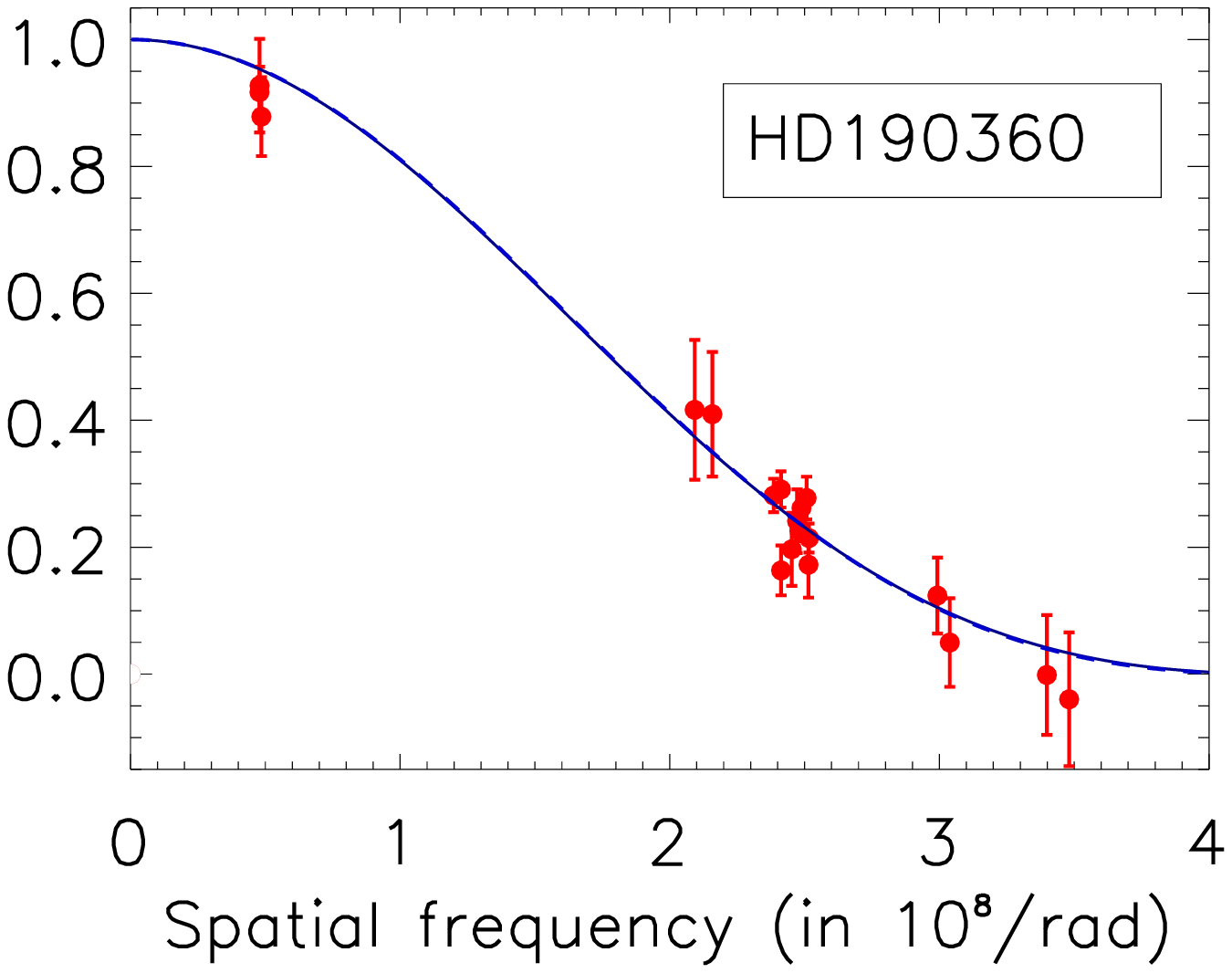}\\
\vspace*{0.2cm}
\end{array}$
\includegraphics[scale=0.5]{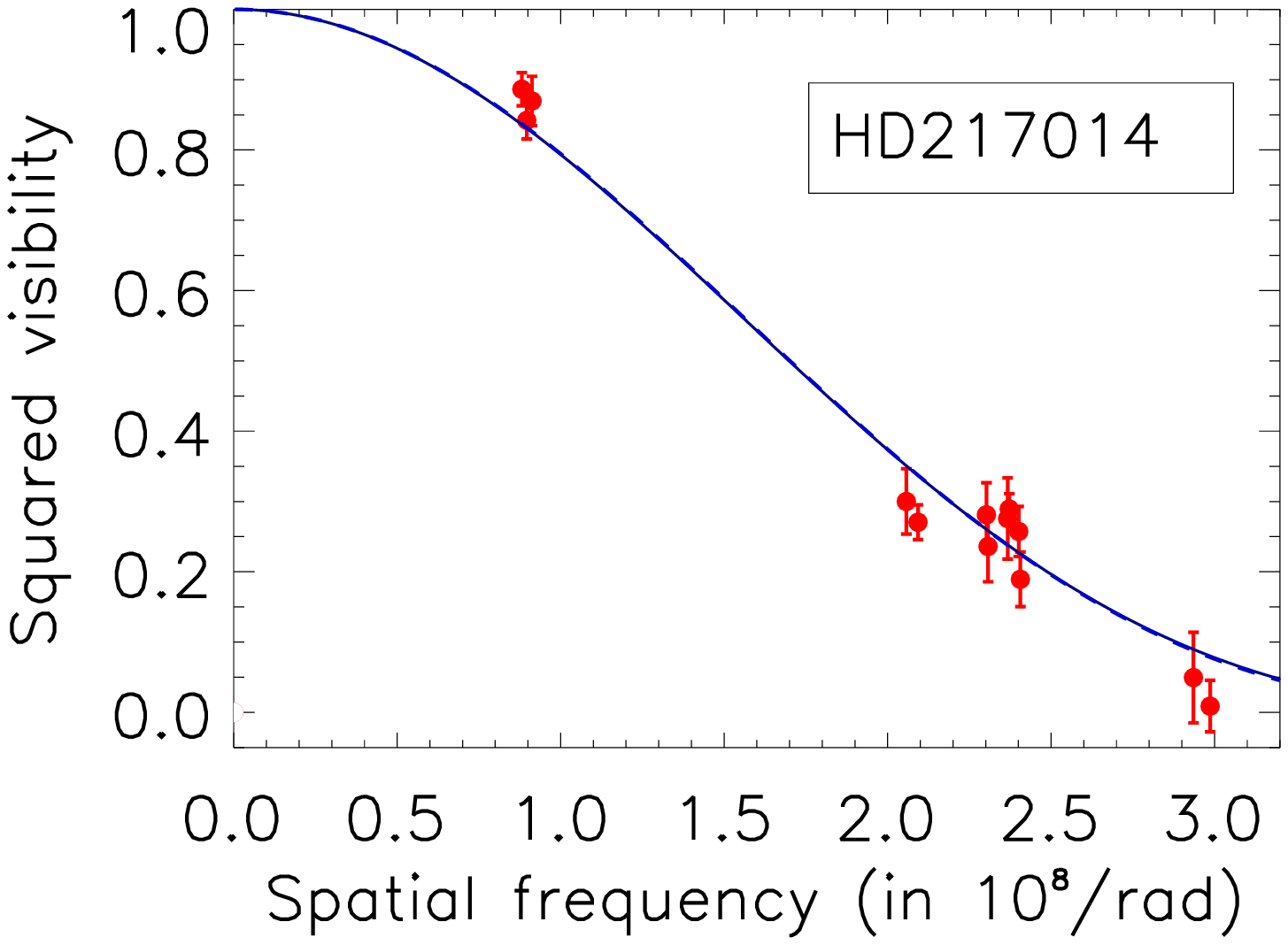}
\end{center}
\caption{Squared visibilities of exoplanet host stars. The solid line represents the model of LD diameter and the dashed line the UD diameter (see Sect.~\ref{sec:Observations}).}
\label{fig:VisHostStars}
\end{figure*}

\begin{figure*}[ht]
\begin{center}$
\begin{array}{cc}
\vspace*{-1.cm}
\hspace*{1cm}
\includegraphics[scale=0.5]{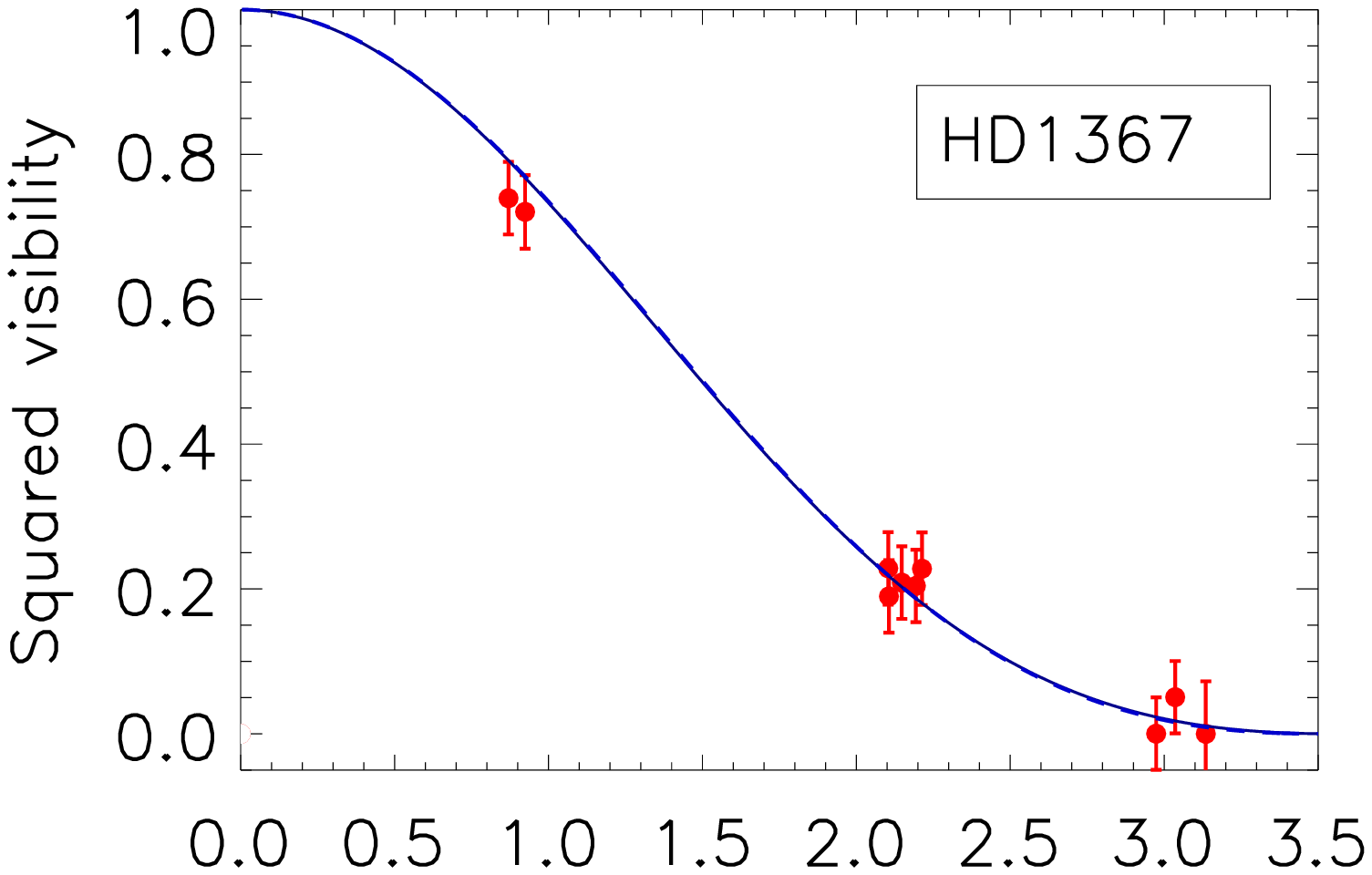}&
\hspace*{-2cm}
\includegraphics[scale=0.5]{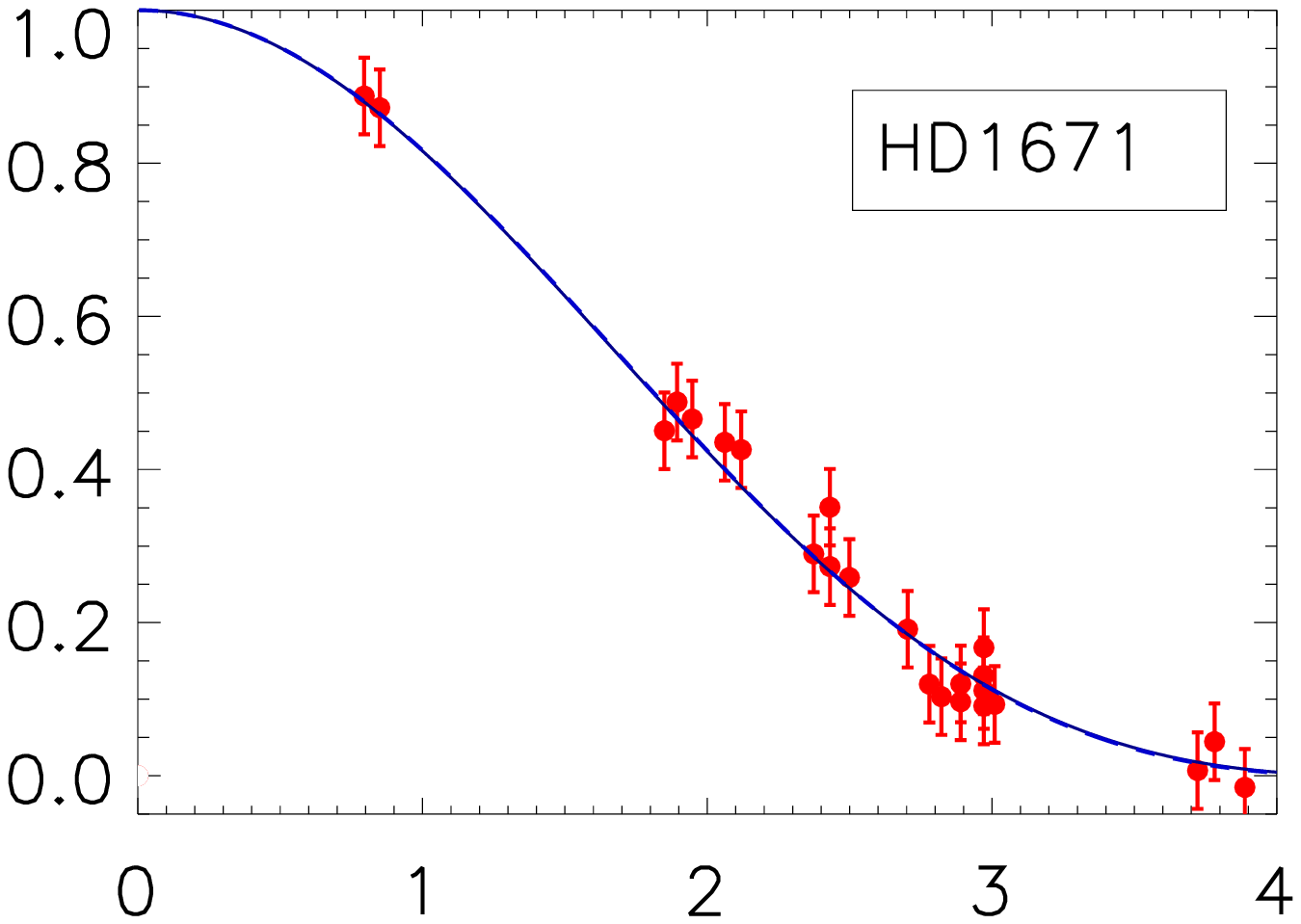}\\ 
\vspace*{-1.4cm}
\hspace*{1cm}
\includegraphics[scale=0.5]{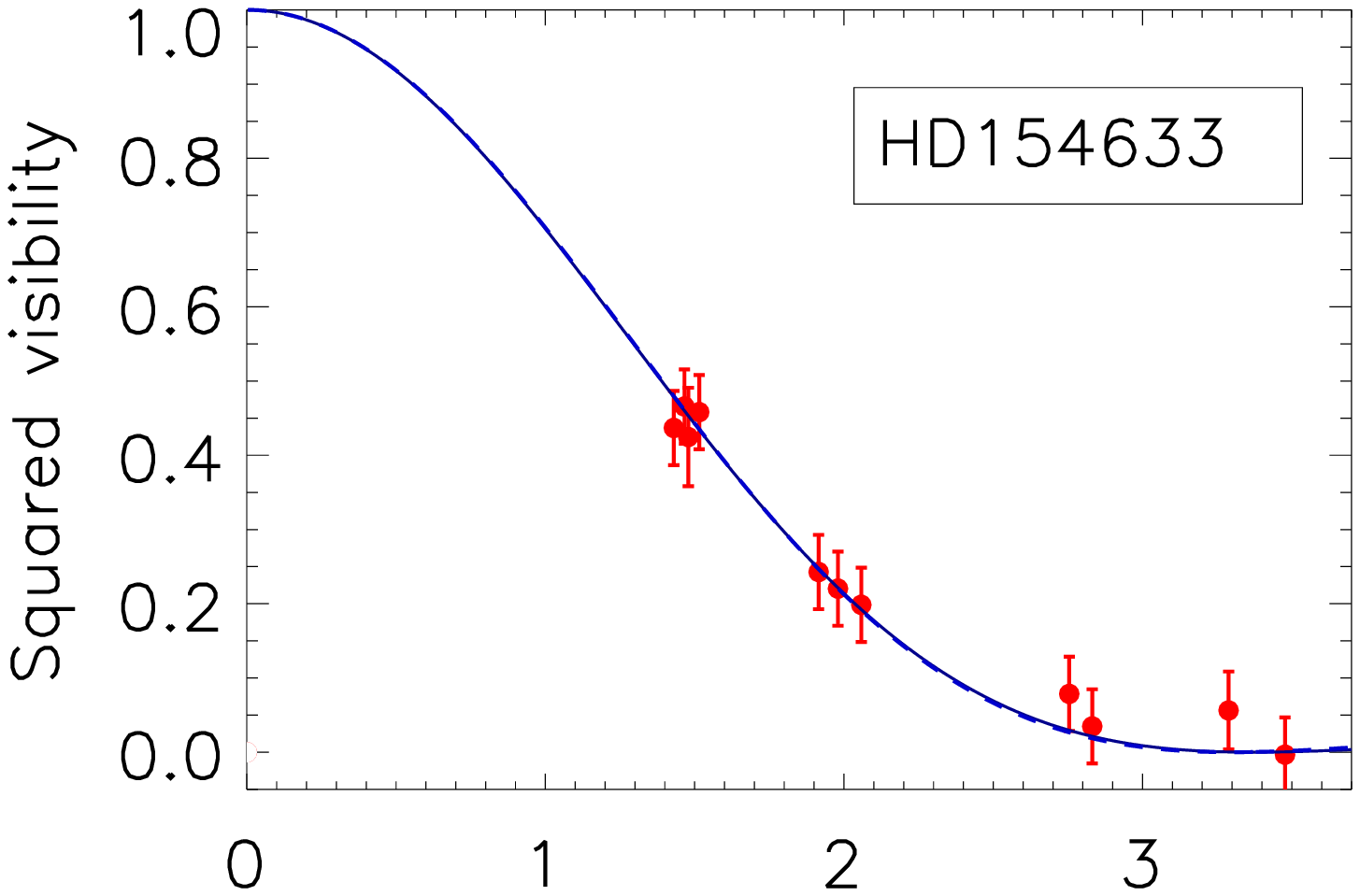}&
\hspace*{-2cm}
\includegraphics[scale=0.5]{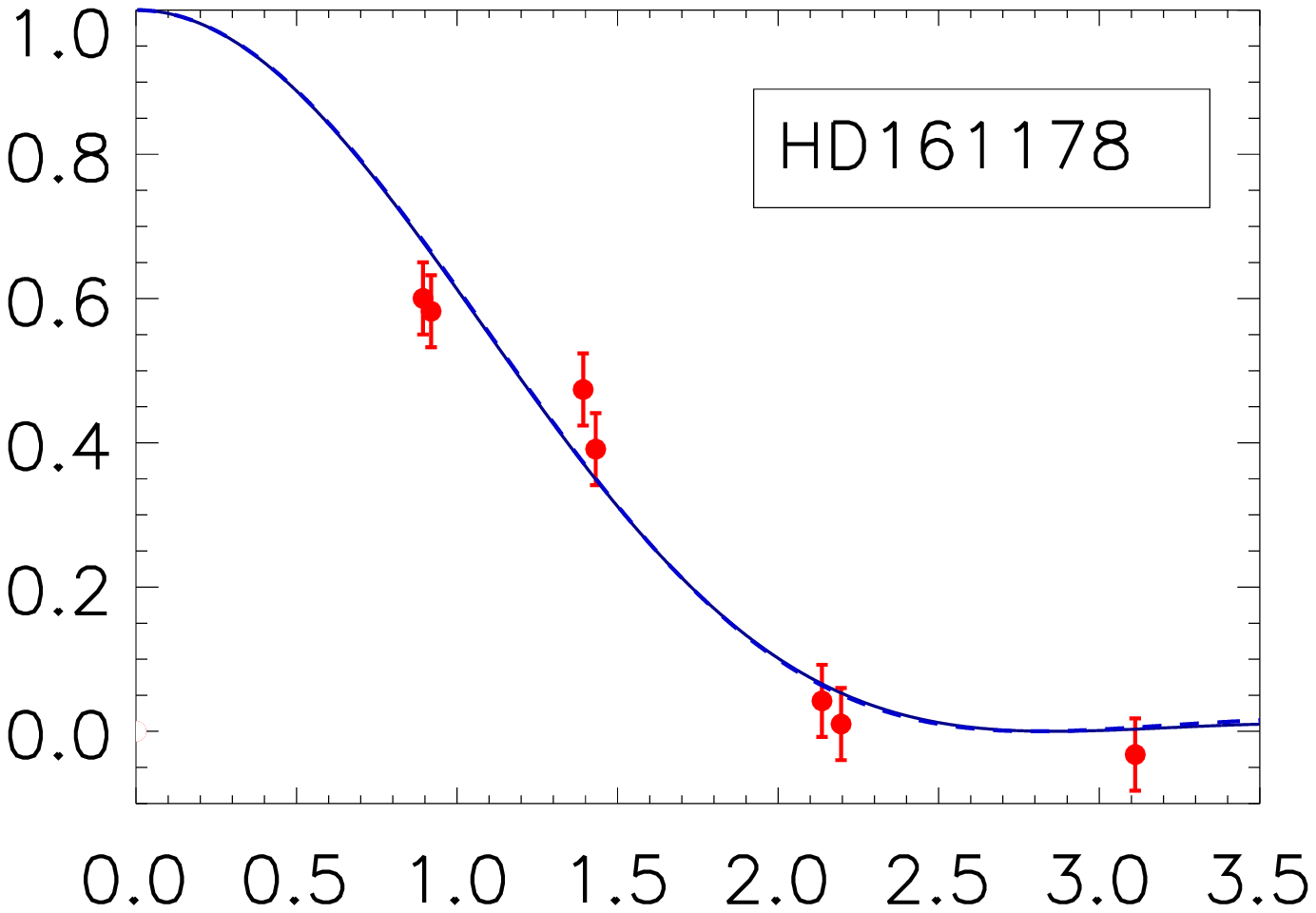}\\
\vspace*{-1.4cm}
\hspace*{1cm}
\includegraphics[scale=0.5]{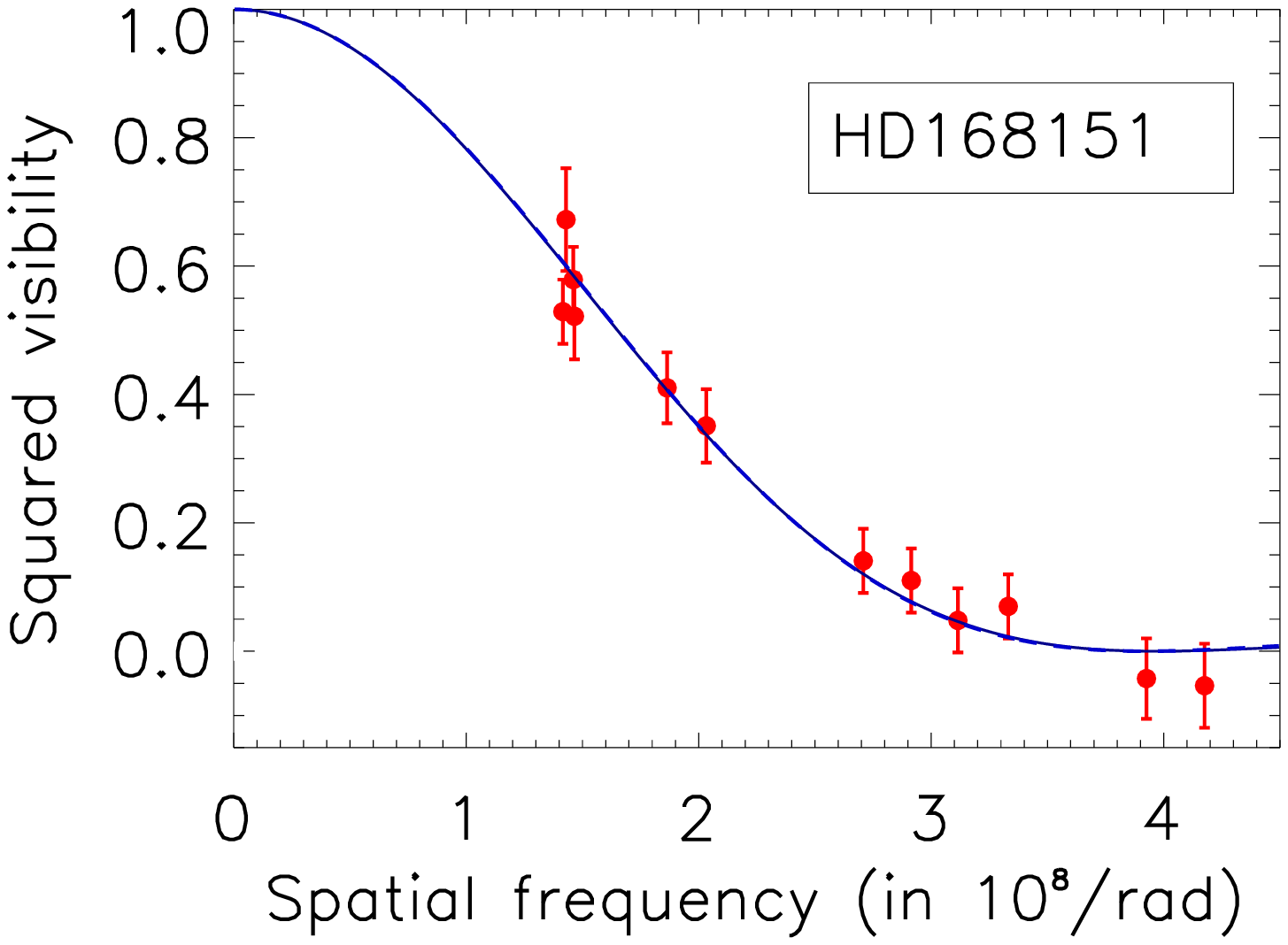}&
\hspace*{-2cm}
\includegraphics[scale=0.5]{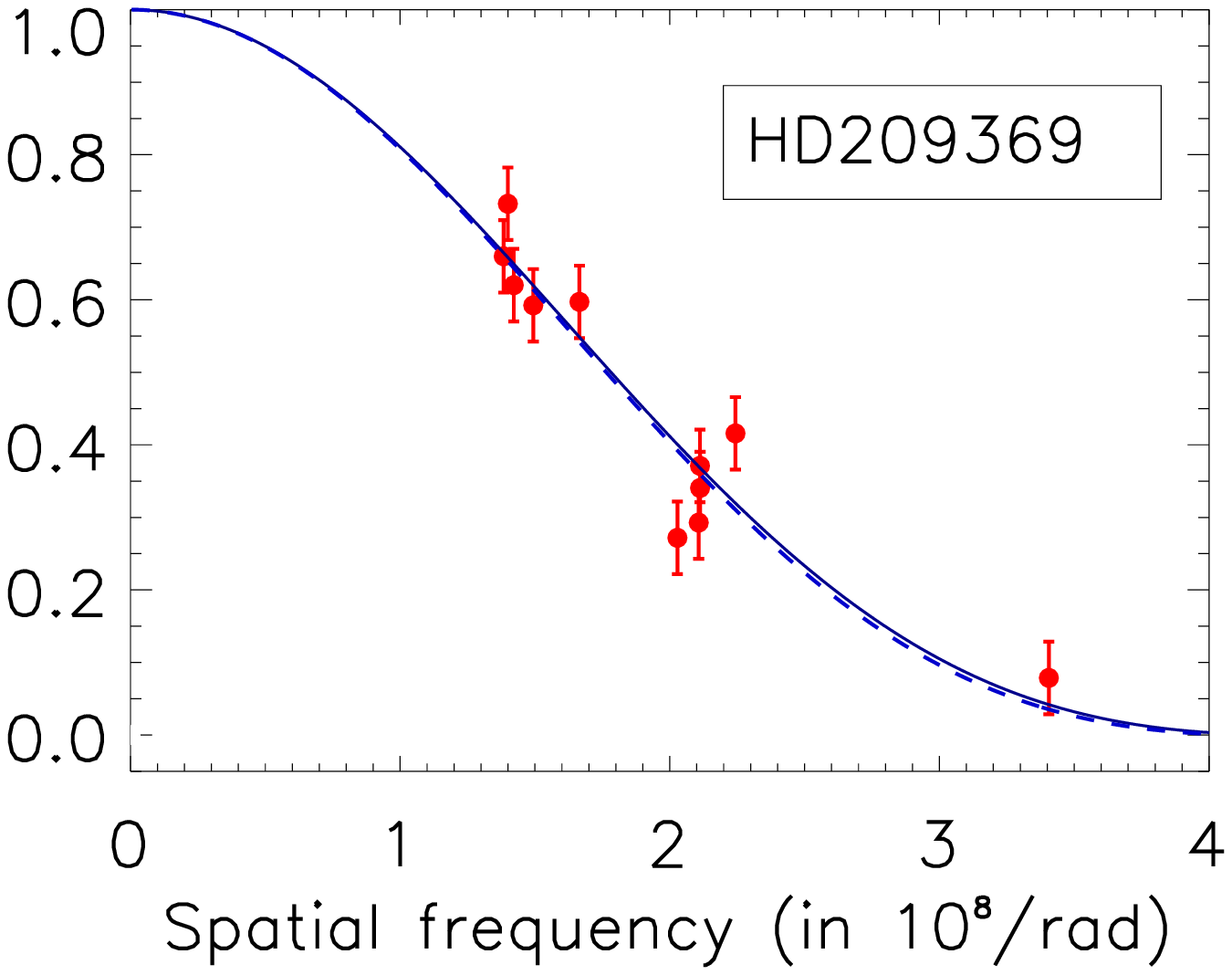}
\vspace*{0.4cm}
\end{array}$
\includegraphics[scale=0.5]{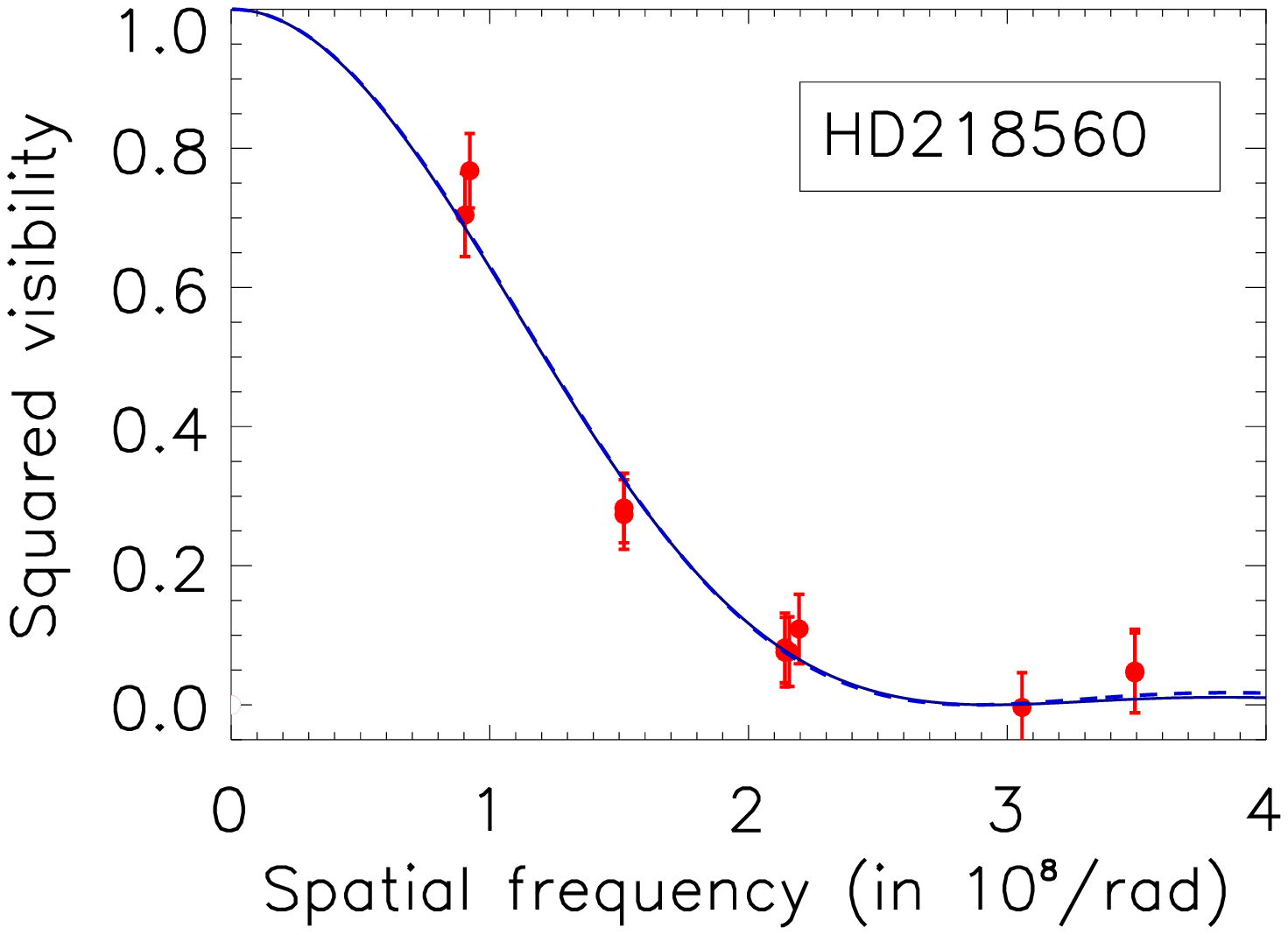}
\end{center}
\caption{Squared visibilities of single stars. The solid line represents the model of LD diameter and the dashed line the UD diameter (see Sect.~\ref{sec:Observations}).}
\label{fig:VisSingleStars}
\end{figure*}

\begin{figure*}[ht]
\centering
$\begin{array}{ccc}
\vspace*{-0.8cm}
\hspace*{-1.cm}
\includegraphics[scale=0.38]{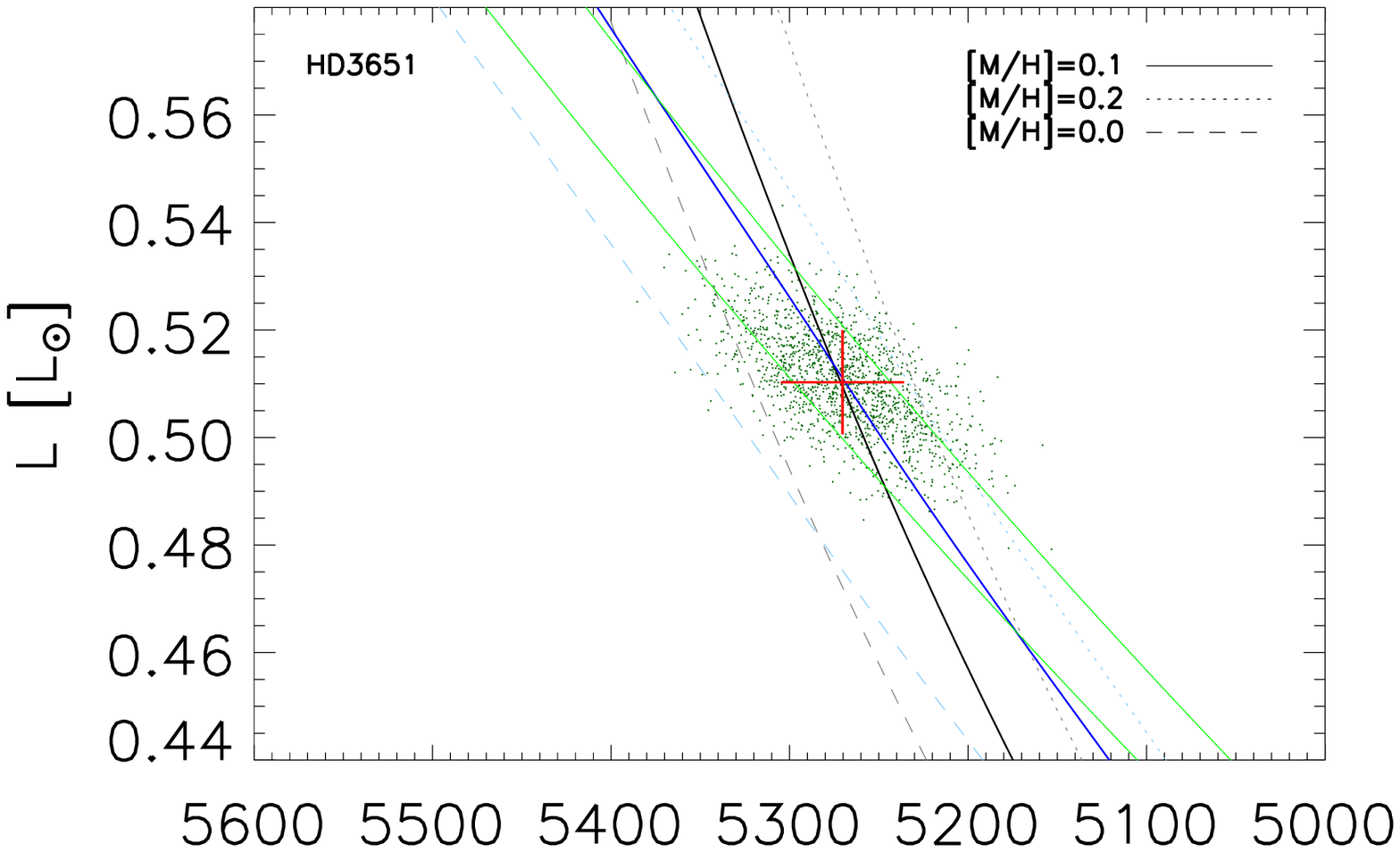} & 
\hspace*{-1.cm}
\includegraphics[scale=0.38]{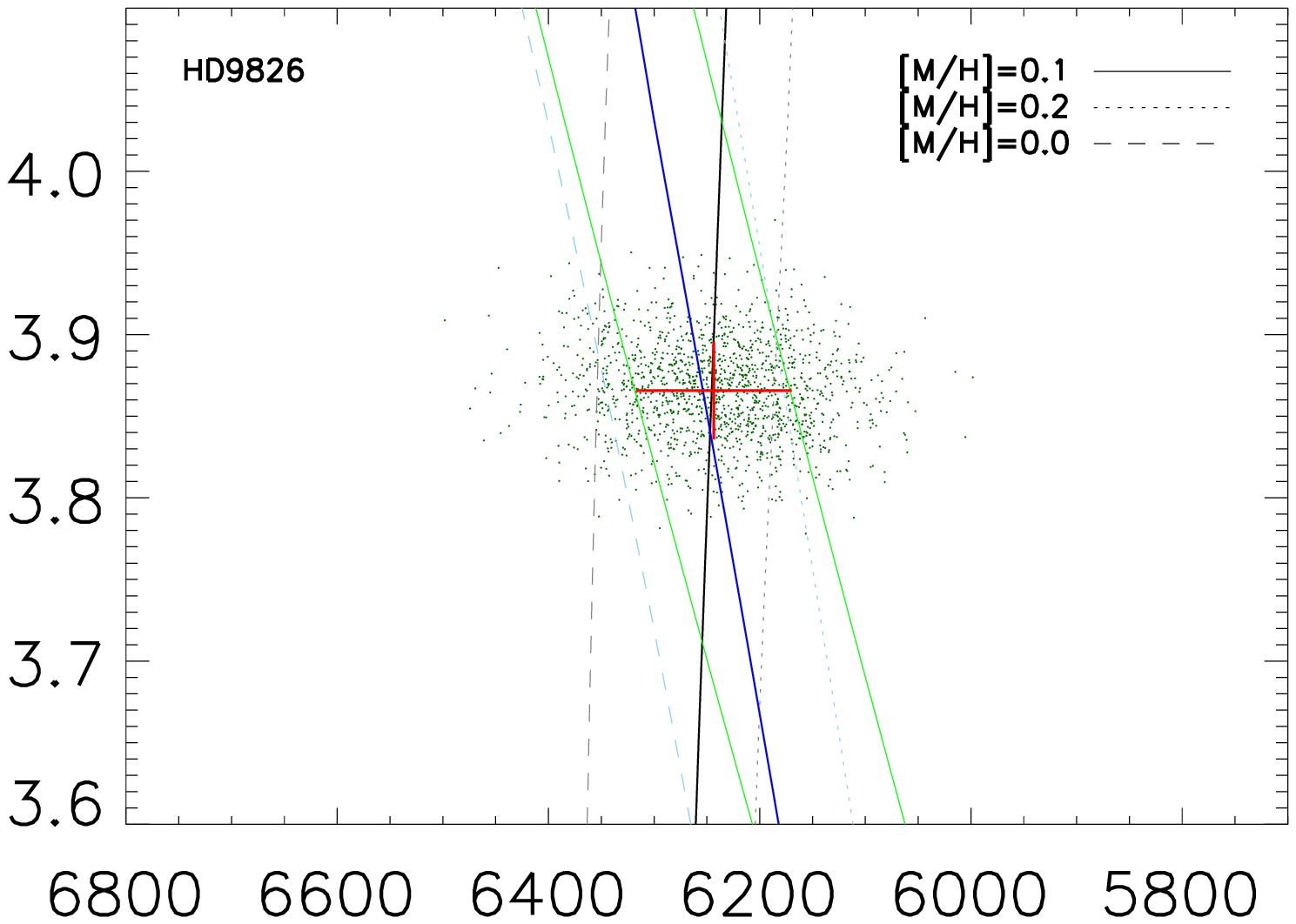} &
\hspace*{-1.cm}
\includegraphics[scale=0.38]{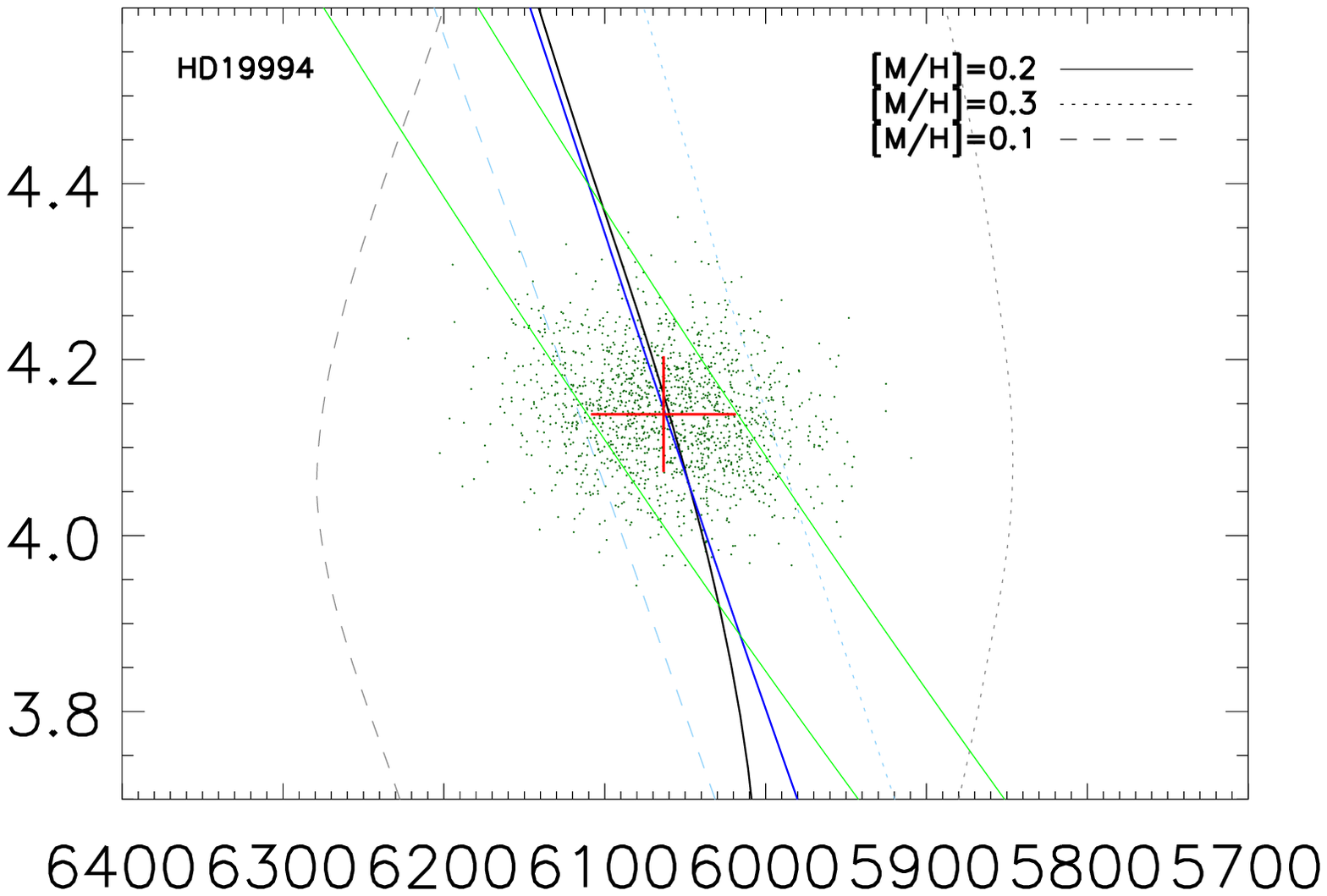} \\
\vspace*{-0.8cm}
\hspace*{-1.cm}
\includegraphics[scale=0.38]{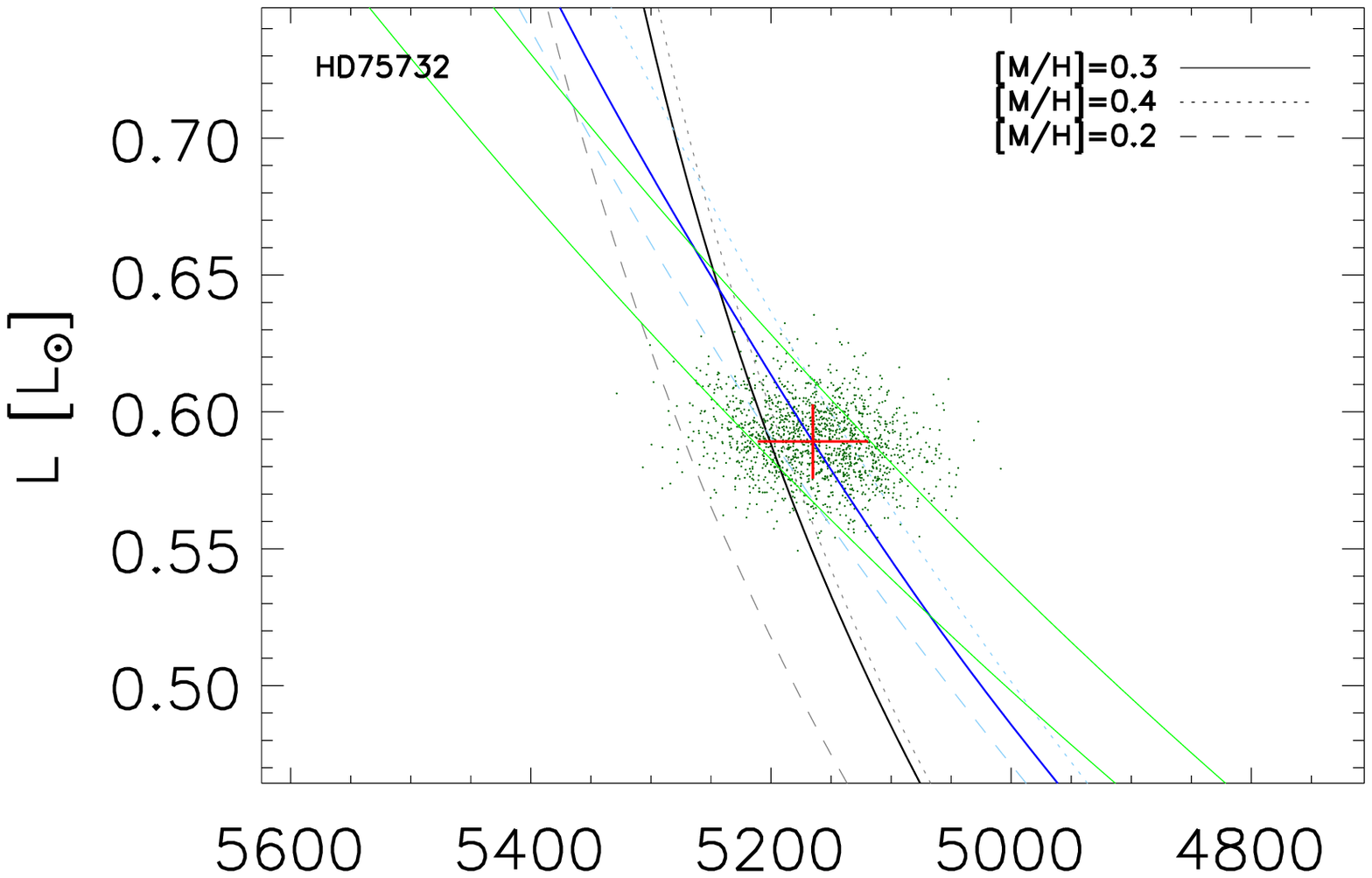} &
\hspace*{-1.cm}
\includegraphics[scale=0.38]{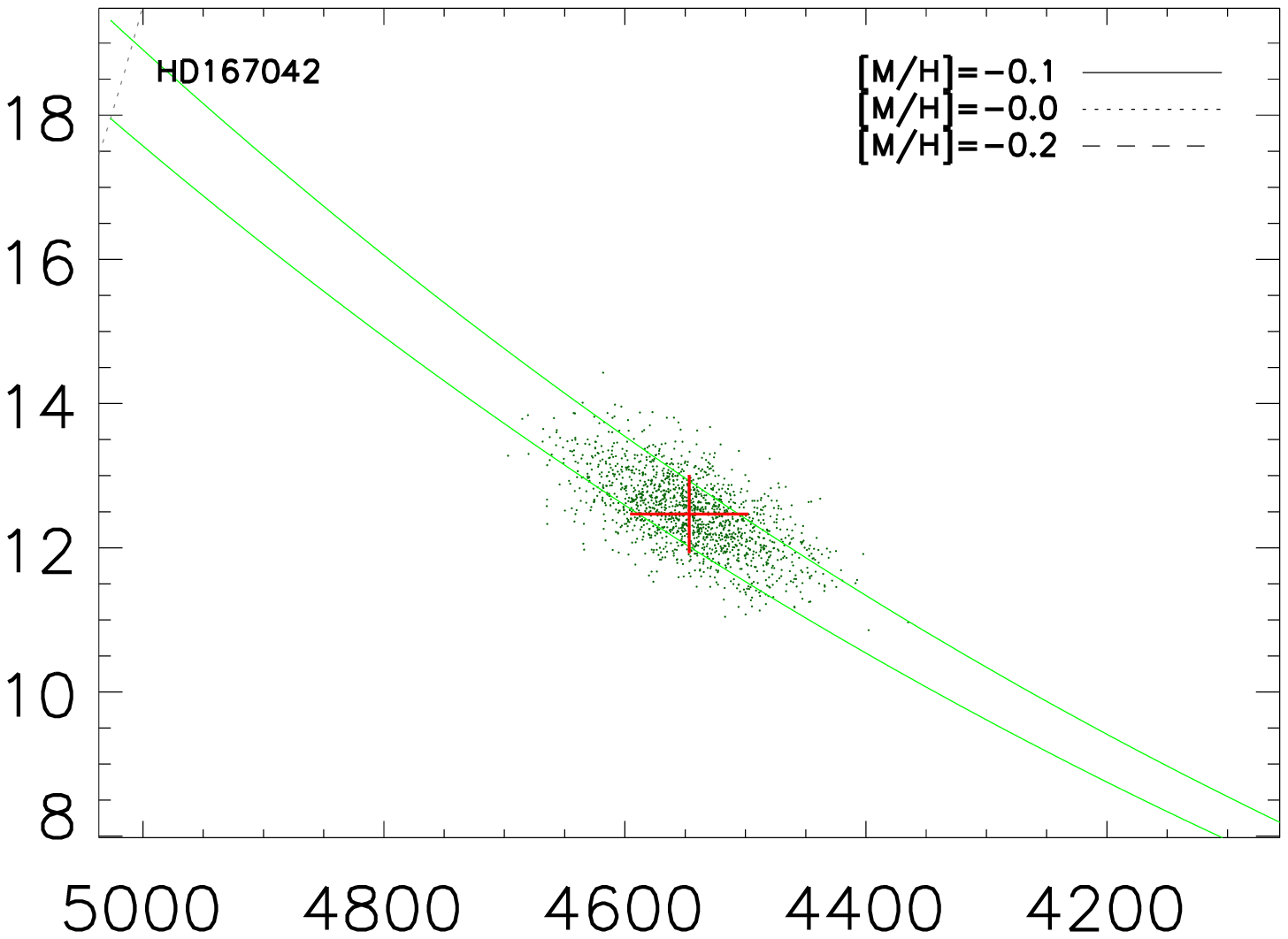} &
\hspace*{-1.cm}
\includegraphics[scale=0.38]{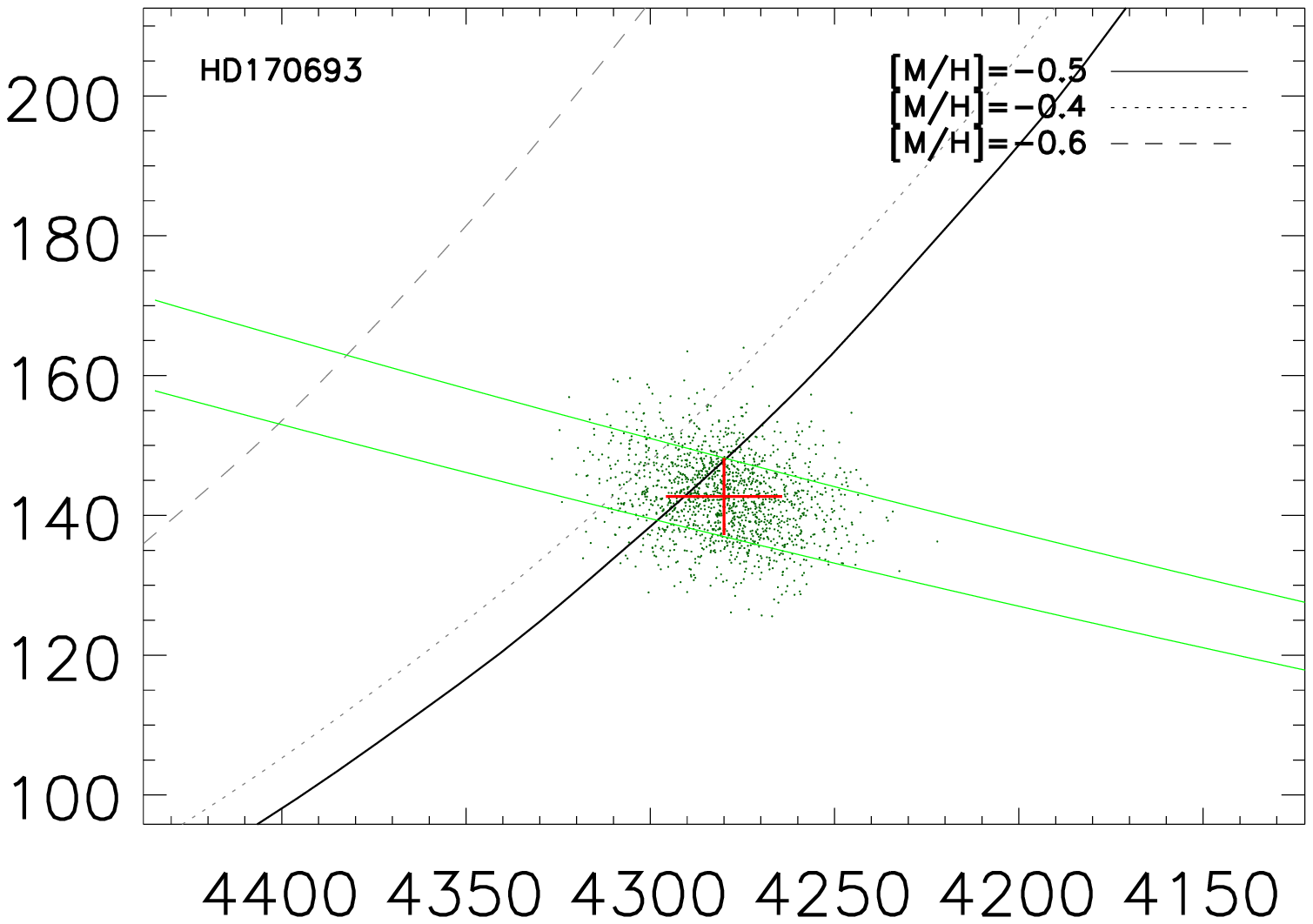} \\
\vspace*{-0.8cm}
\hspace*{-1.cm}
\includegraphics[scale=0.38]{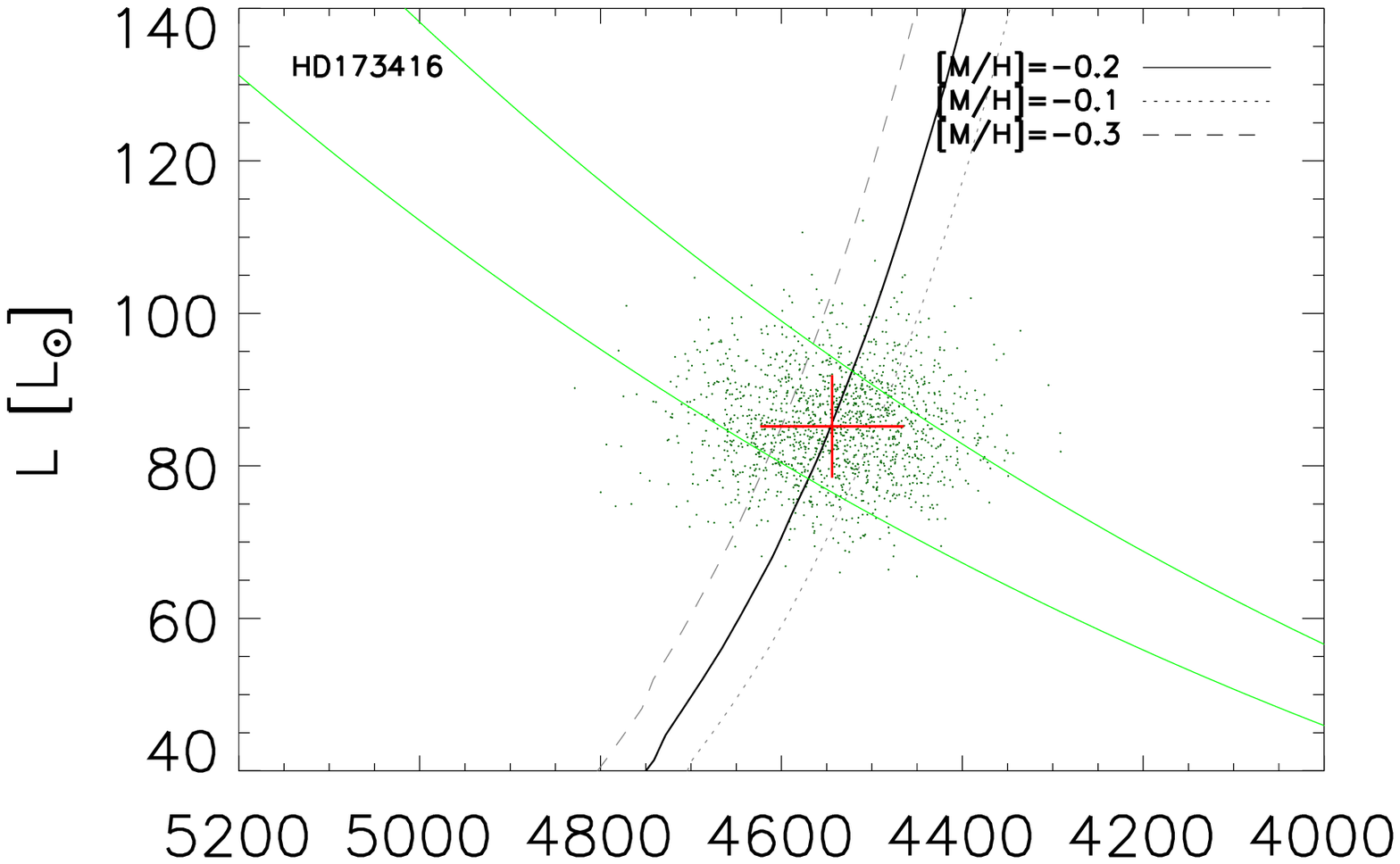} & 
\hspace*{-1.cm}
\includegraphics[scale=0.38]{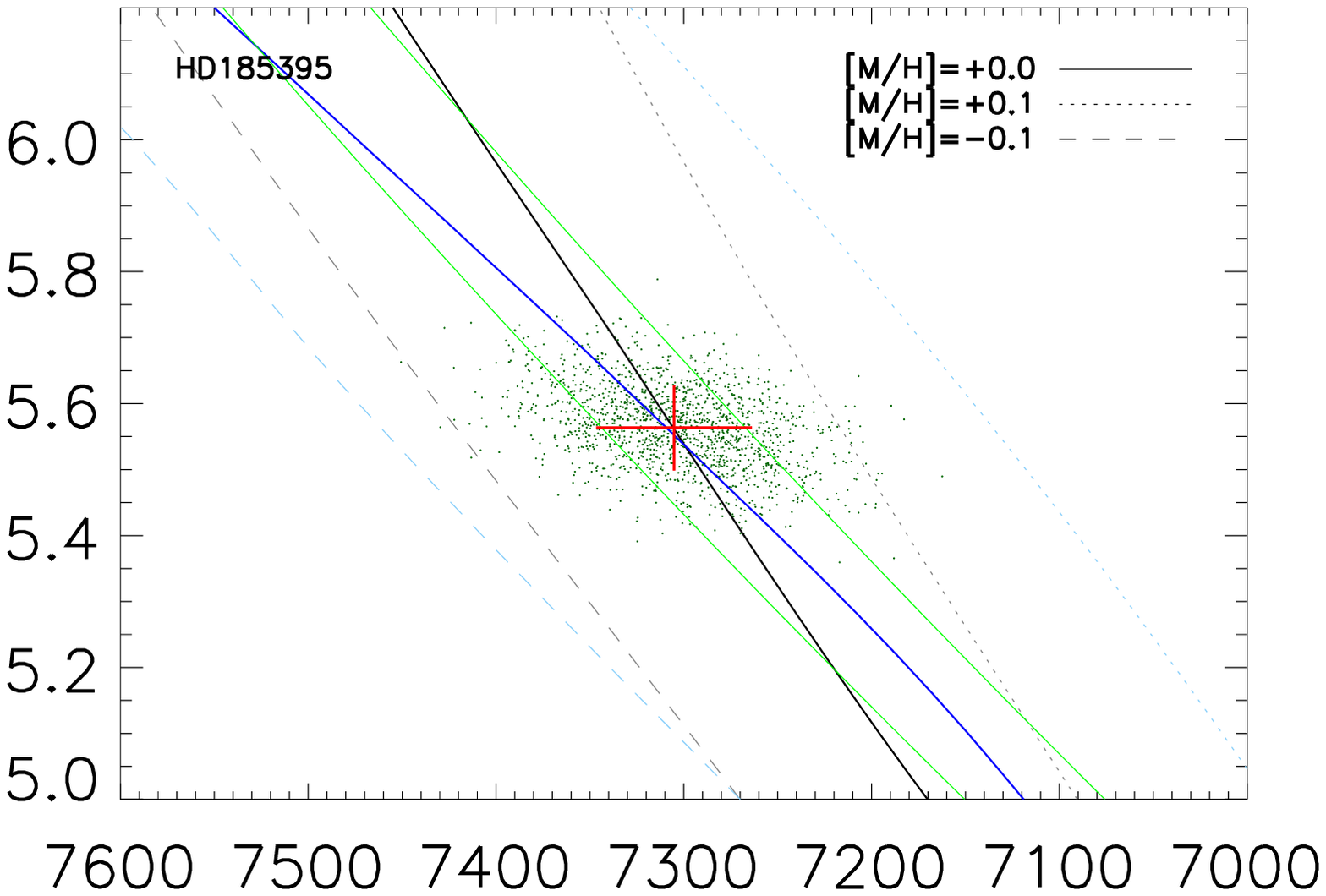} &
\hspace*{-1.cm}
\includegraphics[scale=0.38]{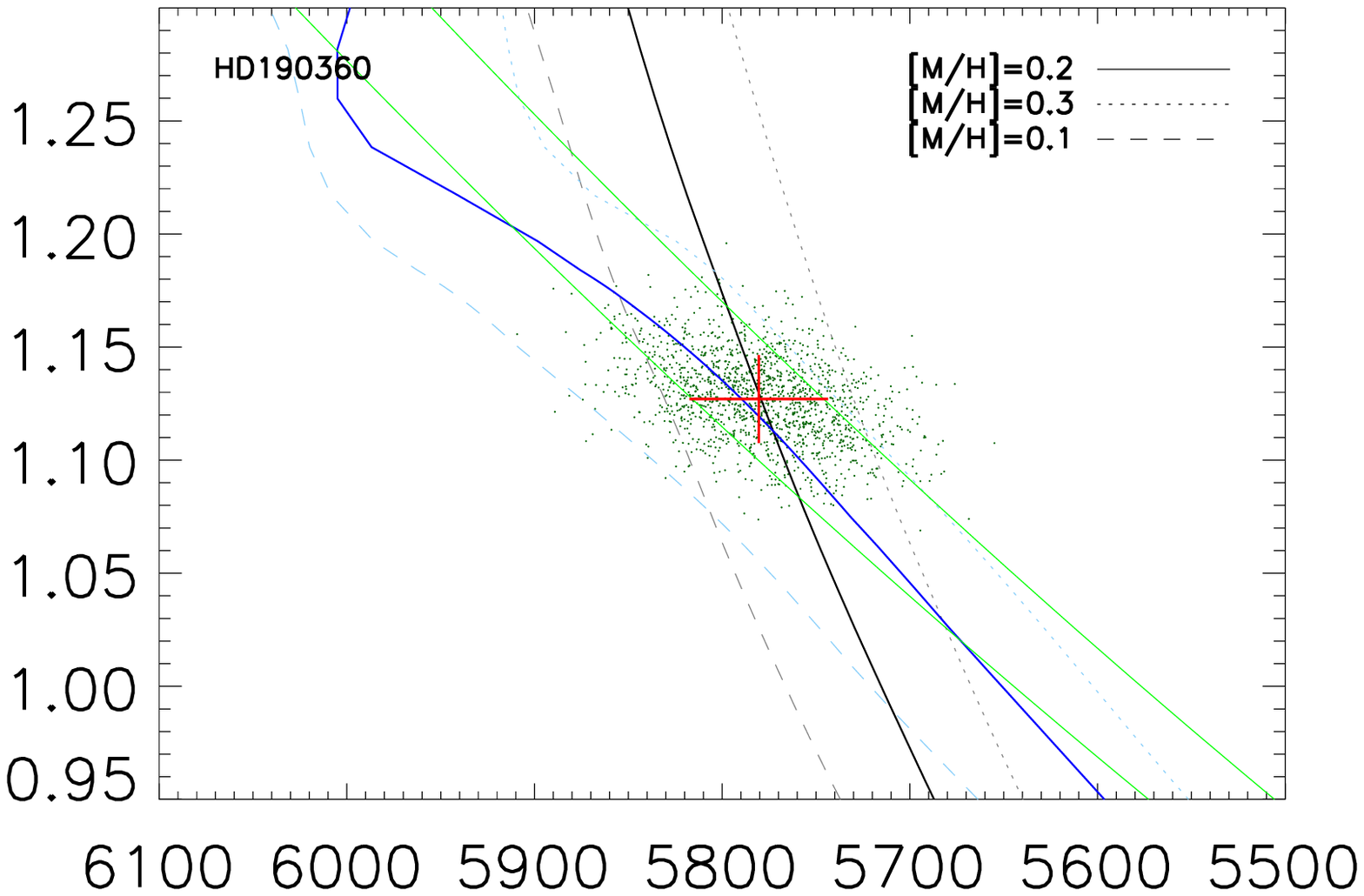} \\
\vspace*{-0.8cm}
\hspace*{-1.cm}
\includegraphics[scale=0.38]{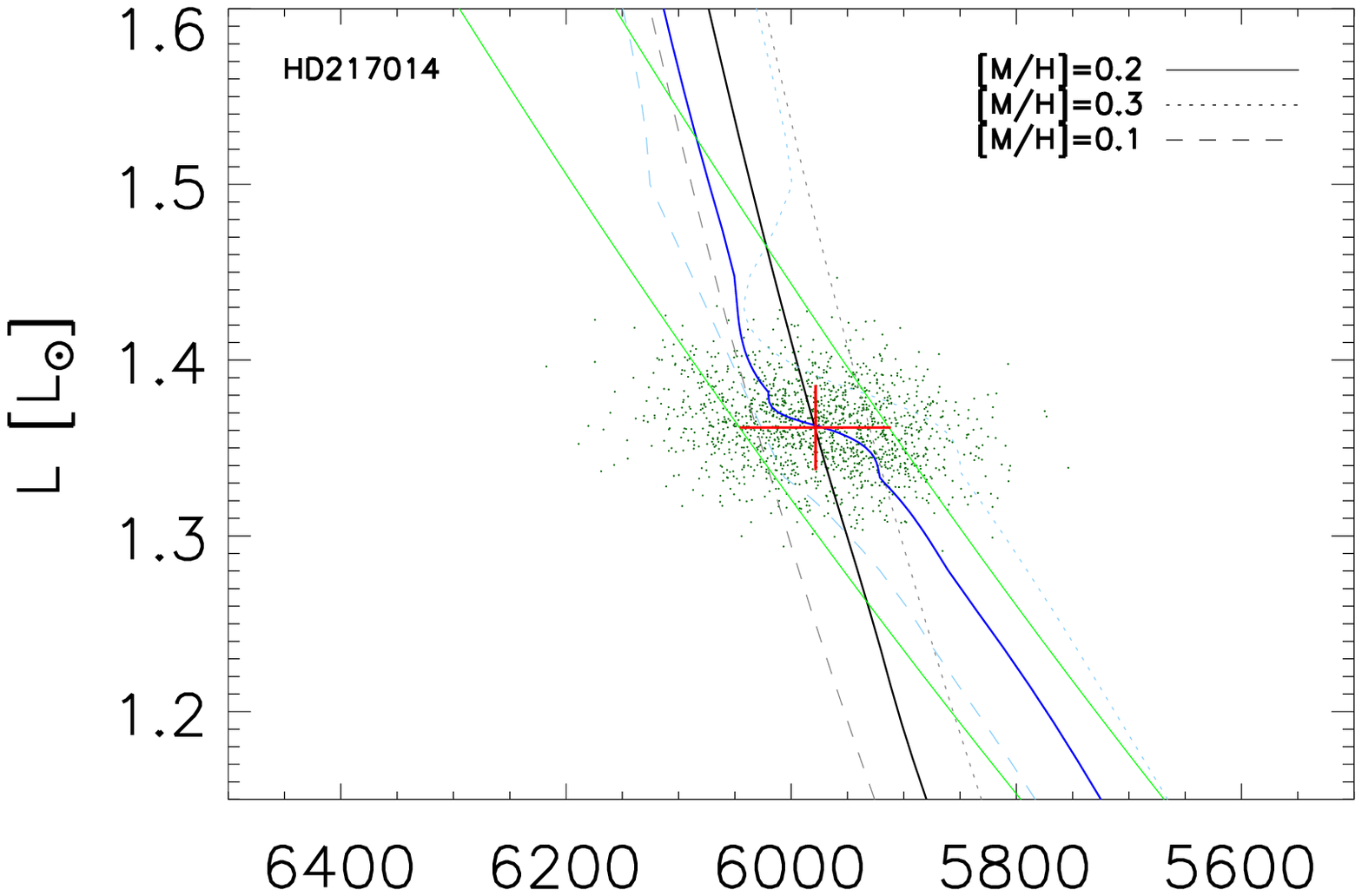} &
\hspace*{-1.cm}
\includegraphics[scale=0.38]{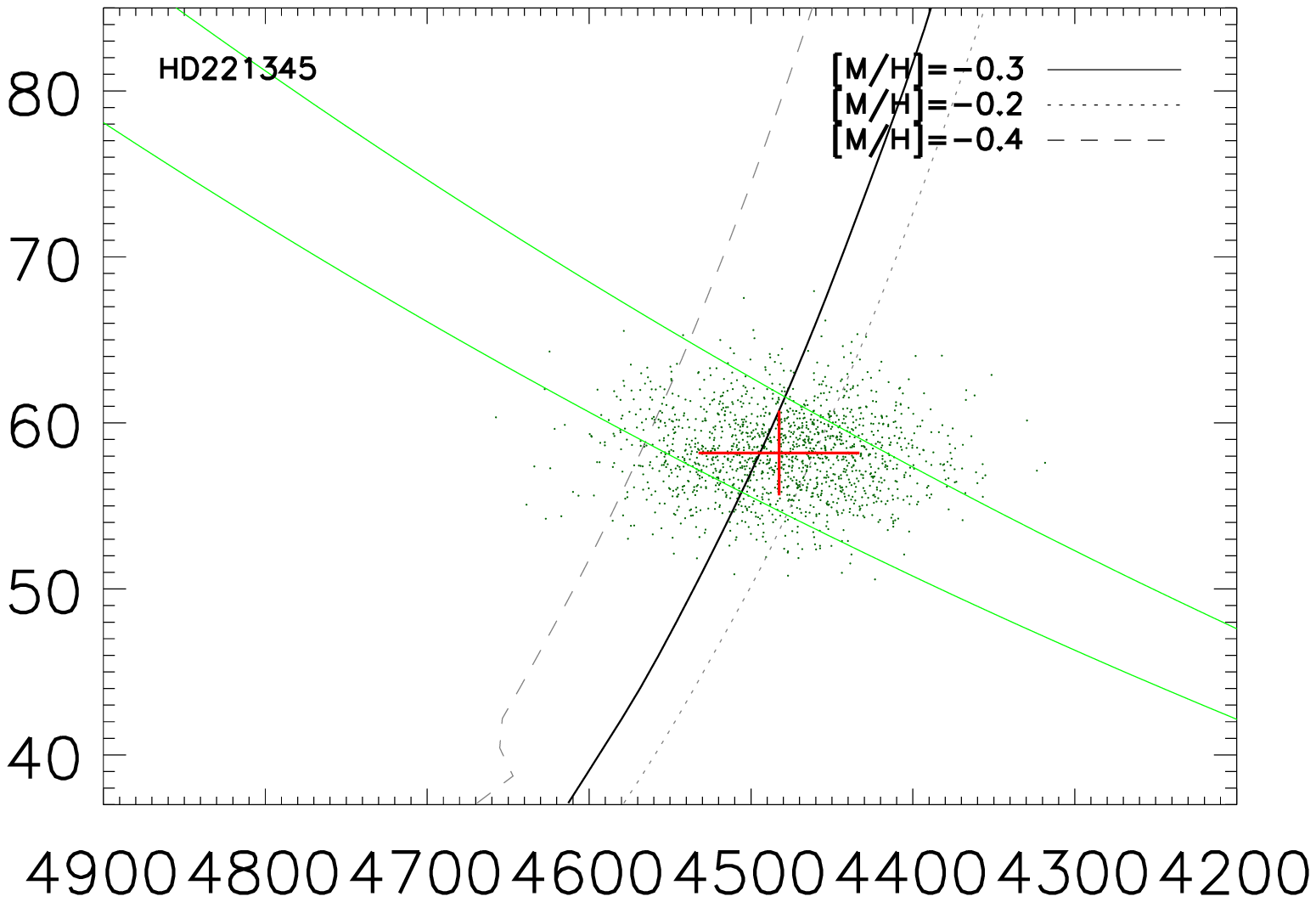} &
\hspace*{-1.cm}
\includegraphics[scale=0.38]{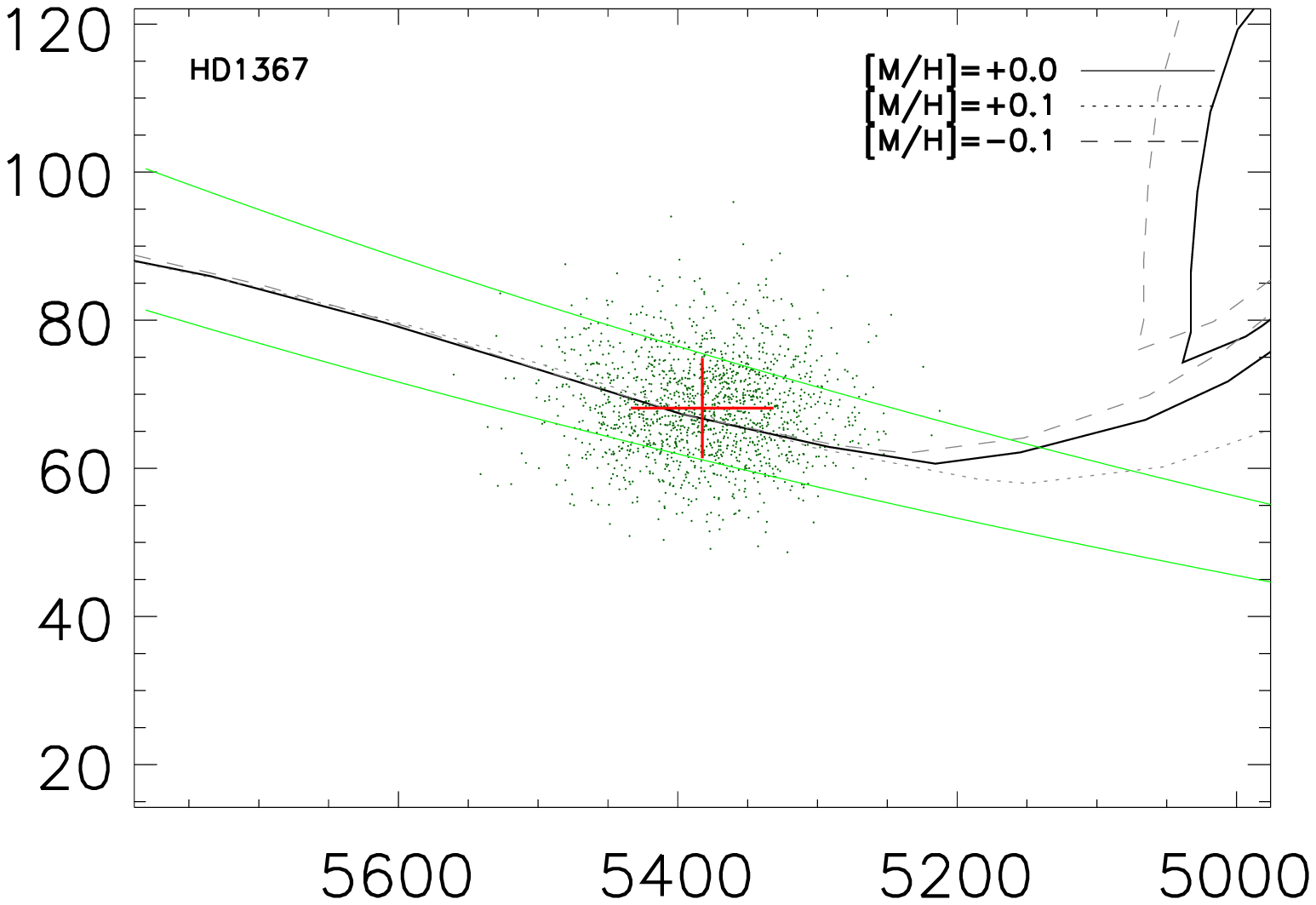} \\
\vspace*{-0.8cm}
\hspace*{-1.cm}
\includegraphics[scale=0.38]{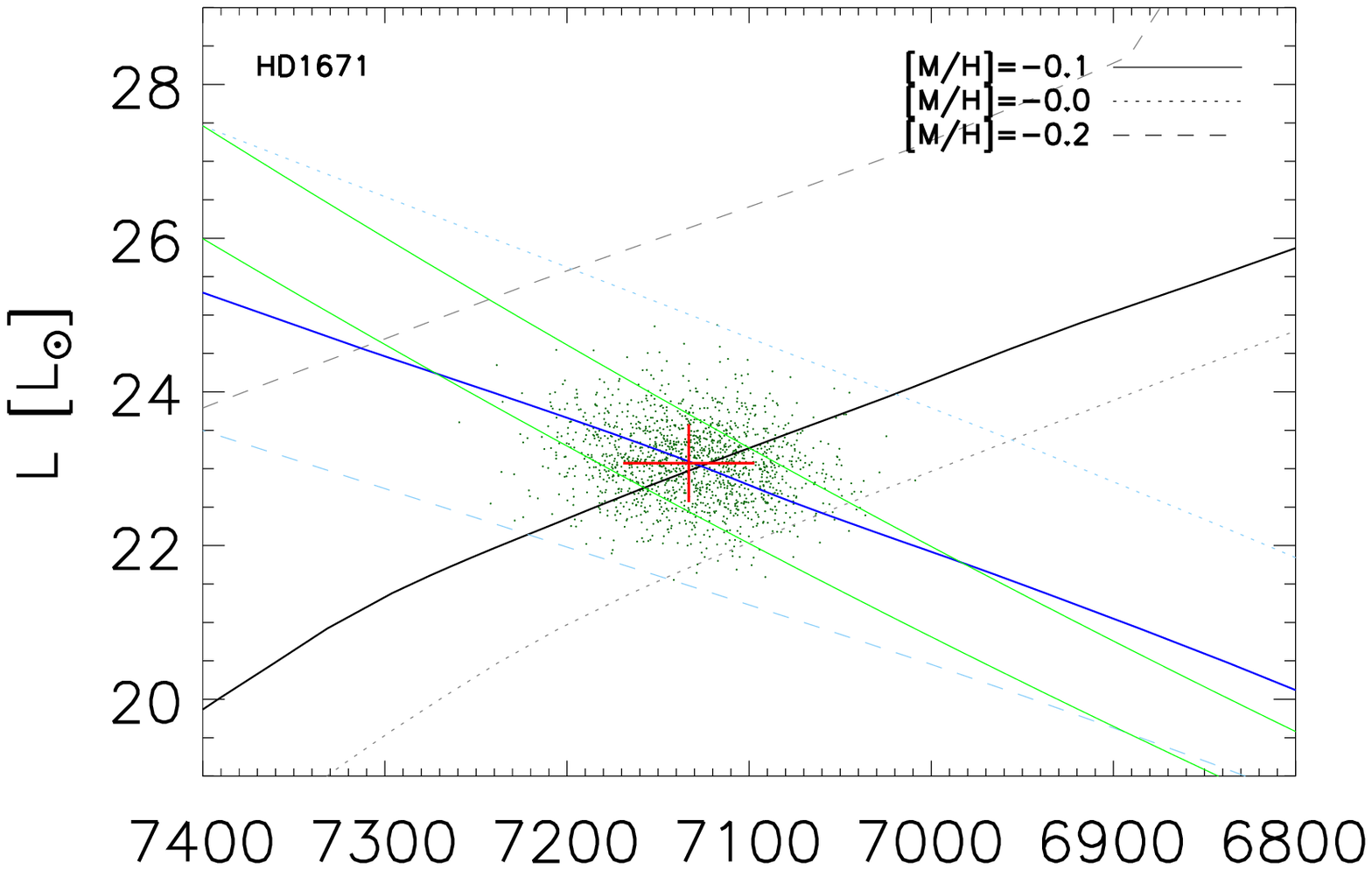} &
\hspace*{-1.cm}
\includegraphics[scale=0.38]{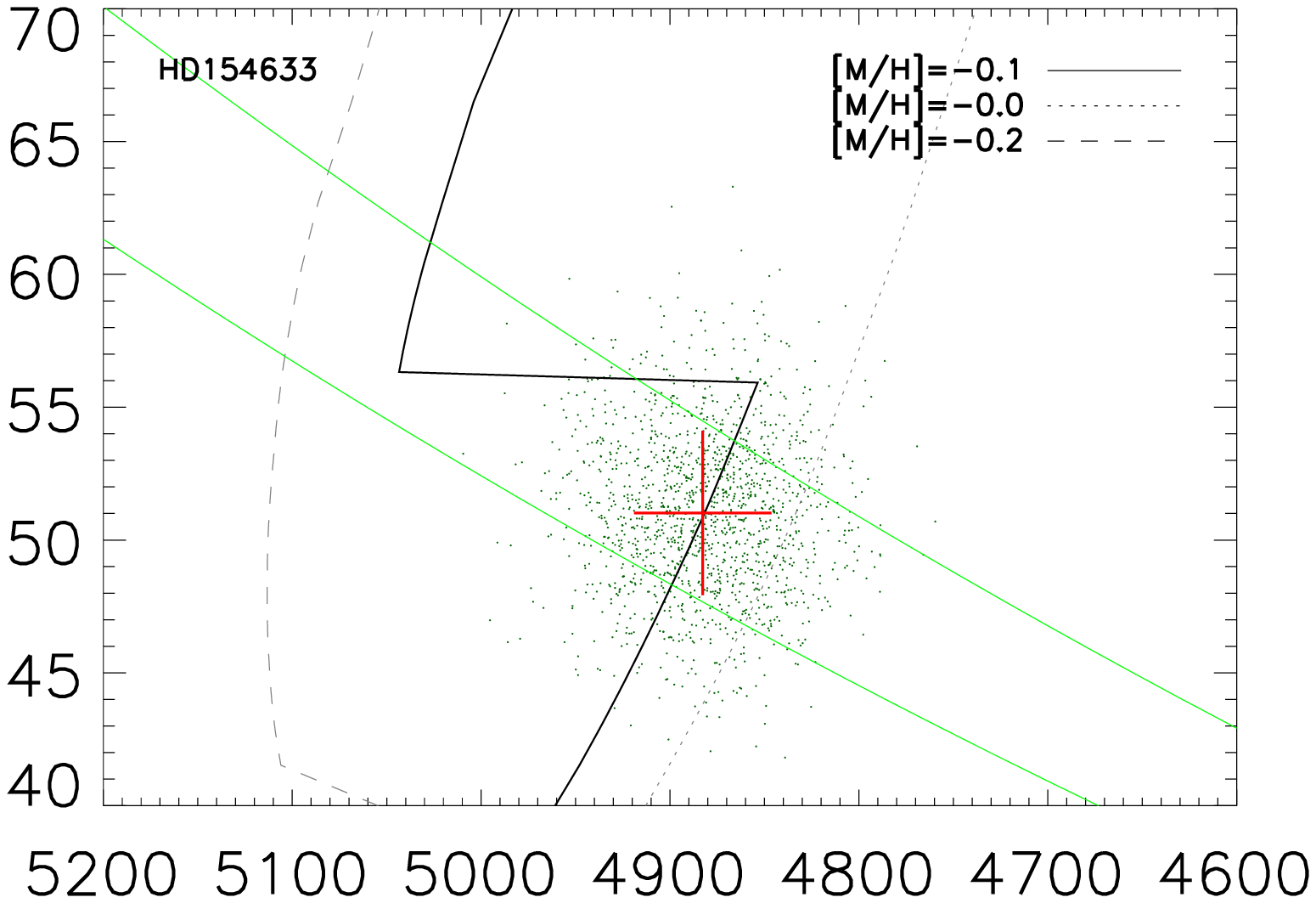} &
\hspace*{-1.cm}
\includegraphics[scale=0.38]{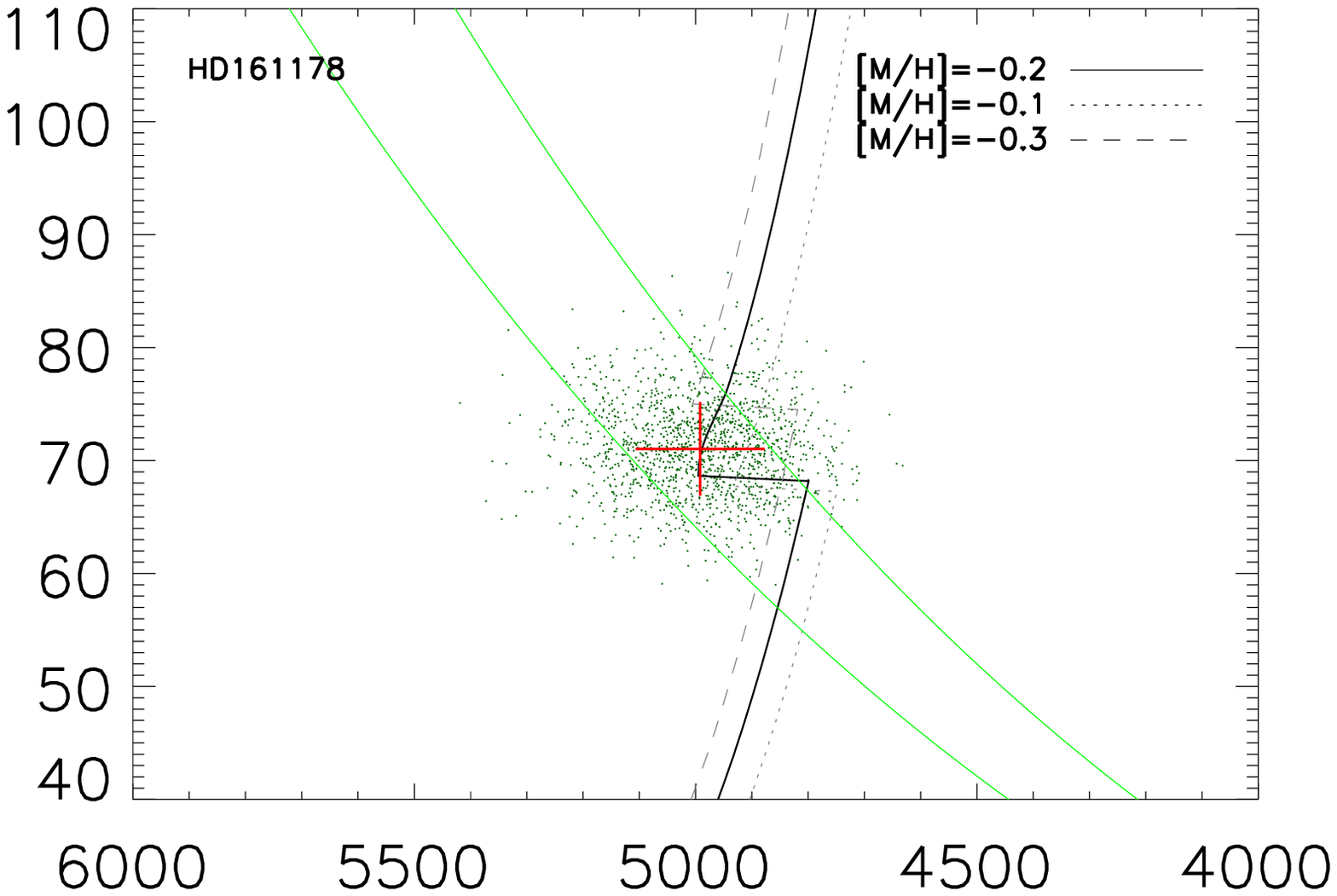} \\
\vspace*{-0.8cm}
\hspace*{-1.cm}
\includegraphics[scale=0.38]{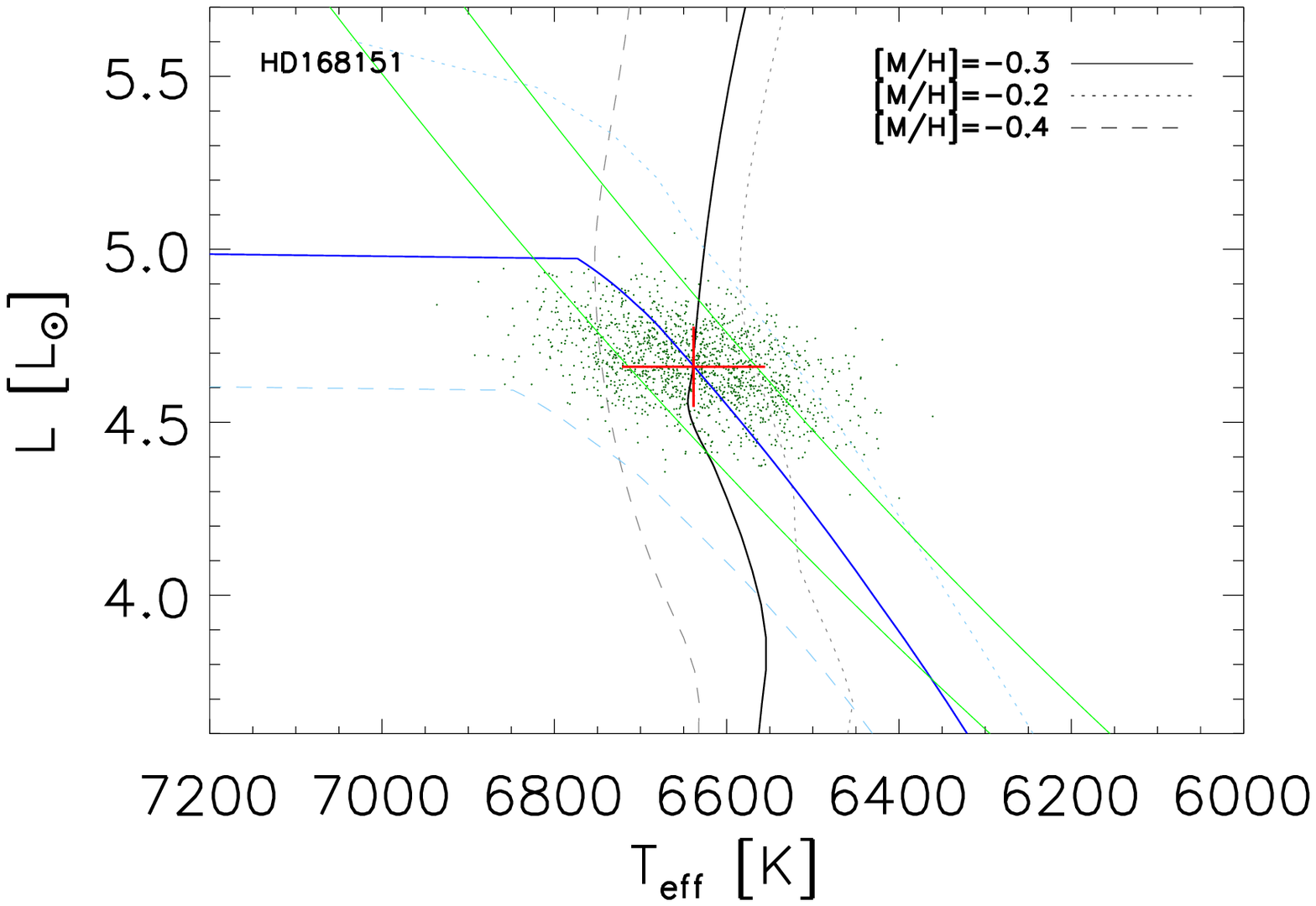} &
\hspace*{-1.cm}
\includegraphics[scale=0.38]{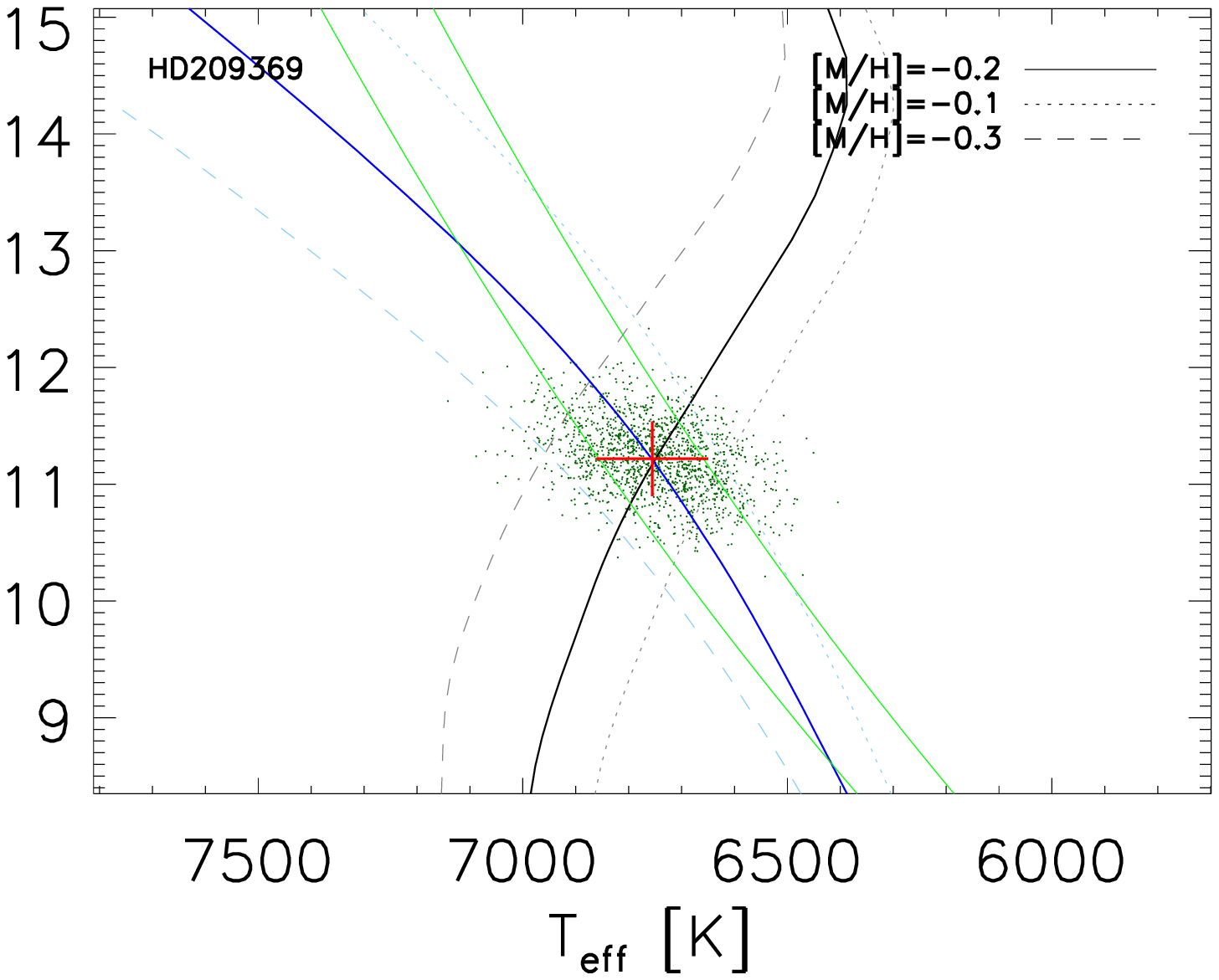} &
\hspace*{-1.cm}
\includegraphics[scale=0.38]{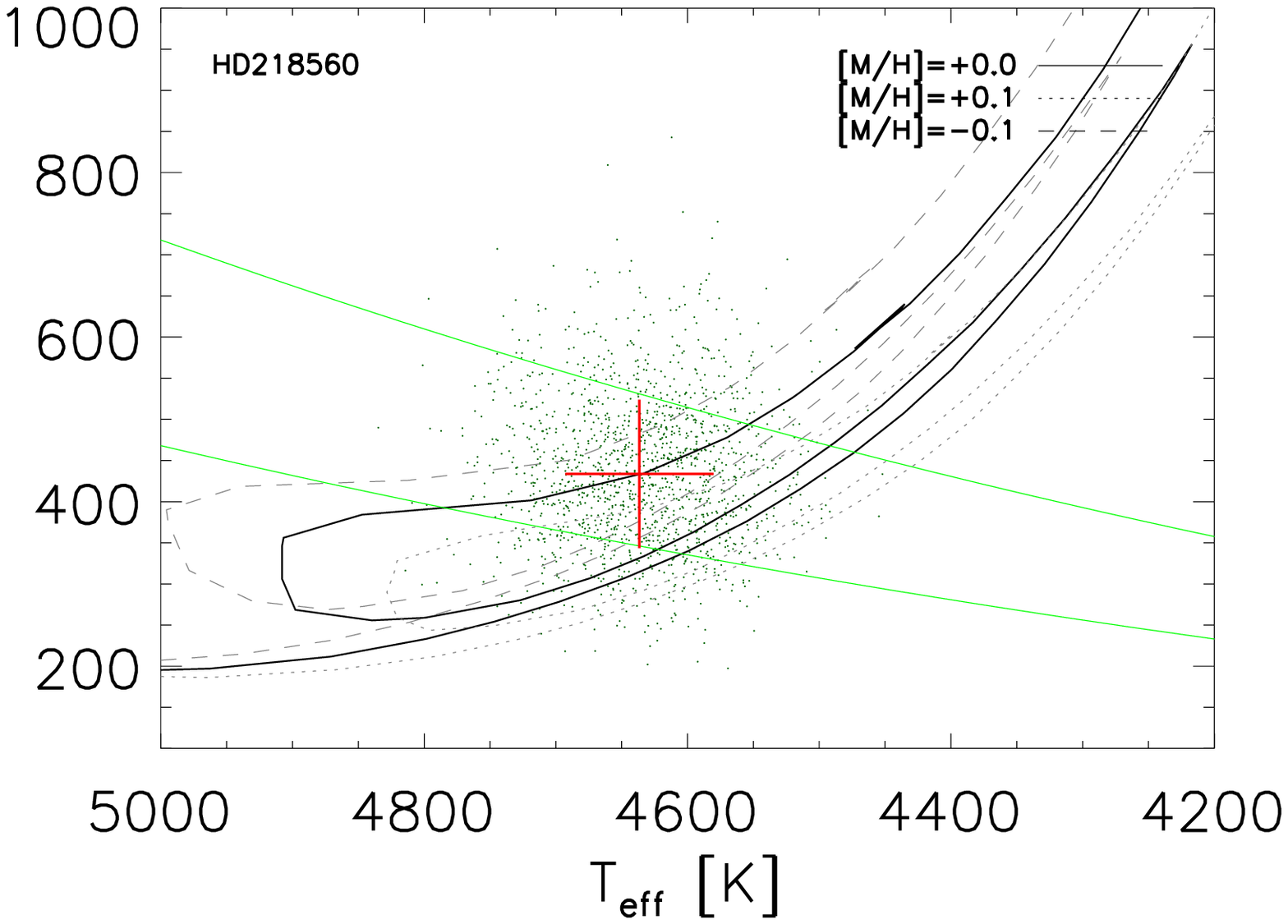} \\
\end{array}$
\vspace*{0.4cm}
\caption{Isochrones from PARSEC models. The clouds of points represent the MC distribution. The black lines are for old solutions and the blue lines for young solutions. The isochrones of the stellar age at metallicity of $\pm$0.1 dex are plotted in lighter colors than the reference (dotted and dashed lines). The solid light green lines represent the contraints of the radius at $1 \sigma$ (see Sect.\ref{sec:MassAge}).}
\label{fig:isochrones}
\end{figure*}

\begin{figure*}[ht]
\centering
\hspace*{-0.5cm}
\includegraphics[scale=1.1]{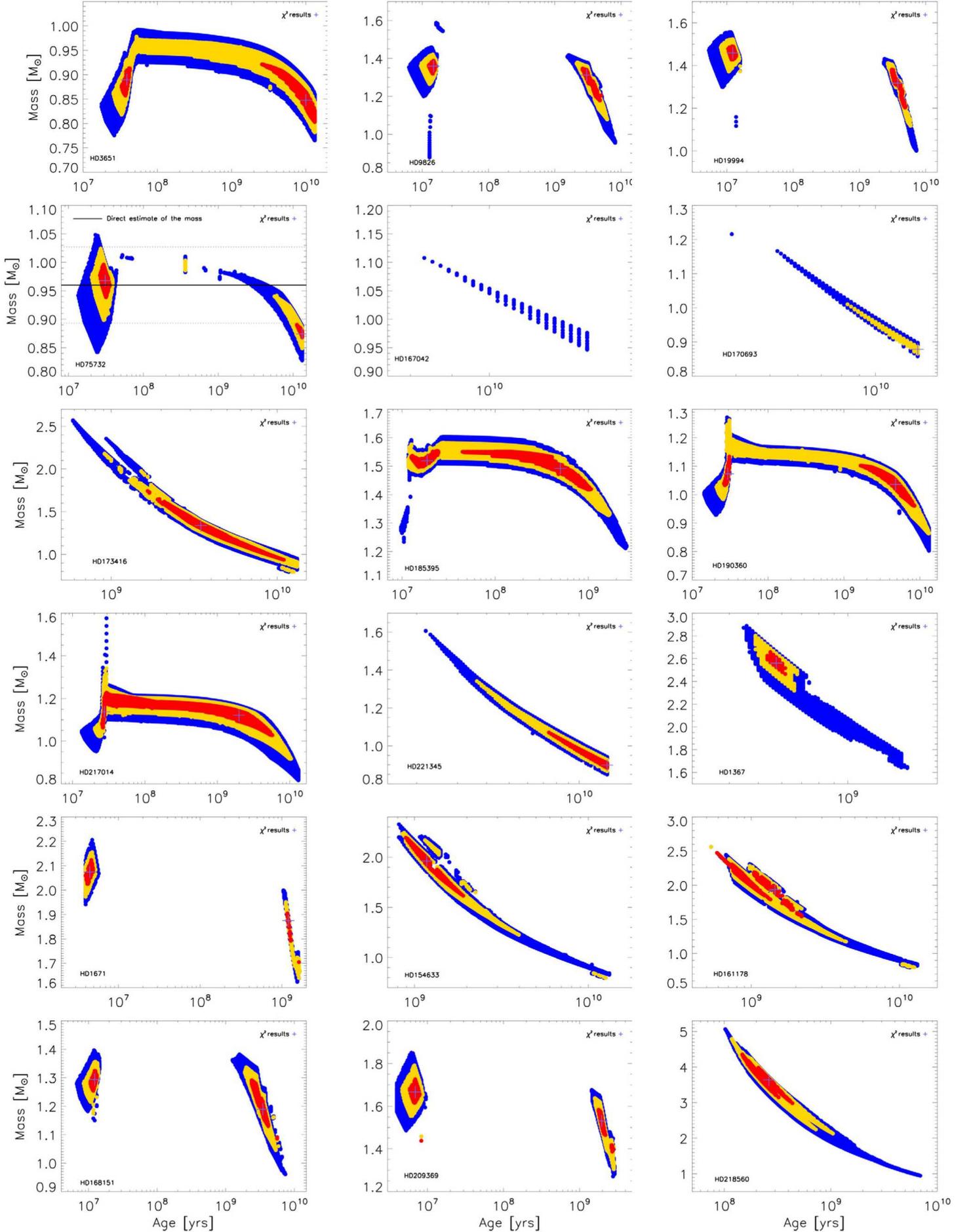}
\caption{ Approximate contour maps of the likelihood $\mathcal{L}$ as a function of the stellar masses and ages. Values within 1, 2, and 3 of each term of Eq.~\ref{eq:Chi2} appear in red, yellow, and blue, respectively. Considering a flat prior (which is reasonable), these maps show the joint PDF of $M_\star$ and the age. The best fit old and young (if any) solutions are represented by crosses. The solid horizontal line in the case of HD75732 corresponds to an independent estimate of the mass, and the dashed horizontal lines correspond to the errobars on $M_\star$ (see Sect.~\ref{sec:Discussion}).  }
\label{fig:patates}
\end{figure*}

\end{appendix}
\end{document}